\documentclass[fleqn,usenatbib]{mnras}

\usepackage[T1]{fontenc}

\usepackage{graphicx}	
\usepackage{amsmath}	
\usepackage{amssymb}	
\usepackage[dvipsnames]{xcolor} 

\usepackage{newtxtext,newtxmath}

\def\msun{\hbox{M$_\odot$}}

\title[Haro 11 - Untying the knots of the nuclear starburst]{Haro 11 - Untying the knots of the nuclear starburst}

\author[M. Sirressi et al.]{
M. Sirressi$^{1}$\thanks{E-mail: mattia.sirressi@astro.su.se},
A. Adamo$^{1}$,
M. Hayes$^{1}$,
A. Bik$^{1}$,
M. Strandänger$^{1}$,
A. Runnholm$^{1}$,
M. S. Oey$^{2}$,
G. Östlin$^{1}$,
\newauthor{
V. Menacho$^{1}$,
L. J. Smith$^{3}$
}
\\
$^{1}$Department of Astronomy and Oskar Klein Centre; Stockholm University; AlbaNova 106 91 Stockholm, Sweden\\
$^{2}$Department of Astronomy, University of Michigan, 1085 South University Ave., Ann Arbor, MI 48109-1107, USA\\
$^{3}$European Space Agency (ESA), ESA Office, Space Telescope Science Institute, 3700 San Martin Drive, Baltimore, MD 21218, USA\\
}

\date{Accepted 2021 December 22. Received 2021 December 16; in original form 2021 October 13}

\pubyear{2021}

\begin{document}
\label{firstpage}
\pagerange{\pageref{firstpage}--\pageref{lastpage}}
\maketitle

\begin{abstract}
Star formation is a clustered process that regulates the structure and evolution of galaxies. We investigate this process in the dwarf galaxy Haro 11, forming stars in three knots (A, B, C). The exquisite resolution of HST imaging allows us to resolve the starburst into tens of bright star clusters. We derive masses between $10^5$ and $10^7\,\rm M_{\odot}$ and ages younger than 20 Myr, using photometric modeling. We observe that the clustered star formation has propagated from knot C (the oldest) through knot A (in between) towards knot B (the youngest). We use aperture-matched ultraviolet and optical spectroscopy (HST + MUSE) to independently study the stellar populations of Haro 11 and determine the physical properties of the stellar populations and their feedback in 1 kpc diameter regions.
We discuss these results in light of the properties of the ionised gas within the knots. We interpret the broad blue-shifted components of the optical emission lines as outflowing gas ($v_{max} \sim 400$ km/s). The strongest outflow is detected in knot A with a mass-rate of $\dot{M}_{out}\sim 10\,\rm M_{\odot}/yr$, ten times higher than the star-formation in the same region. Knot B hosts a young and not fully developed outflow, whereas knot C has likely been already evacuated.
Because Haro 11 has properties similar to high-redshift unresolved galaxies, our work can additionally aid the understanding of star formation at high redshift, a window that will be opened by upcoming facilities.

\end{abstract}




\section{Introduction}


One of the fundamental questions in the field of extragalactic astronomy is what regulates the evolution of the galaxies in the Universe? Many observational and theoretical studies \citep{krumholz2014} have indicated that the interplay between star formation and feedback has an important role in regulating both the structure and the evolution of galaxies. Numerical simulations \citep{kim2017} find that without the action of feedback (i.e. momentum injections into the interstellar medium), galaxies would form stars at unrealistic rates. In particular, massive stars are the most powerful source of stellar feedback and the majority of them (>=70\%) are found in clusters and OB associations \citep{oey2004}. For this reason, young star clusters are the most relevant sources to constrain and evaluate with direct observations the impact of feedback on the host galaxy and in particular on its interstellar medium (ISM).

Haro 11 is one of the most extreme starburst galaxies known in the nearby Universe (z = 0.021, this work) with a young cluster population (1-20 Myr, this work) and a gas consumption timescale of only 50 Myr \citep{oestlin2021}. The central starburst in this galaxy consists of a number of young star clusters, grouped in the three starburst knots called A, B and C, that were presented in e.g. \citet{vader1993,kunth2003,adamo2010} (hereafter AA10) using multi-band photometry. The spectral energy distribution (SED) fitting models revealed cluster ages between one and tens of Myr and masses between $10^4 - 10^7 \rm M_{\odot}$. Haro 11 is, at the same time, the closest confirmed Lyman continuum leaker in the local Universe \citep{bergvall2006,leitet2011} and it is particularly interesting for its similarity to high-redshift Lyman break galaxies, e.g. young stellar population, low metallicity, low stellar mass, \citep[e.g.][this work]{overzier2006,grimes2007,bergvall2002,adamo2010}. Such similarity allows us to investigate the clustered star formation at high-redshift, a window that will be opened by upcoming and future facilities such as the James Webb Space Telescope (JWST) and the Extremely Large Telescope (ELT). Moreover, the confirmed escape fraction of Lyman continuum radiation from Haro 11 is relevant in the cosmological context of re-ionisation of the Universe and its main mechanisms: first stars, star-forming galaxies, Active Galactic Nuclei (AGNs).

The morphology and kinematics of the stars and ionised gas in Haro 11 suggests a resemblance with a dwarf galaxy merger \citep{oestlin2015}. Ejdetjärn et al. (in prep.)  run simulations of a galaxy merger with the goal of reproducing a galaxy as close as possible to Haro 11 in terms of morphology and tidal tails. This type of work has been already experimented by \citet{Renaud2014}, who simulated the famous Antennae galaxy merger. Since these theoretical simulations include the stellar feedback, it is important that observational studies like ours provide an estimate of the physical quantities used by theorists to model the feedback.

\citet{Menacho+2019} used VLT/MUSE integral field spectroscopy to investigate the impact of stellar feedback at large scales in Haro 11. Maps of several emission lines were used to study the kinematics and morphology of the ionised gas that appears structured in bubbles, arcs, filaments and channels. The halo of $H\alpha$-emitting ionised gas extends over an area of 5x5 kpc in size. 10-30\% of the ionised gas mass may escape the gravitational potential of the galaxy. The velocity of the ionised gas spans a broad range from -400 to +400 km/s with the highest-velocity gas being in the central starburst region, where the young star clusters populate knots A, B and C mentioned above. This is the region that we will focus on in our paper. The engine that powers such a strong mechanical feedback is the starburst occurring in the central knots and dusty arm westward of knot B (see Figure \ref{fig:RGB}). 

\citet{gross2021} presented Chandra observations finding that two X-ray sources are detected and overlap with knots B and C. The most intriguing source is the one in knot B as its spatial variability suggests intermediate mass black holes or low-luminosity active galactic nuclei (AGNs) at the centre of the galaxy in the process of merging. The presence of an AGN is not ruled out by the optical emission line diagnostics: optical line ratios are consistent with mixed populations, and could be explained by either stellar or non-thermal and/or shock ionisation \citep{gross2021}. An earlier study with the X-ray observatory Chandra found a diffuse X-ray emission within and around the starburst knots of Haro 11 \citep{grimes2007}. This is produced by a hot gas heated and accelerated by a strong stellar feedback as we will investigate further in this paper.

In this paper we revised the cluster population analysis adding recently acquired imaging data with the WFC3 camera as well as fixing a problem with the calibration of the SBC data that affected the previous analysis. In addition, we independently analysed the spectroscopy of the starburst knots. We inferred the physical properties of the stellar populations by combining the FUV and optical spectroscopy and modelling it with Starburst99 (SB99) spectral libraries. This allowed us to not only derive cluster properties but also to measure their feedback in terms of photo-ionisation and energetics of stellar winds and supernova (SN) explosions. To gain further insights on the impact of such feedback onto the ISM, in this work we expand on the ionised outflow analysis. We measured its energetics and compare it with the quantified stellar feedback expected from the stellar population.

We present the data reduction of the multi-band photometry and the multi-wavelength spectroscopy in Sec. \ref{sec:data}. In Sec. \ref{sec:phot_clusters} we show the results of the SED fitting of the photometry, which reveals the basic properties of the cluster population. In Sec. \ref{sec:spectral_fits} we describe the stellar population models based on SB99 spectral libraries and show which physical parameters of the clusters and their feedback we can be derived from the best fits and with which accuracy. We quantify the stellar feedback budget and study the kinematics of the ionised gas in the three knots in Sec. \ref{sec:feedback+outflow}, where we also estimate its energetics and compare it with the feedback output of the stellar populations. We discuss how we interpret the results in Sec. \ref{sec:discuss} and write our conclusions in Sec. \ref{sec:conclude}

\section{Data reduction}
\label{sec:data}

\subsection{Multi-band HST photometry}
In recent years, deeper HST imaging data have been acquired for Haro11. Moreover, a new calibration for the solar blind channel (SBC) of the advance camera for surveys (ACS) has been performed resulting in a change of the zeropoint (ZP) of about 30\% \citep{avila2019}. We therefore present a new analysis of the stellar cluster population in the galaxy, previously published by AA10. The dataset used in this work is presented in Table~\ref{tab:hstdata}. Differently from the previous study, we replace the ACS/HRC/F330W, WFPC2/F606W and F814W data with the more recent WFC3/F336W, F555W and F814W ones. We also include new band coverage (WFC3/F275W and F665N, the latter centred on H$_{\alpha}$ emission) and correct the ZP of the SBC/F140LP filter. HST WFC/F550M imaging is also used to align the centre of the COS positions in the MUSE data cube as explained below.

The data reduction follows the standard steps. \texttt{FLC} calibrated single frames have been downloaded from the MAST archive and processed with the most recent reference files. Single frames have been drizzled and combined into final science frames using \texttt{Drizzlepac}. The images are drizzled North-up, and covering the same exact footprint on the image.  All the data have been drizzled to the same pixel scale of 0.04\arcsec/pixel, i.e. the native pixel scale of the WFC3/UVIS detector. This pixel resolution correspond to a physical resolution range of $\sim$17 pc/px. \texttt{Geomap/Geotran} tasks in \texttt{Pyraf} have been employed to align the final science frames and register their world coordinate system (WCS) to a common reference system. Since the dataset covers the galaxy from the FUV to the red optical bands, we used the UVIS/F336W as reference frame for the alignment and registration. 

Cluster candidate extraction and photometry have been performed with the pipeline developed to analyse the HiPEEC sample \citep{adamo2020b}. We refer to the latter work for the details of the analysis steps, while we summarise here the most important aspects. The F555W filter has been used to extract compact sources with the SExtractor \citep{sextractor} software. The settings for the extraction are optimised to maximise the detection of compact sources with at least 4 contiguous pixels with a signal--to--noise $\geq 3$. We set a background mesh of 15 pixel. These extracted positions are then used as input to perform aperture photometry in all the available bands. We allow for a centering of 1 px in each band. We use a radius of 4 px and a local skybackground subtraction at 5 px, with 1 px width. We give the ZP in AB magnitudes in Table~\ref{tab:hstdata}. We perform aperture correction using tabulated encircled energy distributions available on the STScI webpages. The corrections, estimated up to 0.8", are listed in Table~\ref{tab:hstdata}. The final magnitude error is the square root of the sum in quadrature of the photometric error and the uncertainty on the ZP of 0.05 mag. This final automatic catalogue contains 270 sources which are detected in at least the F435W, F555W, and F814W filters with a photometric error $\leq1$ mag. Photometry in the other bands is included in the catalogue if it has a photometric error $\leq1$ mag. The final photometry is corrected for Galactic foreground extinction \citep{schlafly2011}. These initial 270 sources have been visually inspected in three bands (F435W, F555W, F814W) by two independent classifiers. Each object has been flagged as class 1, if considered a cluster candidate, or class 2, if a spurious detection. During the visual inspection, missed potential clusters have been flagged and their positions inserted into the pipeline, performing the same photometric steps as in the automatically extracted sources. All the manually extracted sources (39) have been confirmed by the two classifiers and have been assigned a class 1 flag. The updated extended catalogue contains 309 sources in total. AA10, reported to have initially detected 569 systems using the F814W and F606W filters with an error better than 1 mag. The difference with our new analysis can be explained by the fact that we impose detection during the initial detection steps also in a third filter (F435W). 

Following the selection applied to the HiPEEC sample, we will focus our analysis to cluster candidate visually flagged as class 1 and detected with a photometric error better than 0.35 mag in at least 4 filters (F336W, F435W, F555W, F814W). In total, 86 sources satisfy these conditions. This number is comparable to the number of sources reported in Table 1 by AA10 for their high-fidelity photometric detections in the F330W filter (the corresponding $U$ band for the previous analysis). Hence, we consider the quality of the two catalogues comparable.

\begin{table*}
\centering
\begin{tabular}{lllcccccc}

Filter & Camera & prop. ID (PI) & Exp time [s] & ZP (ABmag) & ap. cor. [mag] & N$_{\rm YSC}$ & m$_{\lambda}$(lim) [ABmag] & $\alpha_{\lambda}$\\

\hline

F140LP & ACS/SBC & 9470(Kunth) & 2700 & 23.448 & -0.55 &96&23.5&1.43$\pm$0.02\\
F220W & ACS/HRC & 10575(\"Ostlin) & 1513 & 23.511 & -0.23&65&23.0&1.45$\pm$0.02\\
F275W$^*$ & WFC3/UVIS &15649(Chandar)& 1920 & 24.158 & -0.28&81&23.0&1.52$\pm$0.06\\
F336W$^*$ & WFC3/UVIS &13702(Oey) & 2664 & 24.692 & -0.27&98&23.0&1.54$\pm$0.04\\
F435W & ACS/WFC & 10575(\"Ostlin) &680 & 25.674 & -0.2&111&23.0&1.43$\pm$0.01\\
F550M & ACS/WFC & 10575(\"Ostlin) &471 & 24.862 & -0.21&94&24.0&1.40$\pm$0.03\\
F555W$^*$ & WFC3/UVIS & 15649(Chandar)&1740 & 25.813 & -0.21&113&23.0&1.44$\pm$0.04\\
F665N$^*$ & WFC3/UVIS & 15649(Chandar)& 1800 & 22.736 & -0.19&68&21.0&1.48$\pm$0.03\\
F814W$^*$ & WFC3/UVIS & 15649(Chandar)&1740 & 25.126 & -0.23&109&23.0&1.44$\pm$0.05\\
\hline

\end{tabular}
\caption{Description of the HST dataset used in this work. The asterisk indicate the new imaging data with respect to AA10. The zeropoint (ZP) and the aperture correction estimated in each band are reported in column 5 and 6. The last 3 columns summarise for each filter the number of YSCs (N$_{\rm YSC}$) that have been included in the cluster luminosity function analysis, the limiting magnitude for the fitted magnitude range (m$_{\lambda}$(lim)) corresponding to the peak in the magnitude distributions, and the recovered slope with uncertainties, $\alpha_{\lambda}$.  }
\label{tab:hstdata}
\end{table*}

\subsection{UV COS spectroscopy}
\label{sec:COS} 




We have used the HST/COS observations with gratings G130M and G160M of knots A and B from program 15352 \citep[PI Östlin,][]{oestlin2021}, and COS observations with the same gratings of knot C from the program 13017 (PI Heckman). Individual observations were reduced with CALCOS Version 3.3.4. Extracted spectra were combined using custom scripts and resampled to 0.4Å, matching the resolution of the Starburst99 models of stellar populations that we used in our analysis (see Sec.\ref{sec:analysis}).

Because the CALCOS pipeline overestimates the errors of low S/N sources \citep[][]{Henry2015}, we estimated the errors by calculating the standard deviation in a window of 50 spectral pixels around each wavelength and fitting a high order polynomial (degree=7) to the vector of standard deviations. We then use this to rescale the actual error vector from the pipeline to the appropriate level. The pipeline error vector does keep track of pixel flags, and has the correct structure, however its level needs to be corrected. In this computation we masked out bright geocoronal lines and absorption lines. We estimated errors up to a factor of 2 smaller than the CALCOS pipeline errors.

\begin{figure}
    \centering
    \includegraphics[width=0.46\textwidth]{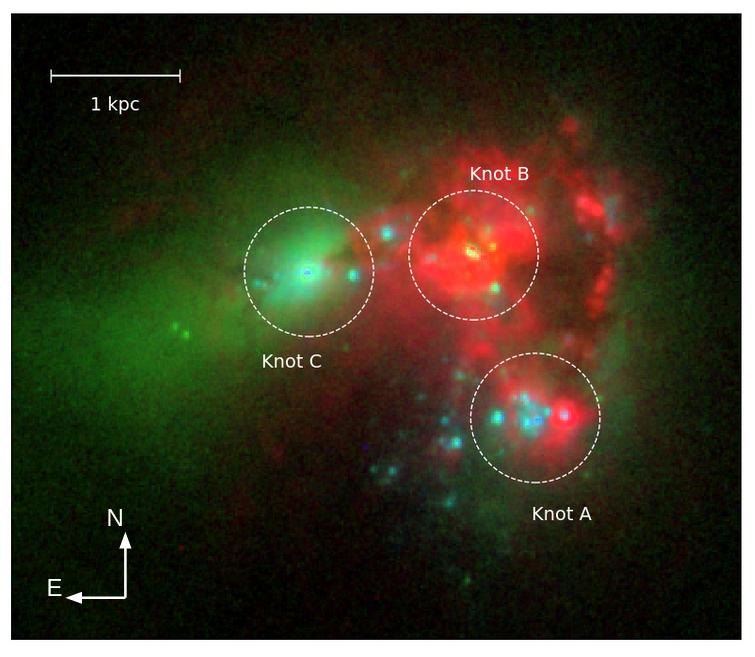}
     \caption{Color composite image obtained using three HST images (red = FR656N, green = F550M, blue = F336W) }
    \label{fig:RGB}
\end{figure}

In Figure \ref{fig:RGB} we show the physical sizes corresponding to the COS apertures centred on the three starburst knots, plotted on top of a composite color image.
Table \ref{tab:COS} reports the details of the FUV spectroscopy of each knot together with the redshift and FWHM measured by fitting a Gaussian to the FUV stellar photospheric absorption line CIII1247 for knots A and C and CIII1176 for knot B. The shift in wavelength of the line centroid with respect to the restframe value gives the redshift; the FWHM of the line gives the stellar broadening. The spectral resolution for the three knots is approximately $R\sim10000$ \citep{Emil2017,oestlin2021} considering the complex UV morphology of the source.

\begin{table*}
	\centering
	\caption{Properties of FUV COS observations of Haro 11.}
	\label{tab:COS}
	\begin{tabular}{lccc} 
		\hline
		      & knot A & knot B & knot C\\
		\hline
		grating & G130M + G160M & G130M + G160M & G130M + G160M\\
		wavelength setting & 1222/1291Å + 1577/1611Å & 1222/1291Å + 1577/1611Å & 1309Å + 1623Å\\
		exposure time (s) & 1929 + 2558 & 1929 + 2558 & 1536 + 2364\\
		measured redshift & $0.02083 \pm 0.00004$ & $0.02058 \pm 0.00004$ & $0.02060 \pm 0.00003$\\
		stellar broadening (Å) & $0.44 \pm 0.07$ & $1.31 \pm 0.10$ & $0.84 \pm 0.05$\\
		\hline
	\end{tabular}
\end{table*}

\subsection{Optical MUSE spectroscopy}

Haro 11 has been observed with the optical integral field spectrometer MUSE \citep{bacon2010} installed at the Very Large Telescope (ESO) in the extended wavelength setting ($\sim$ 4650 - 9500 Å). The data reduction of this observation is described in \citet[][]{Menacho+2019}. From the MUSE cube, we extracted a 1-D spectrum using an aperture of the same size of the COS aperture (i.e. 2.5\arcsec\, in diameter) and taking into account the vignetting function that describes the sensitivity of the COS spectrograph as a function of radius in the aperture. The purpose of the latter operation is to obtain an optical spectrum of the same regions of Haro 11 (knot A, knot B and knot C) for which we have FUV COS spectroscopy. By doing so we can join the two spectra (FUV and optical) and model them simultaneously with a stellar population code based on the Starburst99 spectral libraries as described in Sec. \ref{sec:analysis}. The errors of the 1-D extracted optical spectra have been derived from the variance MUSE cube using the same extraction technique used for the fluxes and corrected by a factor of 15\% \citep[][]{Bacon+2017}. However, the signal-to-noise ratio (SNR) obtained in this way is of the order of 1000 and accounts for only the statistical error. We therefore set a maximum SNR of 20 in order to account for the absolute photometric accuracy of MUSE, i.e. $\sim 5\%$. We measured this accuracy in our data by following five steps: (1) we convolved the HST image in the F550M filter with the MUSE seeing of FWHM=0.8"; (2) we rebinned the convolved HST image to the pixel size of MUSE equal to 0.2"; (3) we performed photometry for each of the three knots using the COS apertures; (4) we created a MUSE continuum image in the same wavelength range of the HST filter F550M; (5) we performed MUSE photometry for each of the three knots using the COS apertures and we compared the fluxes obtained with HST photometry: the offset between the two gives the relatvie accuracy of the MUSE flux measurements and ACS/F550M phtotometric calibrations.

The wavelength range of the MUSE spectroscopy analysed in our work extends up to $\sim$ 7000 Å since this is the upper limit of the spectral coverage of the stellar populations models that we deployed (Starburst 99, see Sec. \ref{sec:spectral_fits}). The spectral resolution of the MUSE instrument in the wavelength region used in our analysis ranges between 2.5 - 2.9 Å.





\section{Different approaches to dissect the starburst in Haro 11}
\label{sec:analysis}

\subsection{Cluster population based on HST photometry}
\label{sec:phot_clusters}
\subsubsection{Multi-band photometric SED analysis}
Following the SED fitting analysis performed for the HiPEEC galaxies \citep[see][for details]{adamo2020b}, we fit cluster candidates detected in the 4 reference bands with a photometric error better than 0.35 mag and flagged as class 1. The fit includes photometry in other bands if the photometric error is below 0.35 mag. We use Yggdrasil single stellar population models \citep{zackrisson2011}, including the treatment for nebular emission. We assume stellar and gas metallicities of $Z=0.004$. For the gas we use average values of electron densities of 100 $\rm e^-/cm^{3}$, and two assumptions for the covering fractions, i.e., 1 (all ionising photons will ionise the gas within the HII regions) and 0.5 (50\% of ionising photons will escape the region). The latter models are used as reference for the analysis presented in this work. The set of models with covering fraction of 1 are employed to facilitate the comparison with the analysis by AA10. The model grid includes age steps from 1 Myr to 14 Gyr and extinction steps from E($B-V$)$=0$ to 1.5 mag in steps of 0.01 mag. Attenuation is applied to the stellar continuum and the emission line modeled spectrum separately following the prescription by \citet{calzetti2000}, before the convolution with the filter transmission.

\subsubsection{A map of the starburst propagation traced by star clusters}
\begin{figure*}
    \centering
    \includegraphics[width=0.38\textwidth]{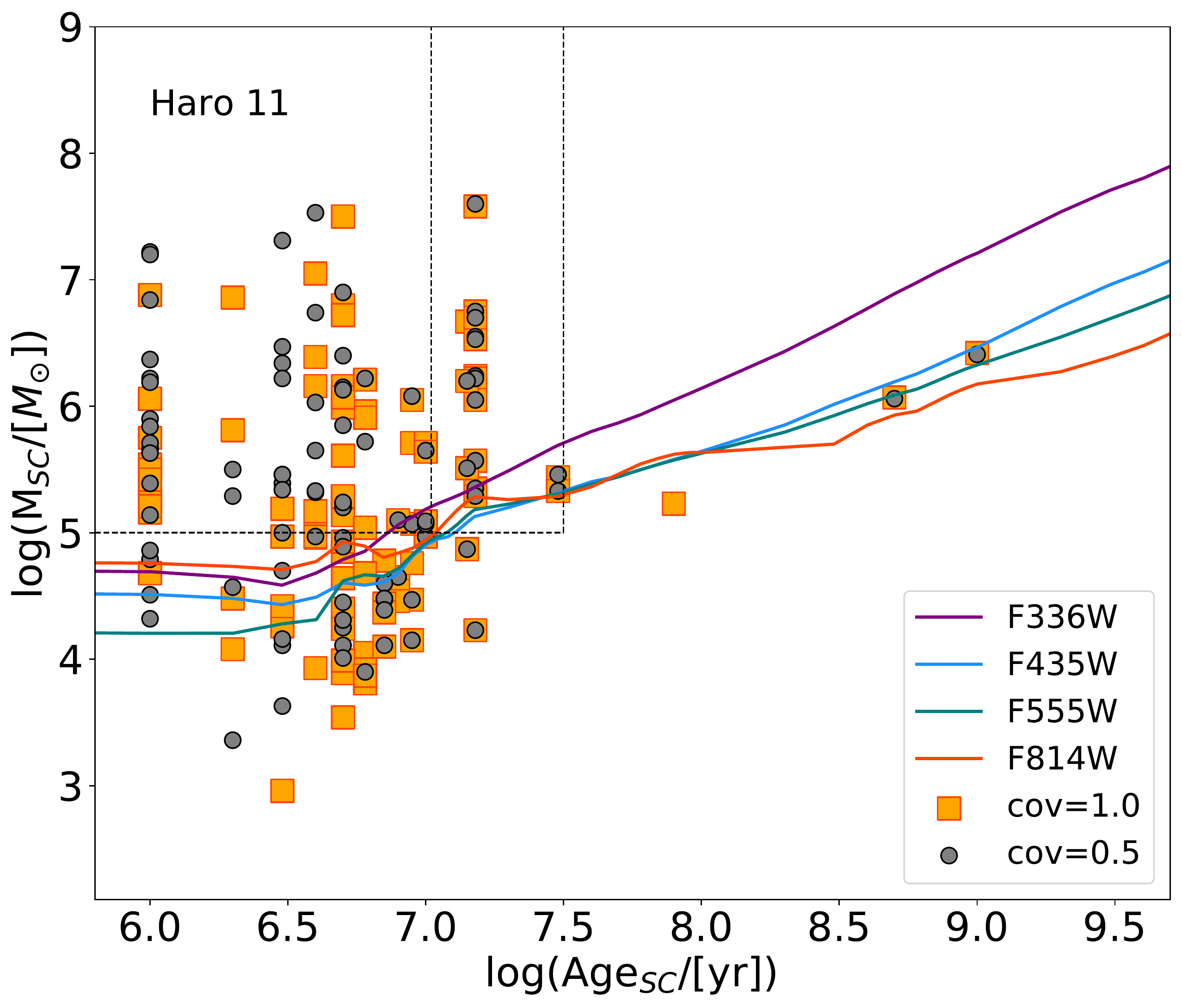}
    \includegraphics[width=0.61\textwidth]{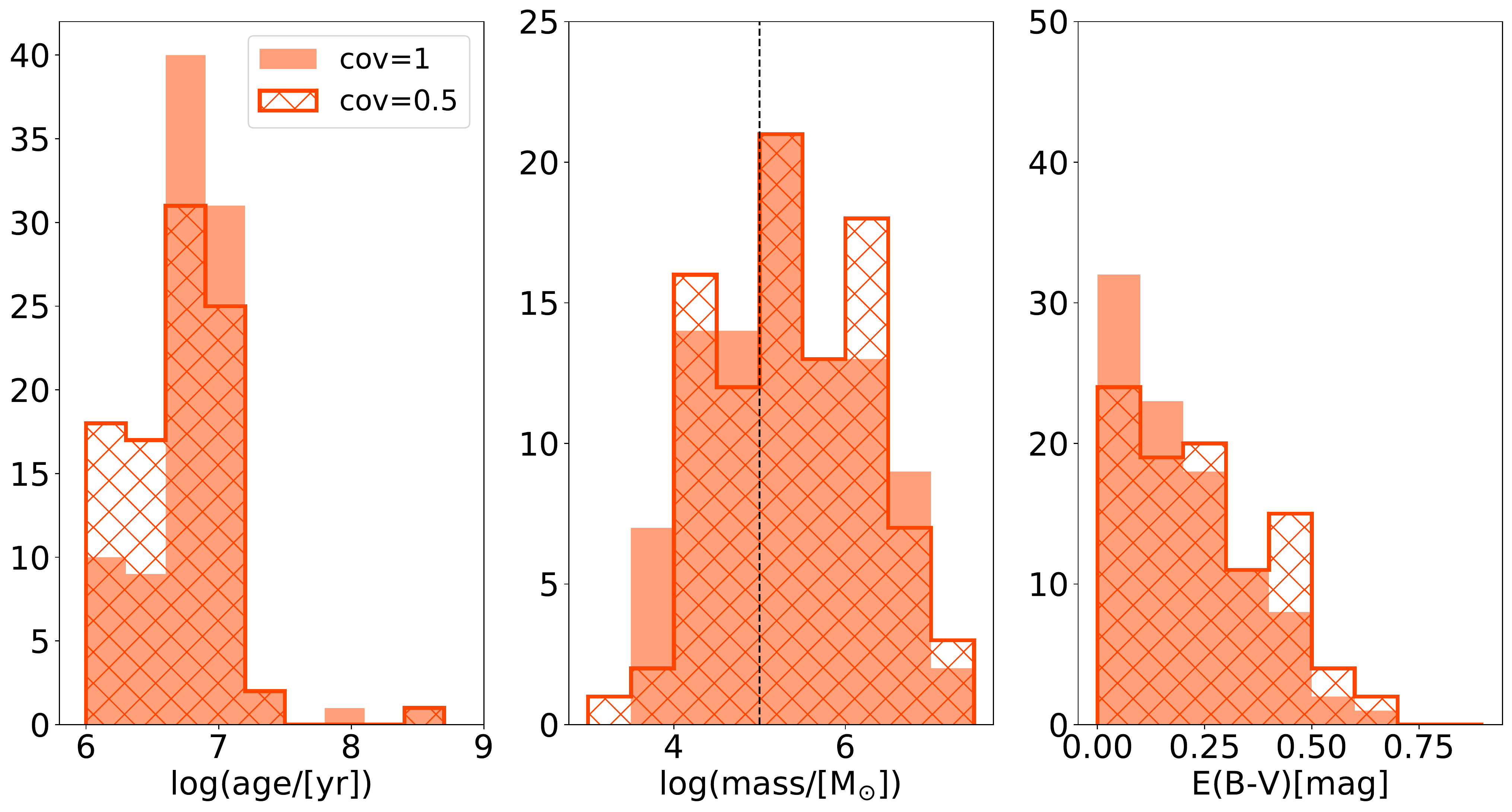}\\
    \includegraphics[width=0.324\textwidth]{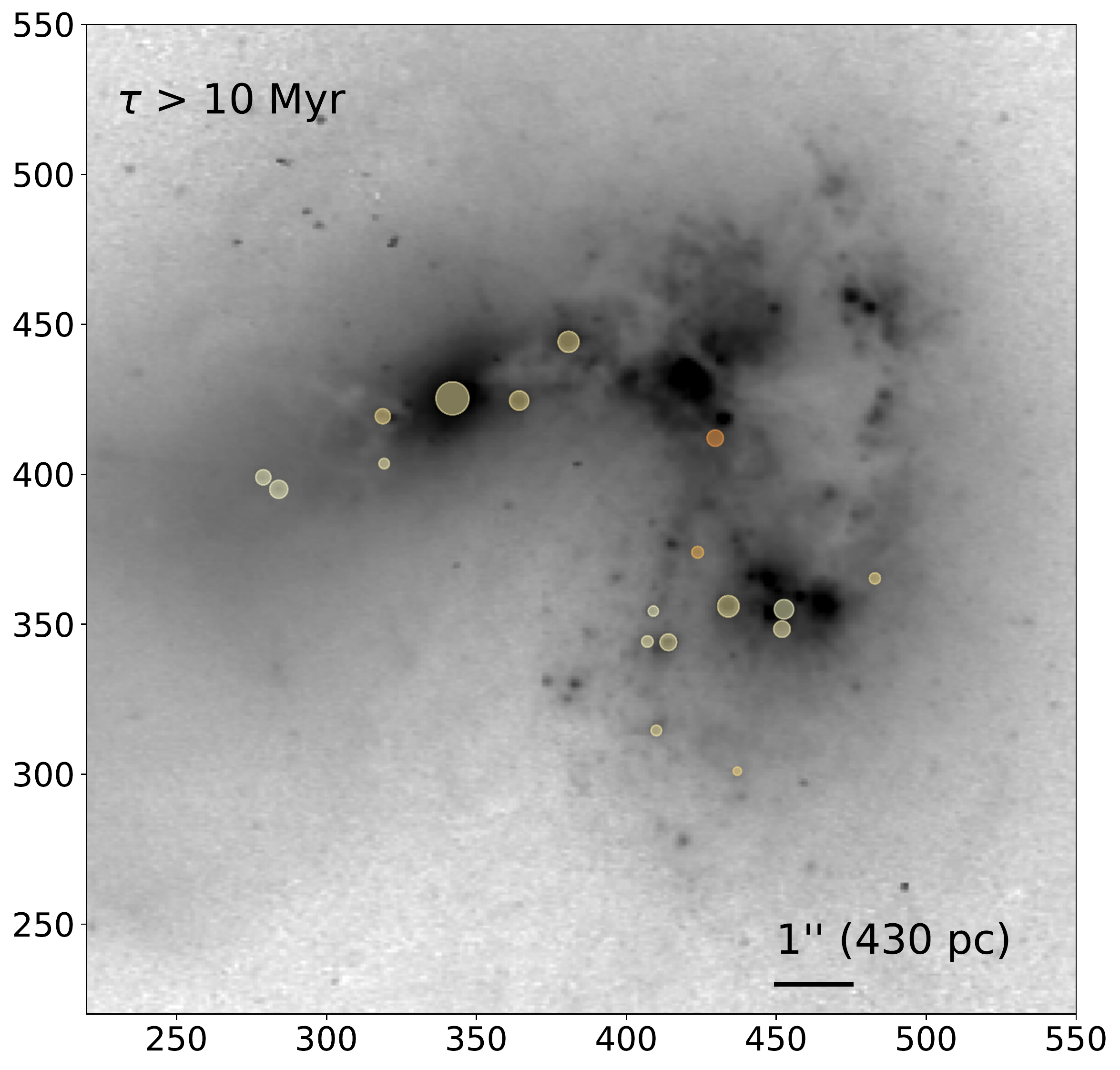}
    \includegraphics[width=0.305\textwidth]{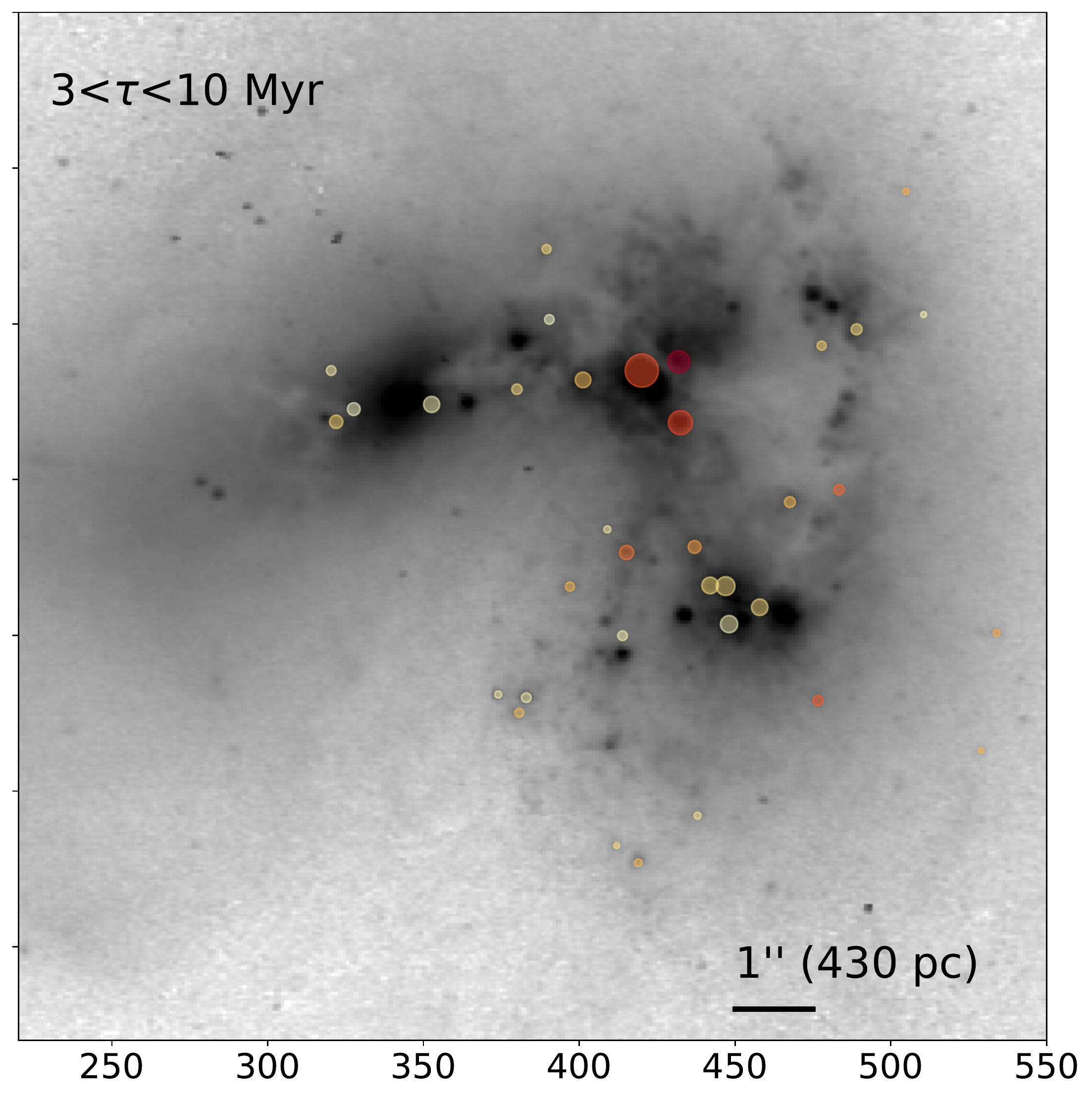}
    \includegraphics[width=0.360\textwidth]{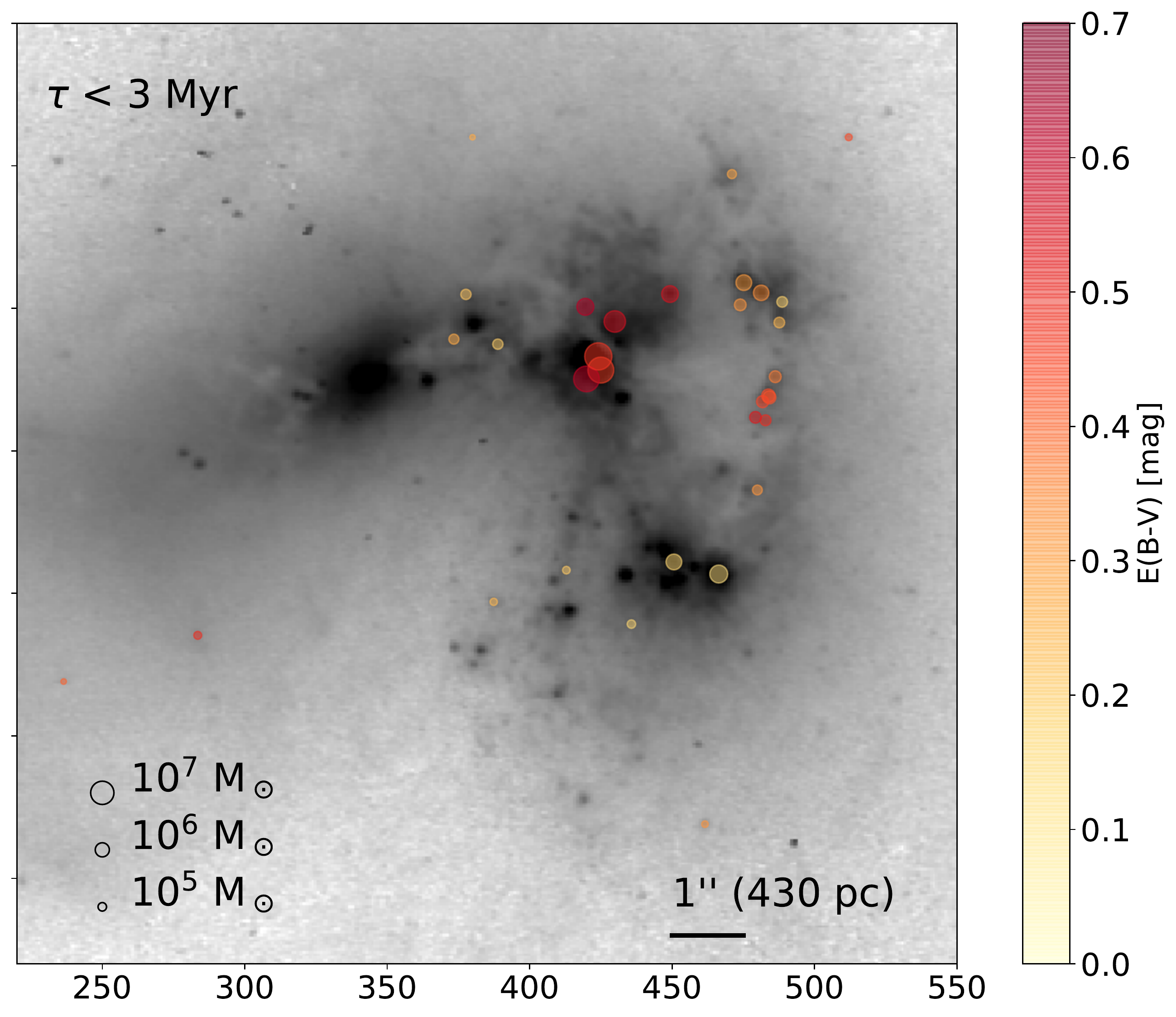}
     \caption{\emph{Top left:} Age-mass diagram of the cluster population in Haro11. Ages and masses derived with the two sets of models (covering fraction 1 and 0.5) are shown with different symbols. The solid lines, color-coded accordingly to the corresponding filter, show the age and mass limits for a given magnitude limit established from the luminosity distribution in each band (see main text). The dotted lines outline clusters that will be used to study the mass function (masses above $10^5$ \msun\, and ages below 30 Myr. \emph{Top right:} The histograms show the cluster age, mass, and extinction distributions. The results achieved with the two sets of models are outlined. \emph{Bottom:} Starburst propagation within the galaxy as timed by cluster formation history ($\tau$). The clusters are plotted on the F555W frame (north-up and log scales) as a function of their age (oldest to the left, intermediate central, youngest right panel), mass (size of the circle), extinction (color of the circles). We notice that the burst has propagated from east to west. Extinction is highest in the most recently formed clusters (located in knot B and dusty arm). The most massive clusters are preferentially located in the knots, covered by FUV spectroscopy.}
    \label{fig:ysc_prop}
\end{figure*}

\begin{figure}
    \centering
    \includegraphics[width=0.46\textwidth]{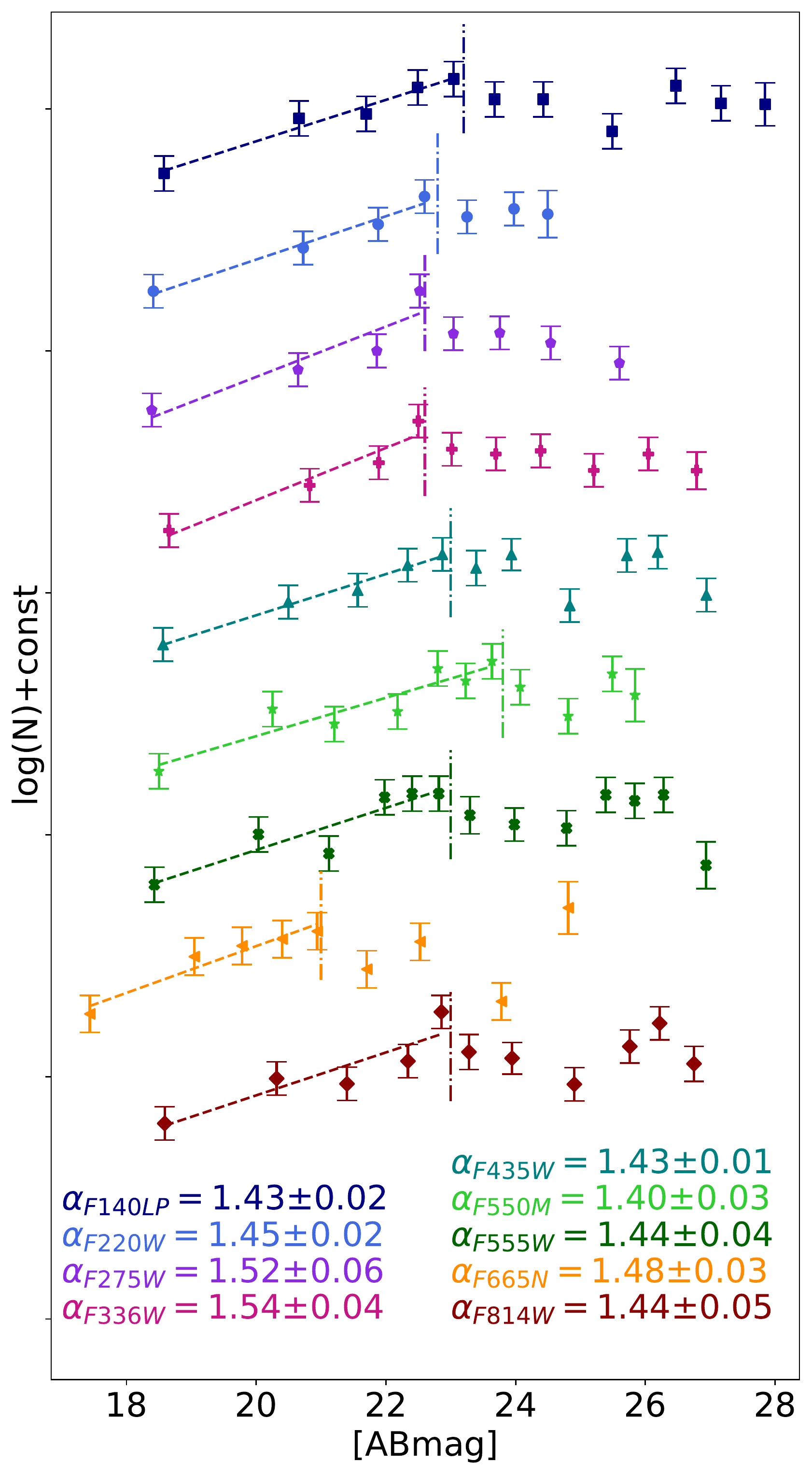}
     \caption{Multi-band luminosity functions of the cluster population of Haro 11.}
    \label{fig:ysc_lf}
\end{figure}

In Figure~\ref{fig:ysc_prop} we provide an overview of the physical properties of the cluster population. The age-mass diagram (top left panel) confirm the presence of very massive YSCs. Using the bend in the magnitude distributions (see Figure \ref{fig:ysc_lf}) as an indication of the 90\% completeness in the catalogue \citep[e.g.,][]{bastian2012}, we determine that the catalogue is complete for clusters more massive than $10^5$ \msun\, (similar detection limits are found for the HiPEEC systems). Indeed, $10^5$ \msun\, coincide with the peak in the mass distribution histogram on the top right panel of the same figure. The difference in the recovered mass distribution with respect to AA10 (see their figure 8) is due to the different distance assumed for our revisited study (88.5 instead of 82.3 Mpc). The recovered extinction distribution agree very well with the previous study. We notice, however, that the bulk of the clusters have ages younger than 20 Myr. Using different covering fraction for the nebular emission component does not significantly change this general trend, as confirmed by the distributions in the top central histogram. This is also confirmed by the top right histograms. Using models with covering fractions of 1, clusters result slightly older, but the effect is limited to ages below 10 Myr. In spite of the model used, the revisited analysis does not include clusters with ages between 20 and 40 Myr reported in AA10. In Figure 20 of AA10, one can notice that these intermediate age clusters are mainly located around knot A and C. As discussed below, we find trace of a diffuse stellar population in this age range in both knots (see Section~\ref{sec:spectral_fits}). It is important to notice that the cluster catalogue published in AA10 was based on error cut in V and I bands and not on compact morphological appearance. In this work we include a third filter in the initial extraction of the source catalogue. As shown in Figure~\ref{fig:ysc_prop}, the detection limits in the B band (F435W) would have not prevented us from detecting clusters more massive than $10^5$\msun\, in Haro11. We speculate that many of these "older" cluster candidates could be spurious detections in the WFPC2 data within the diffuse stellar component that our updated analysis approach has removed. We speculate that the lack of clusters at ages larger than 20 Myr could suggest either rapid disruption rates, typical in merger systems \citep[][]{whitmore2010} or that clusters have formed with lower masses than in more recent times, remaining undetected. If the latter scenario is true, it would imply dramatic changes in the gas conditions over the last tens of Myr that have favoured the formation of very massive clusters. 

In the bottom panel of Figure~\ref{fig:ysc_prop}, we map the starburst propagation across the galaxy using the star clusters as tracers. Cluster formation has preferentially started in the eastern side of the galaxy, with the nuclear star cluster in knot C being the most massive cluster, formed in this age range. The extinction in these clusters is generically low. Clusters in the age interval $3<\tau<10$ Myr are observed in all the three knots, with preference for the most massive clusters forming in knot B. Clusters formed in the latter region are also the most reddened. Currently ($\tau<3$ Myr) the youngest and most extinguished clusters are preferentially forming in knot B and in the dusty arm (the most eastern tidal feature visible in optical). Clusters have continued to form at a lower rate also in knot A, which is the only region in the galaxy where cluster formation has been going on throughout the entire age interval (propagating east-west in a similar fashion as on the overall galaxy).

\subsubsection{Analysis of the cluster luminosity and mass functions}
To conclude, we present the luminosity and mass function of the detected cluster population. Differently from AA10, who limited the analysis to the F814W filter, we extend here the luminosity function analysis to all the 9 filters. In each band, we include all the systems that have been detected in that band with a photometric error better than 0.35 mag and flagged as class 1. The slope $\alpha$ of the luminosity function is defined as: $dN/dL_{\lambda}  \propto L_{\lambda}^{\alpha}$. The fit for the luminosity function is done by constructing variable bin sizes containing equal number of objects per bin so that each bin has the same weight in the fit \citep[e.g.][for comparisons with other techniques]{adamo2017}. Moreover, we fit in log space, i.e.,  $d\log(N)/dm_{\lambda} = const+\beta\times m_{\lambda}$, where $\alpha = -(2.5\times\beta+1.0)$.  We show the fit in Figure~\ref{fig:ysc_lf} from the FUV (F140LP) to the NIR (F814W). Only bins up to the peak of the distributions are included in the fit (as indicated by the vertical lines). Typically, we construct bins with $\sim 10$ objects. We report in Table~\ref{tab:hstdata} the number of objects that are included in the luminosity function analysis, the magnitude limit corresponding to the peak of the distributions and the index $\alpha_{\lambda}$. The latter are also included in Figure~\ref{fig:ysc_lf}, where we show the multi-band luminosity functions. The recovered slopes have index values ranging between 1.4 and 1.5. These slopes are in good agreement with the one reported by AA10 ($\alpha_{F814W}=1.52\pm0.05$). 

Using the same analysis tools developed for the study of the cluster mass function in the HiPEEC galaxies \citep[see][for a detailed description of the formalism and the fitted functions]{adamo2020b} we perform a Bayesian inference analysis to determine the slope and possible evidence of a truncation at the high-mass end of the mass distribution. In the Appendix A, we include the corner plots, probability distribution functions and the goodness of the fit in reproducing the observed mass distributions. Overall, the median and 1$\sigma$ confidence level of the resulting probability distribution returns $\beta_{\rm PL}=-1.47^{+0.07}_{-0.08}$ for the index of a single power-law function (see Figure \ref{fig:PDF_PL} in the appendix). This value is in very good agreement with the values derived for the luminosity function. In a similar way, the analysis of the mass distributions with a Schechter function returns $\beta_{\rm Sch} = -1.31^{+0.12}_{-0.11}$ and a truncation mass $\log($M$_c)=7.6^{+0.5}_{-0.3}$. The corner plot in Figure \ref{fig:PDF_Sch} show that there is convergence in the determination of these two parameters, and indeed a Schechter function with these value is a better representation of the observed mass function than a single power-law (Figure \ref{fig:cumulative_MF}). However, the statistical significance of this result is not strong, as already observed for the HiPEEC sample.

The slope of the luminosity (mass) distribution of the cluster population of Haro11 are shallower than the typical value of $-2$ observed in luminosity and mass function of local spiral galaxies  \citep[see][for a recent review]{adamo2020a}. Blending and incompleteness effects can in part explain the flattening of the power law index \citep[e.g.][]{Dessauges2018}. However, we notice that a flattening of the power-law index of the cluster populations forming during starburst episodes has been reported for other merging galaxies \citep{adamo2020b} and even for nearby blue compact galaxies \citep[e.g. in NGC4449 at 4 Mpc, ][]{annibali2011}, suggesting that starburst events might favour the formation of more numerous high-mass clusters (e.g, a top-heavy mass function) with strong impact on the stellar feedback originating from these massive clusters.

We focus for the rest of this work on the analysis of the stellar and cluster populations within the three knots. In these regions we find the largest concentration of massive young star clusters. Using independent analysis approaches we will determine the properties of the stellar populations (including star clusters) within the knots and constraint their stellar feedback on the neutral warm and ionised ISM. 

\subsection{Stellar populations of the three knots based on spectroscopy}
\label{sec:spectral_fits}

We combined the FUV and optical spectroscopy to study the properties of the stars populating the three starburst knots in Haro 11. We modeled the combined spectra of each knot using models built with the Starburst99 \citep[SB99, ][]{leitherer1999,leitherer2014} spectral libraries. In particular, we used the synthetic spectra that have been obtained with the Geneva stellar evolutionary tracks with high mass-loss rates, Salpeter IMF with cut-offs at 0.1 and 120 $\rm M_{\odot}$, non-rotating stars and with a single-burst star-formation history. These models include not only the continuum emission produced by stars but also the continuum emission produced by the nebular gas. Our code creates a continuous function interpolating the grid of the parameter space defined by discrete values of ages and metallicity of the stellar populations, and fits the function to the observed combined (FUV + optical) spectrum. Dust extinction is applied using the Calzetti law \citep{calzetti2000}. The best model is found using the minimiser \texttt{"differential evolution"} of the python package \texttt{lmfit}. 

With UV only data it is quite difficult to accurately derive the attenuation reddening since the range in wavelength in too short. By adding the optical data we had much stronger constraints on the overall spectral shape, and reddening. This improves the mass estimates dramatically, because the optical data severely limits how much obscured mass can be buried behind the dust.

We re-sampled the spectra to 0.4 Å to match the resolution of the SB99 models and we convolved the synthetic spectra using a Gaussian of FWHM equal to the value measured for the photospheric FUV lines CIII 1176 and CIII 1247. We used a carefully designed mask to exclude from the fit the detectors gaps, the geocoronal lines, the Ly-$\alpha$ emission, the ISM absorption lines and everything else that from a visual inspection was identified to be of non-stellar origin, since the SB99 model only implements the stellar physics and nebular continuum. We applied a redshift to the models equal to the one measured in the FUV data and we corrected the observed spectra for the reddening attenuation of the Milky Way using the Cardelli, Clayton and Mathis law \citep[][CCM89]{Cardelli1989}.

The FUV spectra feature a broad Ly-$\alpha$ absorption around 1200 Å that can affect the NV P-cygni line at 1241.5 Å (wavelengths given in the restframe). The NV feature and a few other P-cygni lines (e.g. SiIV 1402Å, CIV 1550Å) contain most of the diagnostic power for determining the age of the stars at times below 4 Myr, because only young stellar populations (1-3 Myr) can power the strong winds producing such spectral features. Therefore we decided to correct for the Ly-$\alpha$ absorption to improve the fit around the NV feature. We fit the Ly-$\alpha$ absorption with an interactive tool used in \cite{oestlin2021} that allows to determine the continuum level and fit a Voigt profile describing the HI absorption. After fitting the Voigt profile, we normalised the spectrum, we divided it by the normalised Voigt model and then rescaled it using the same normalisation constant, obtaining in this way a rectified spectrum.

We fitted the combined spectra of each knot with a model given by three stellar populations. Each modeled population has 4 fitting parameters: stellar age, metallicity Z, mass and reddening attenuation E(B-V). Single- and two-population fits have also been tested, but the returned $\tilde{\chi}^2$ was significantly higher reflecting the inability of these model assumptions to properly fit both the stellar features and continuum of the combined FUV and optical spectrum. The addition of a third older population (>40 Myr) is necessary to fully reproduce the optical continuum of each region.

The three populations have age priors that will reflect specific feedback: the first population is between 1-4 Myr and will provide the bulk of the pre-SN feedback (photoionisation and stellar winds); the second population between 4-40 Myr will account for SN feedback; the third population between 40-100 Myr is included to account for a wider SFH to which the optical spectroscopy is more sensitive to. We decided to use a narrow range at low values of metallicity as indicated by the 20\%-solar oxygen abundance \citep[][]{bergvall2002}. Table \ref{tab:SB99fit} lists the priors of all parameters for each of the stellar populations used to model each knot as well as the best-fit values and $\tilde{\chi}^2$ of each fit. The uncertainties on the best-fit values have been determined with a classic Monte Carlo approach. The metallicity of the first population derived from the spectral fits are close to 0.004 (20\% solar) confirming the similarity of Haro 11 to high red-shift galaxies. The masses found for each stellar population range between tens to hundreds of million solar masses. We notice that the spectral fit is not able to constrain the parameters of the models in some cases, being the error intervals as large as the given prior (see Table \ref{tab:SB99fit}). One of these cases is the metallicity of the second population in all the three knots, for instance). Another case is the reddening attenuation of the second and third population, which is often unconstrained. The age and the mass of the stellar populations are better defined within the interval they are allowed to change during the fit. These two parameters have however large errors that will propagate into the estimate of feedback quantities, described in section \ref{sec:feedback}.
Using as models stellar libraries with standard mass-loss rates, rather than the high mass-loss rates, would slightly affect the age of the third population but leave the rest similar within uncertainties or even poorer in the reproduction of the P-Cygni line shapes.
\begin{table}
	\centering
	\caption{Stellar population fits of the three knots of Haro 11}
	\label{tab:SB99fit}
	\renewcommand{\arraystretch}{1.5} 
	\begin{tabular}{lccc} 
		\hline
		      & knot A & knot B & knot C\\
		\hline
		age priors & Myr & Myr & Myr\\
		\textit{Pop 1} & 1-4 & 1-4 & 1-4\\
		\textit{Pop 2} & 4-40 & 4-40 & 4-40\\
		\textit{Pop 3} & 40-100 & 40-100 & 40-100
		\vspace{1.0mm}\\
		Z priors & 0.004-0.008 & 0.004-0.008 & 0.004-0.008
		\vspace{1.0mm}\\
		mass priors & $<10^9\, \rm M_{\odot}$ & $<10^9\, \rm M_{\odot}$ & $<10^9\, \rm M_{\odot}$
		\vspace{1.0mm}\\
		E(B-V) priors & 0.01-0.4 & 0.01-0.4 & 0.01-0.8
		\vspace{1.0mm}\\
		best-fit age & Myr & Myr & Myr\\
		\textit{Pop 1} & $3.1^{+0.0}_{-1.0}$ & $2.8^{+0.2}_{-0.0}$ & $2.0^{+1.2}_{-0.1}$\\
		\textit{Pop 2} & $12^{+28}_{-0}$ & $13^{+27}_{-2}$ & $37^{+0}_{-22}$\\
		\textit{Pop 3} & $53^{+41}_{-0}$ & $72^{+22}_{-17}$ & $100^{+0}_{-58}$
		\vspace{1.5mm}\\
		best-fit Z &  &  & \\
		\textit{Pop 1} & $0.0043^{+0.0011}_{-0.0002}$ & $0.0054^{+0.0002}_{-0.0005}$ & $0.0063^{+0.0007}_{-0.0018}$\\
		\textit{Pop 2} & $0.008^{+0.000}_{-0.004}$ & $0.008^{+0.000}_{-0.004}$ & $0.0040^{+0.0032}_{-0.0000}$\\
		\textit{Pop 3} & $0.004^{+0.000}_{-0.000}$ & $0.0040^{+0.0009}_{-0.0000}$ & $0.007^{+0.000}_{-0.003}$
		\vspace{1.5mm}\\
		best-fit mass & $ 10^6\, \rm M_{\odot}$ & $ 10^6\, \rm M_{\odot}$ & $ 10^6\, \rm M_{\odot}$\\
		\textit{Pop 1} & $9^{+6}_{-0}$ & $5.7^{+1.7}_{-0.0}$ & $15^{+1}_{-5}$\\
		\textit{Pop 2} & $27^{+25}_{-24}$ & $27^{+57}_{-19}$ & $170^{+6}_{-117}$\\
		\textit{Pop 3} & $10^{+88}_{-0}$ & $280^{+55}_{-256}$ & $207^{+258}_{-58}$
		\vspace{1.5mm}\\
		best-fit E(B-V) &  &  & \\
		\textit{Pop 1} & $0.06^{+0.04}_{-0.01}$ & $0.09^{+0.01}_{-0.00}$ & $0.15^{+0.01}_{-0.05}$\\
		\textit{Pop 2} & $0.40^{+0.00}_{-0.39}$ & $0.18^{+0.22}_{-0.10}$ & $0.4^{+0.4}_{-0.1}$\\
		\textit{Pop 3} & $0.01^{+0.39}_{-0.00}$ & $0.4^{+0.0}_{-0.3}$ & $0.80^{+0.00}_{-0.78}$
		\vspace{1.5mm}\\
		$\tilde{\chi}^2$ & 2.17 & 1.89 & 2.29\\
		
		\hline
	\end{tabular}
\end{table}
\begin{figure*}

	    \includegraphics[width=\columnwidth]{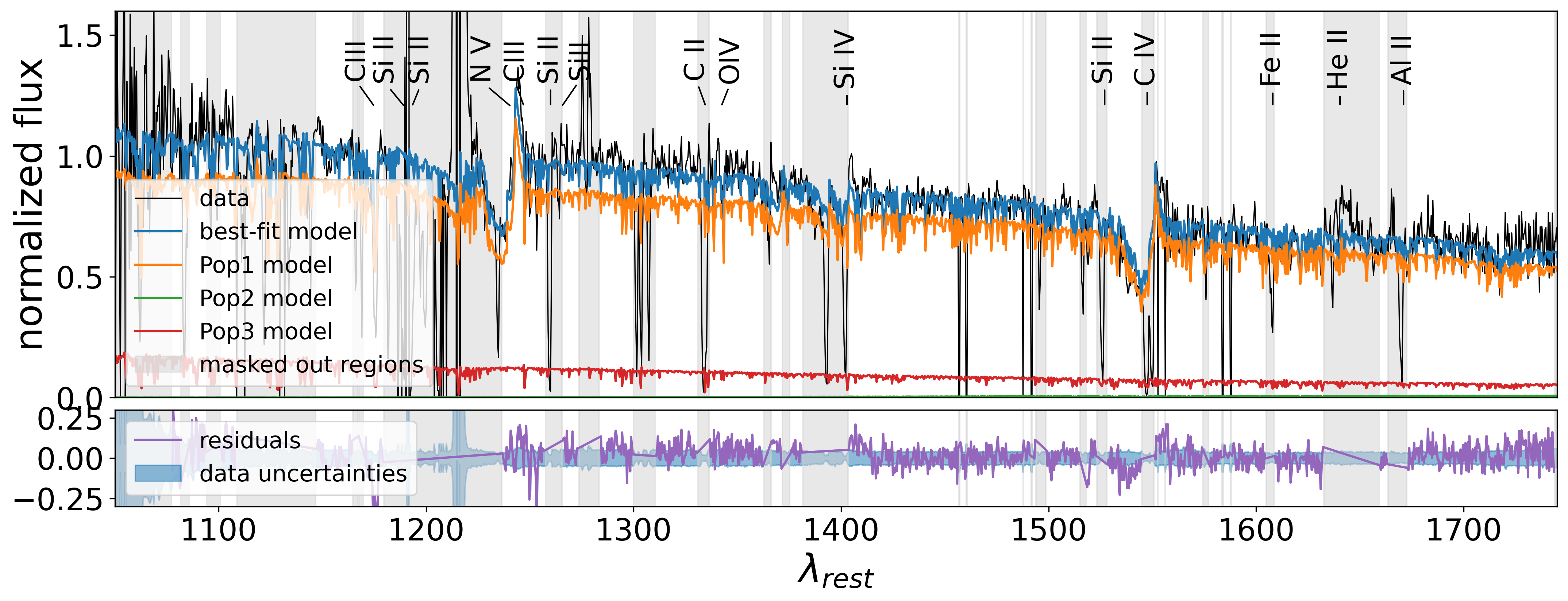}
	      \label{fig:fit_UV_knotA}
	    \includegraphics[width=\columnwidth]{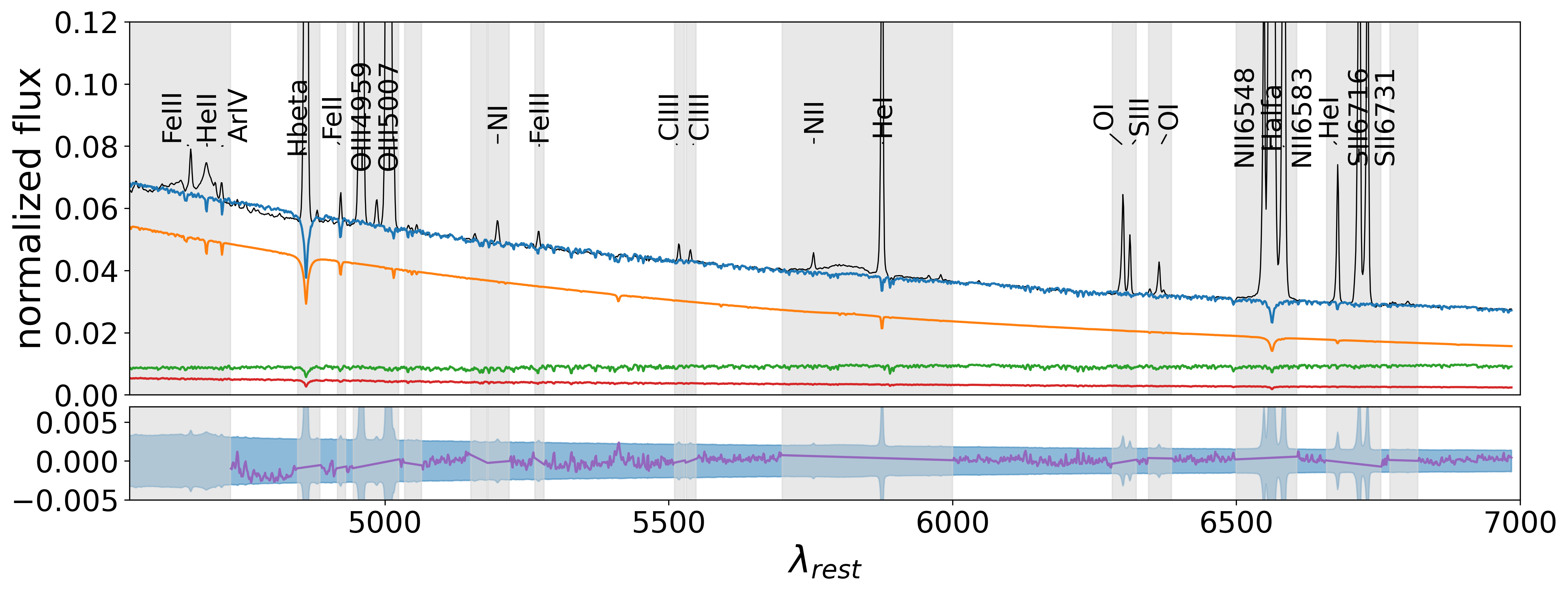}\\
	      \label{fig:fit_Opt_knotA}
	      
	    \includegraphics[width=\columnwidth]{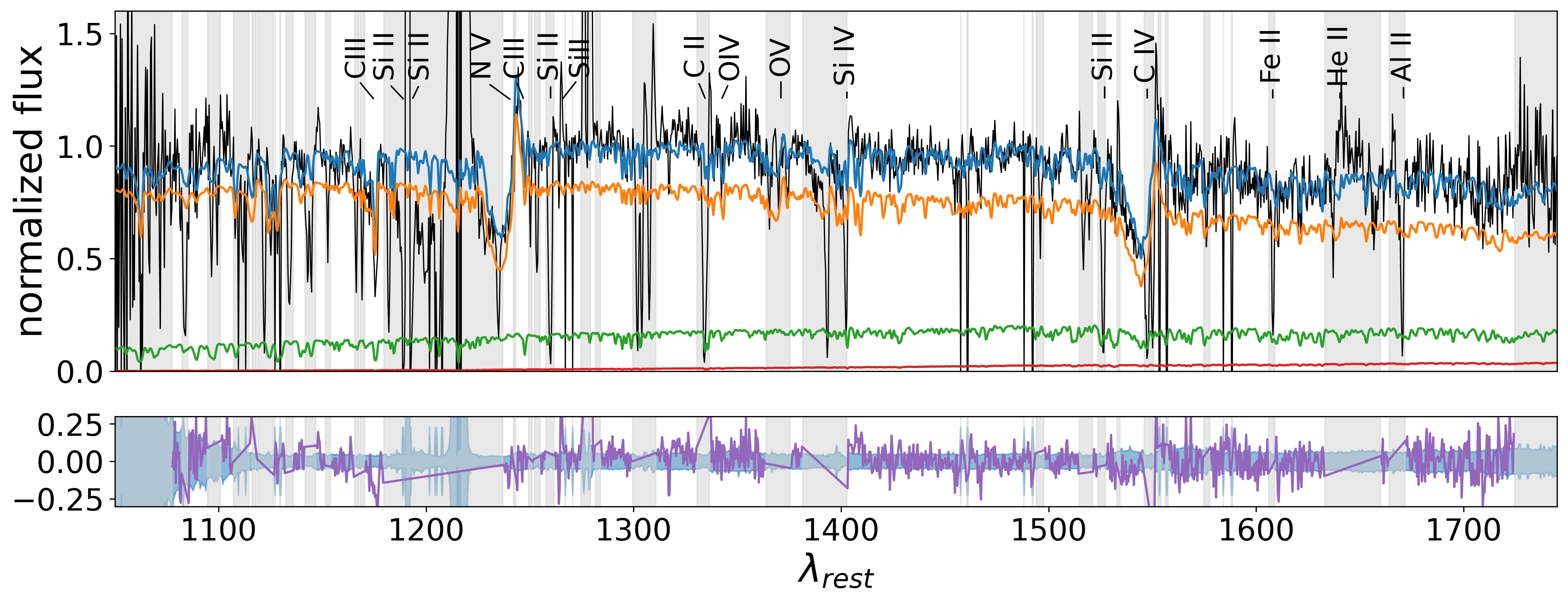}
	     \label{fig:fit_UV_knotB}
	    \includegraphics[width=\columnwidth]{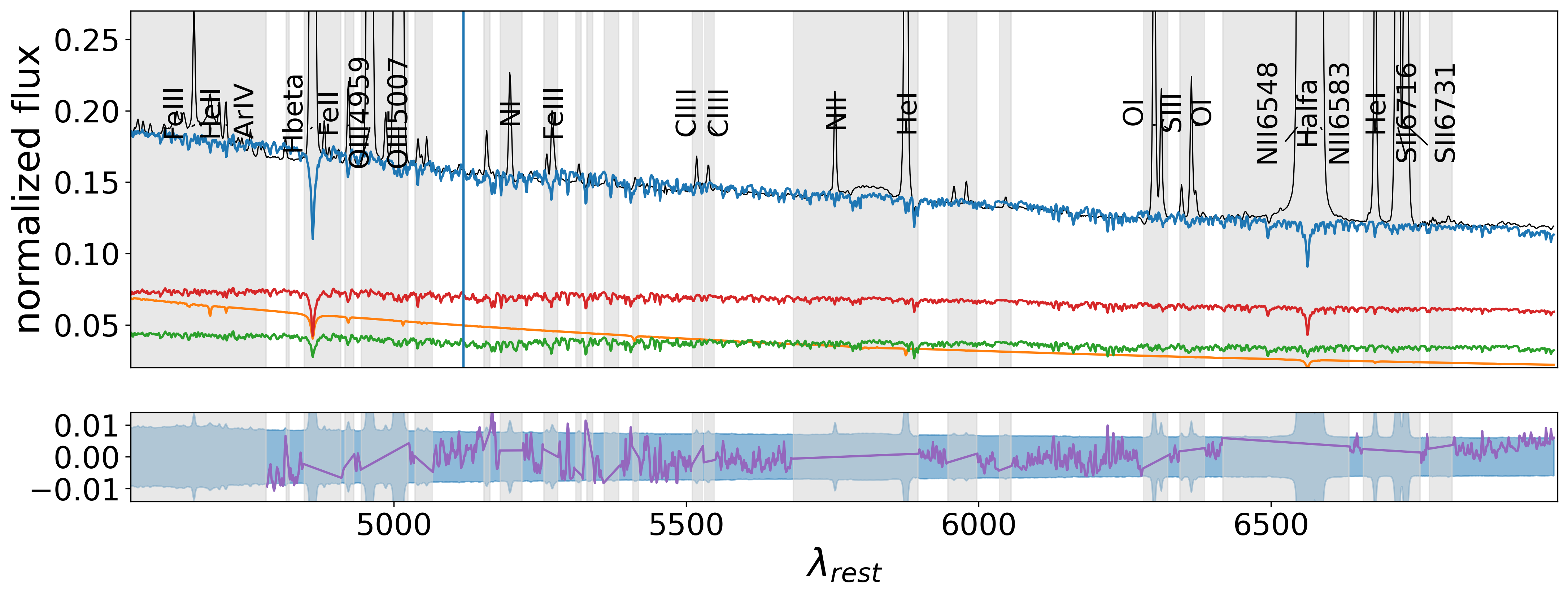}\\
	     \label{fig:fit_Opt_knotB}	     	     

	    \includegraphics[width=\columnwidth]{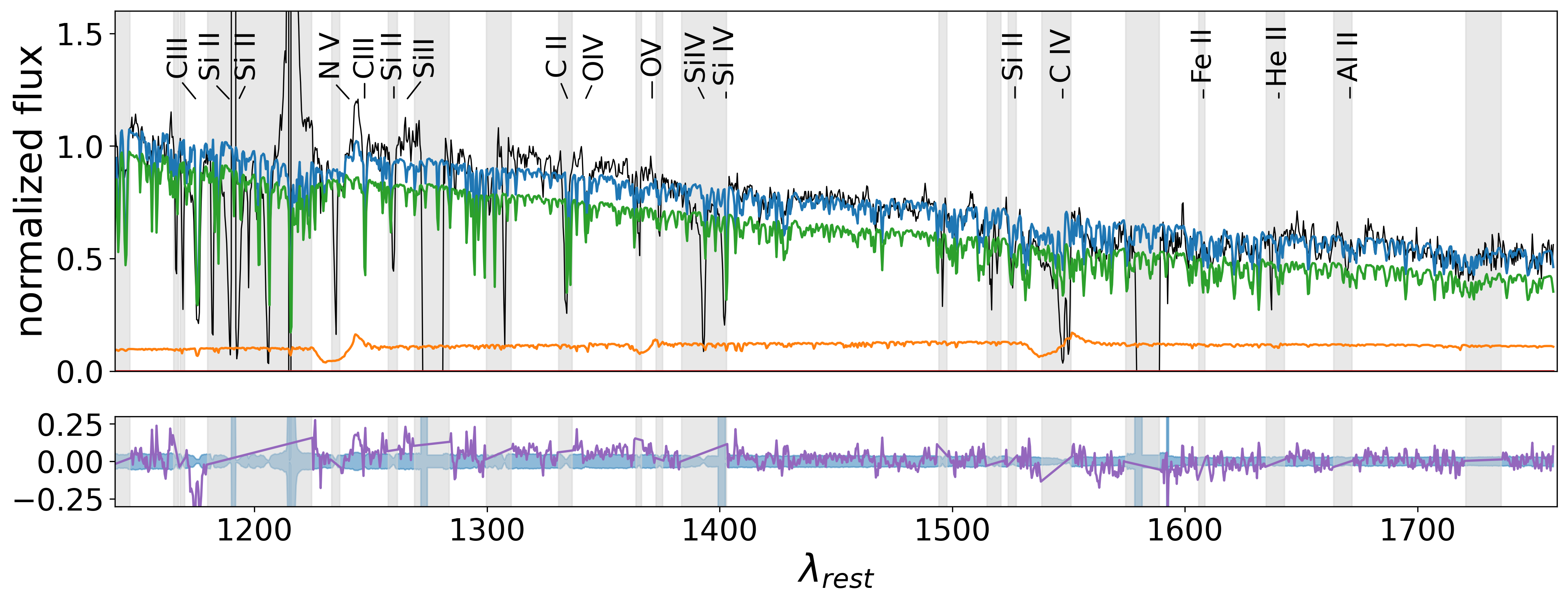}
	     \label{fig:fit_UV_knotC}
	    \includegraphics[width=\columnwidth]{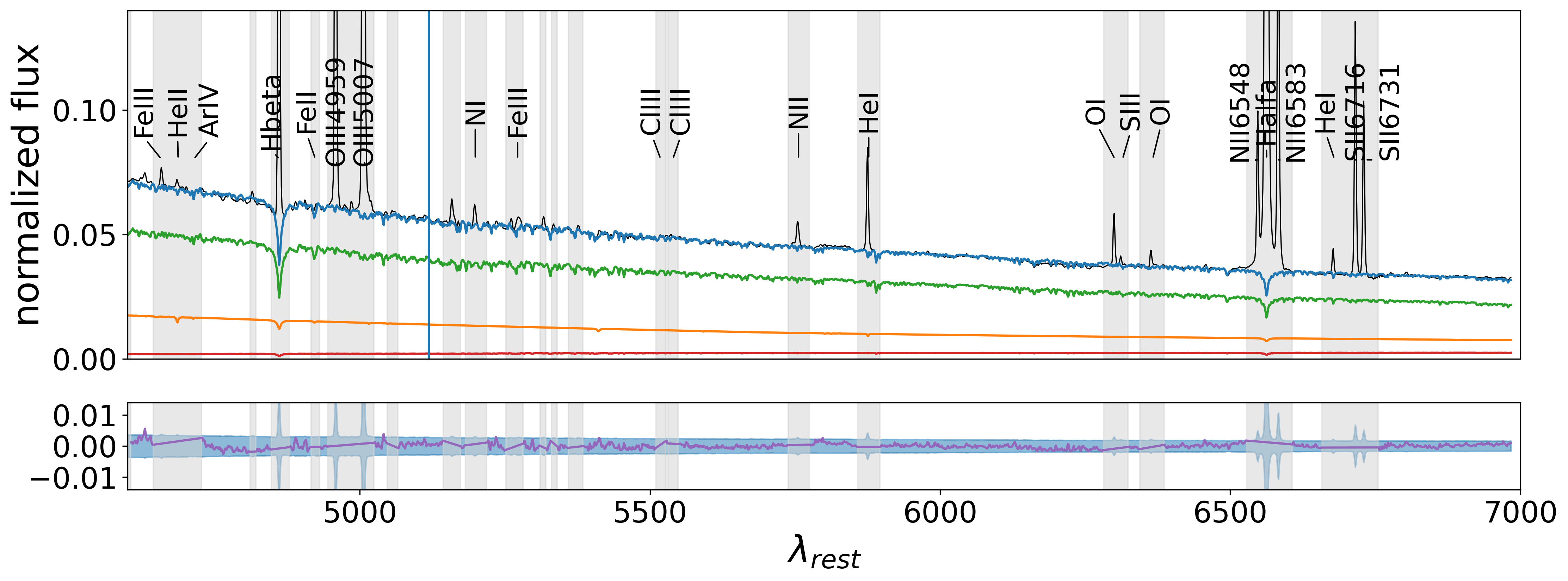}\\
	     \label{fig:fit_Opt_knotC}
	     
    \caption{FUV and optical spectroscopy of knots A, B and C (top to bottom) fitted simultaneously by our SB99-based model consisting of three stellar populations for the age intervals: 1-4 Myr, 4-40 Myr, 40-100 Myr. In black the normalised de-reddened data, in blue the best-fit model, in orange, green and red the models of the three single stellar populations. The gray vertical bands mark the spectral regions that have been masked out before fitting the data. Below each fitted spectrum we show a panel with the residuals in purple as a function of wavelength compared to the data uncertainties (light blue region).}
    \label{fig:fit_knots}
    
\end{figure*}

\begin{figure}

	    \includegraphics[width=0.5\columnwidth]{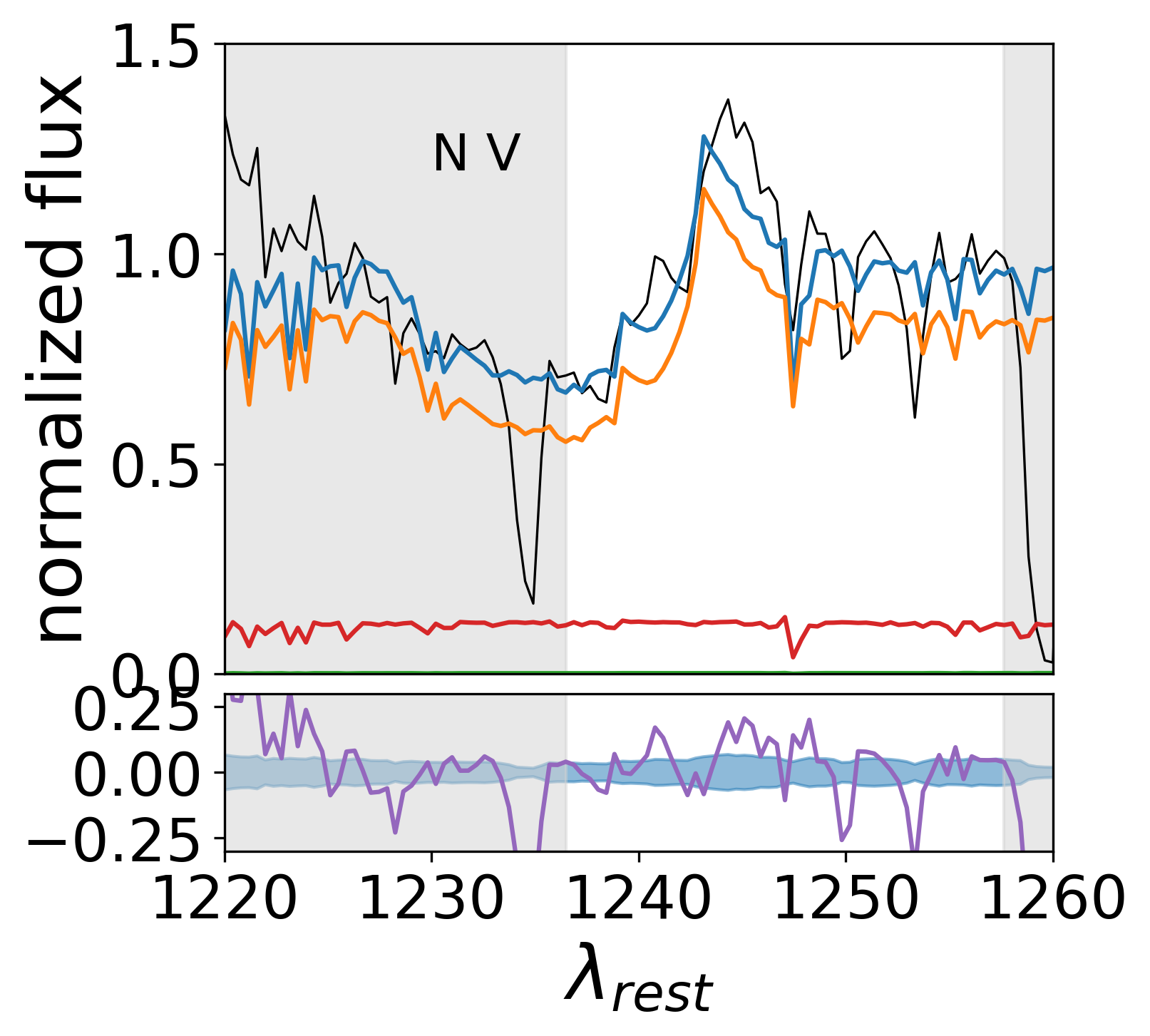}
	      \label{fig:fit_UV_knotA_NV}
	    \includegraphics[width=0.5\columnwidth]{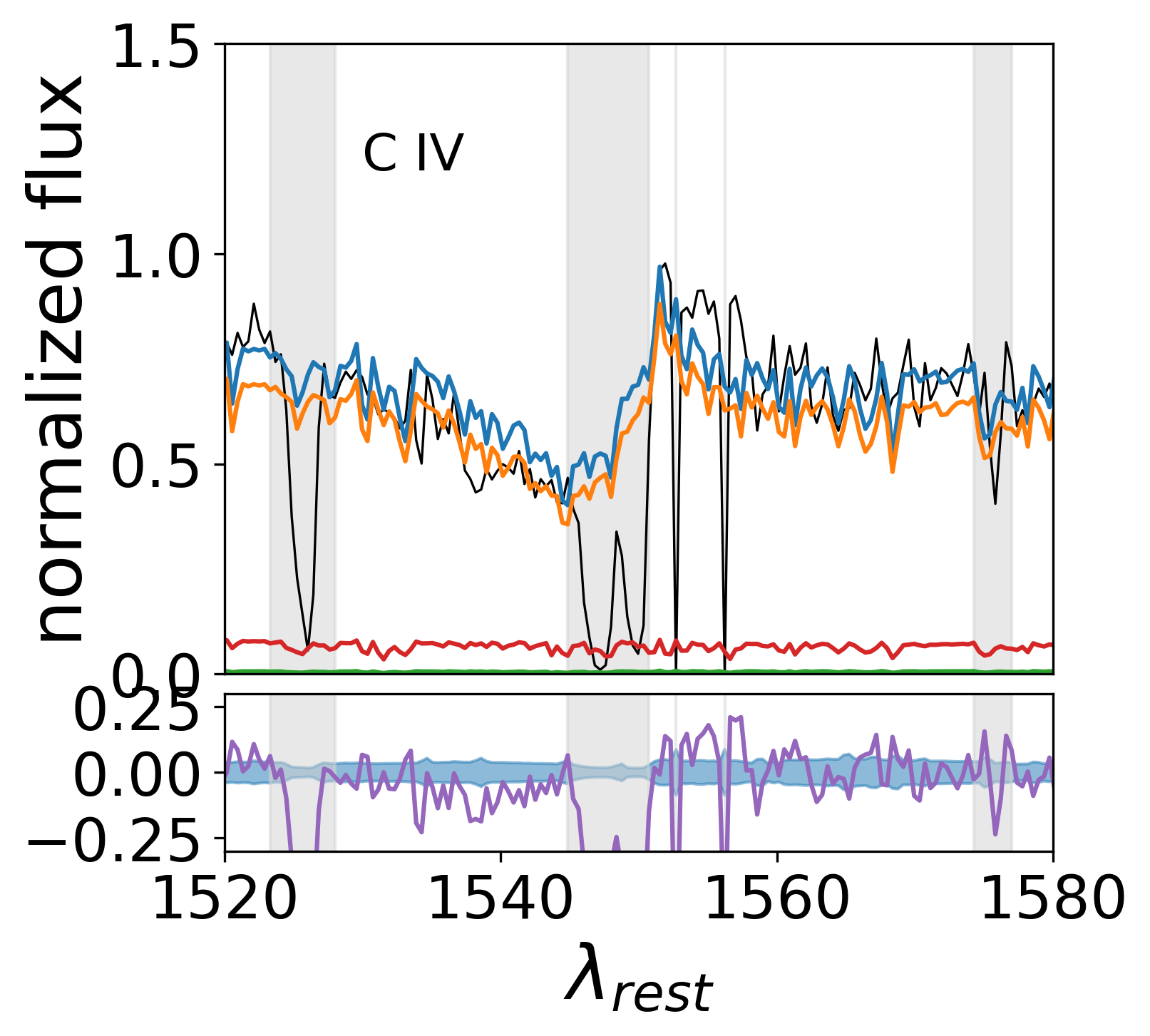}\\
	      \label{fig:fit_UV_knotA_CIV}
	      
	    \includegraphics[width=0.5\columnwidth]{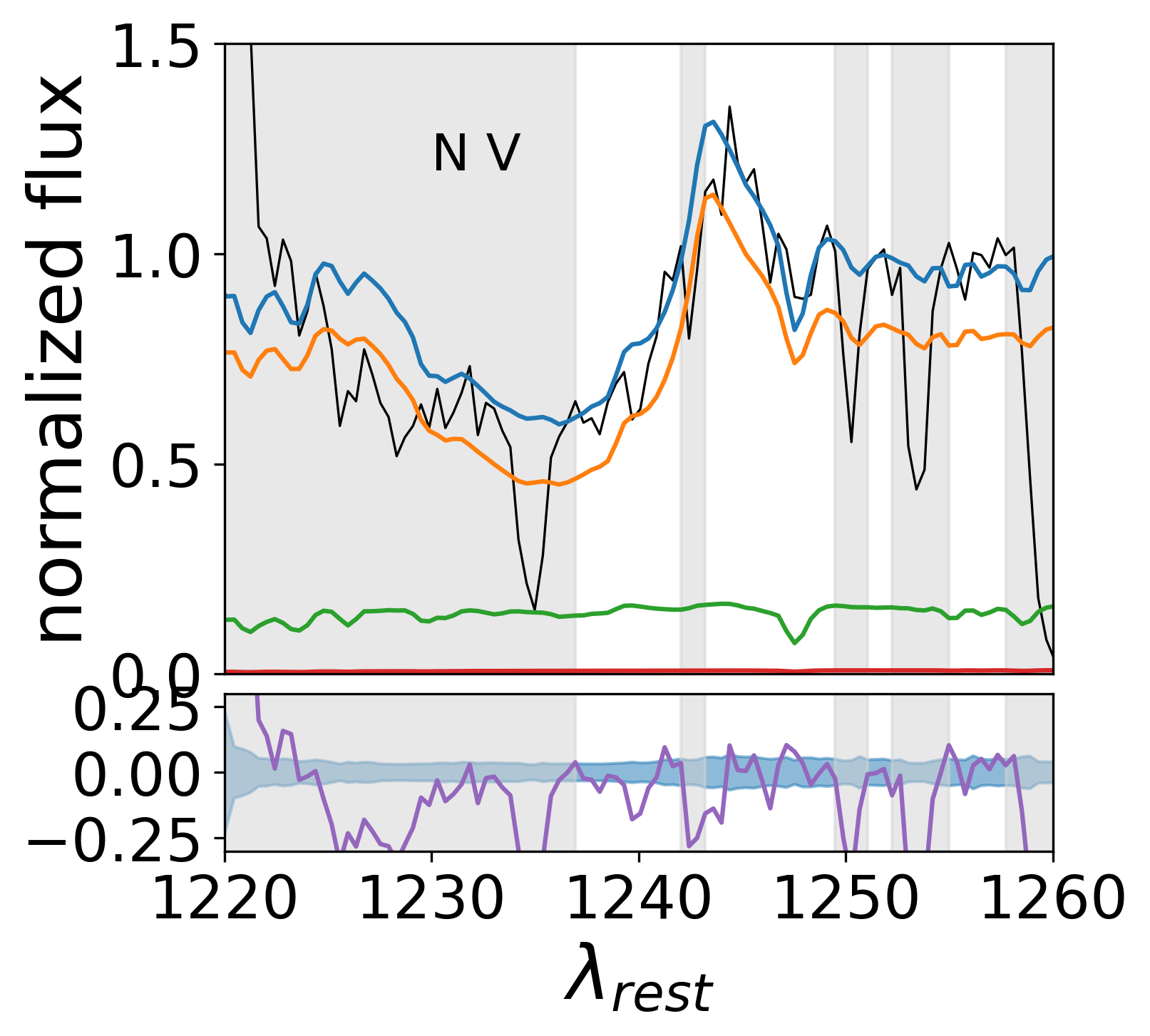}
	     \label{fig:fit_UV_knotB_NV}
	    \includegraphics[width=0.5\columnwidth]{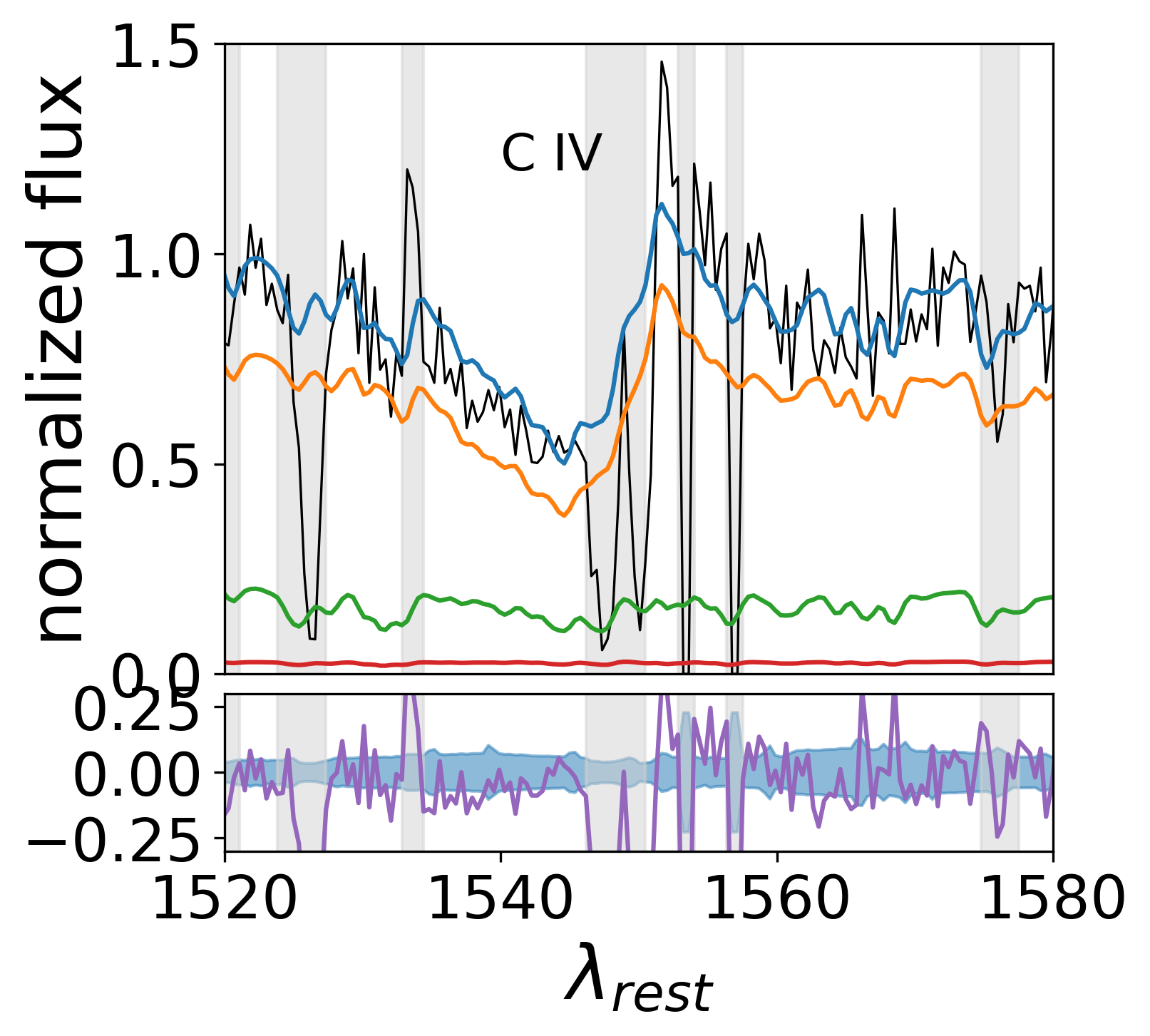}\\
	     \label{fig:fit_UV_knotB_CIV}	        

	    \includegraphics[width=0.5\columnwidth]{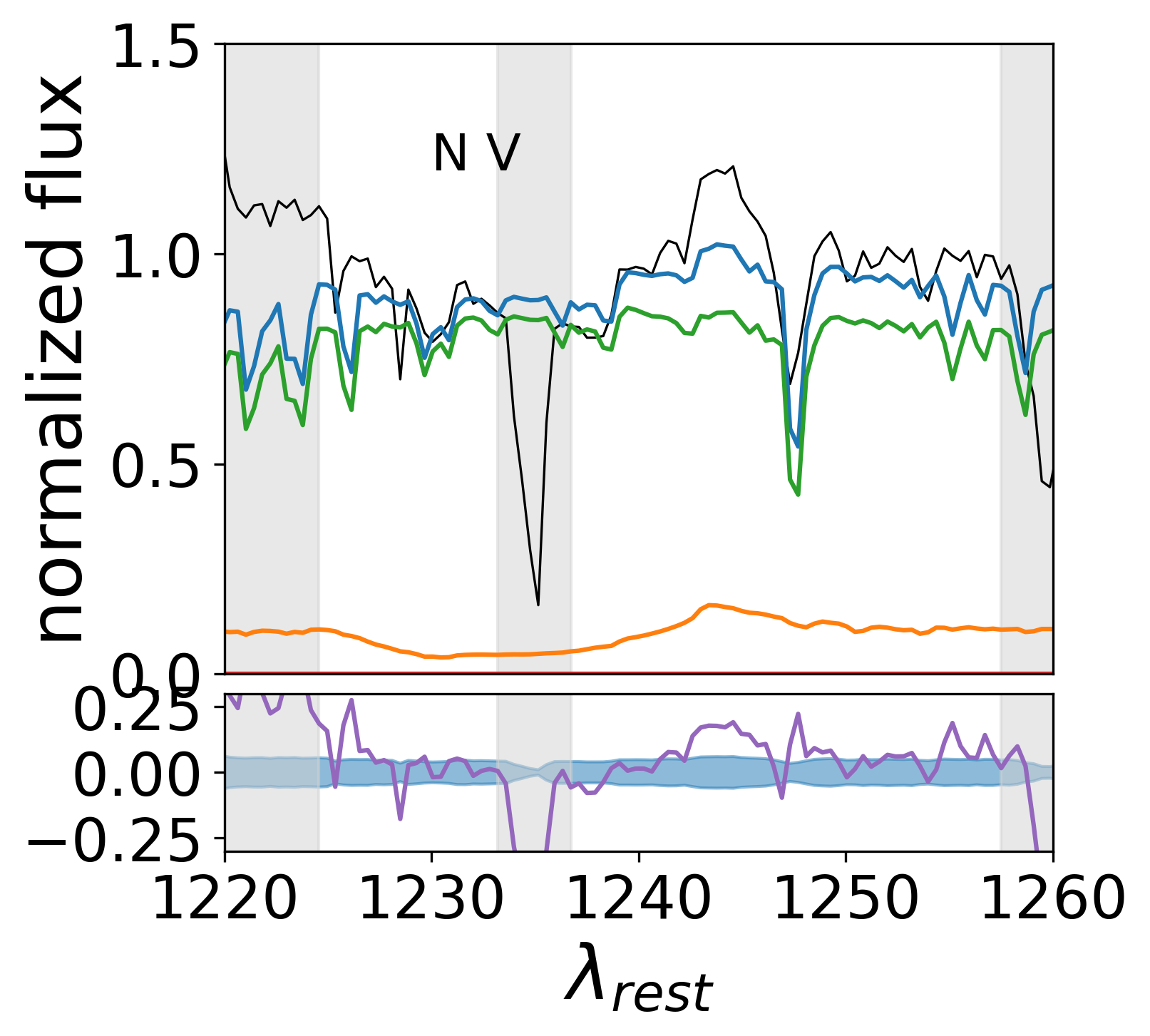}
	     \label{fig:fit_UV_knotC_NV}
	    \includegraphics[width=0.5\columnwidth]{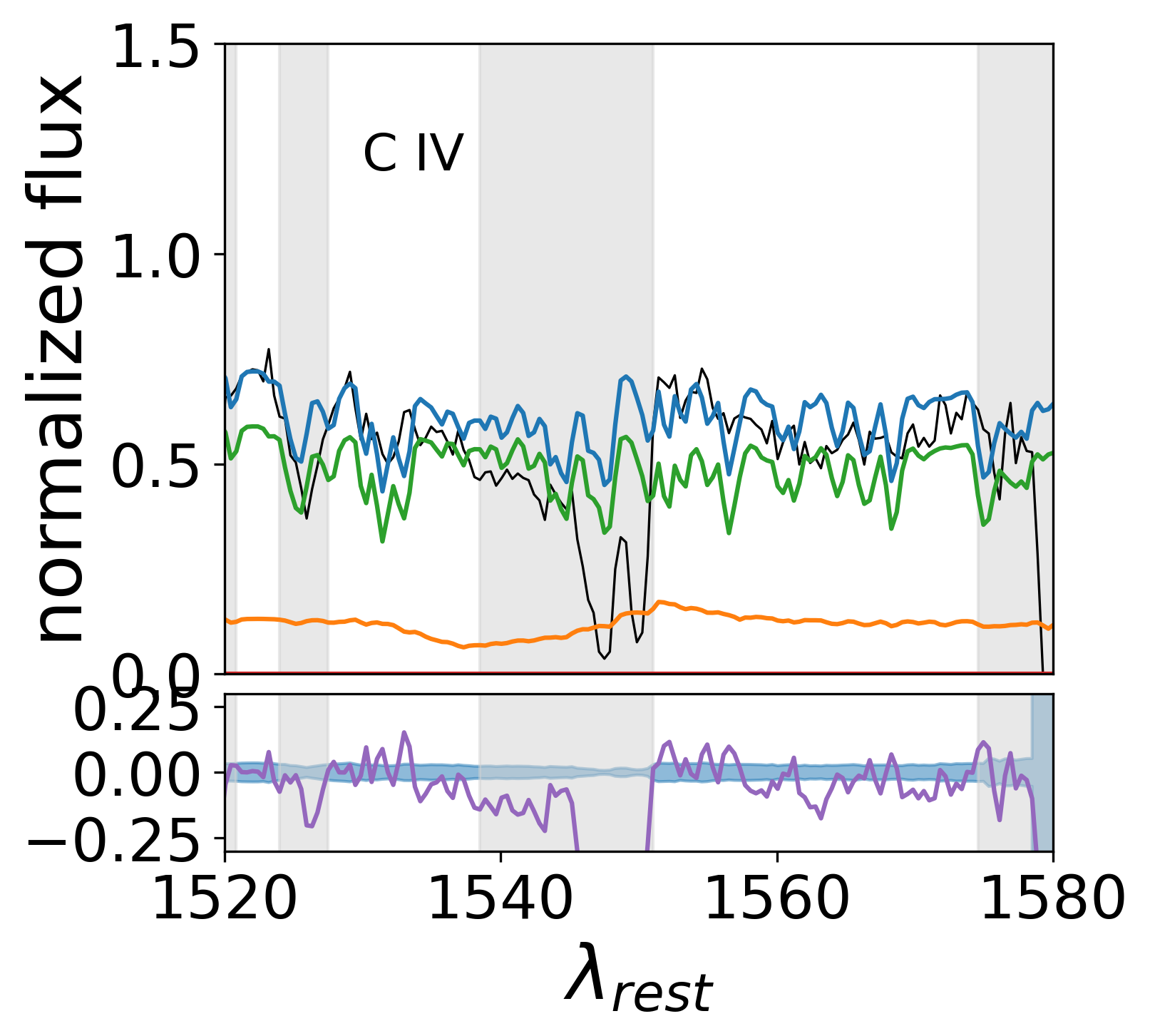}\\
	     \label{fig:fit_UV_knotC_CIV}
	     
    \caption{Zoom-in plots of the P-Cygni features (N V and C IV) of the FUV spectra of the Haro 11 knots A, B and C (top to bottom). Same legend applies as in Figure \ref{fig:fit_knots}}
    \label{fig:fit_zoom}
    
\end{figure}

The combined UV and optical spectra of the starburst knots in Haro 11 are well fitted both in terms of stellar continuum and spectral features ($\chi^2$ = 2.27, 1.89, 2.29 for knots A, B and C respectively). The P-Cygni lines NV 1241.5Å and CIV 1550Å, signatures of stellar winds of 3 Myr old stars, are most prominent and well reproduced by the model in knot A and B. In knot C, the young population has an age of 2 Myr and has not yet developed strong stellar winds. The model of knot C is not able to exactly reproduce the P-Cygni lines, likely because of its assumptions and limitations when treating stellar atmosphere and stellar winds \citep[see e.g. fig. 7 in ][]{Senchyna2021}. The residuals of the UV continuum are comparable with the statistical errors of the data. The optical spectra (obtained with MUSE) have a much lower statistical error and their continuum is well modeled in all knots with small residuals. In knots A and B, we find the signatures of Wolf-Rayet (WR) stars both in the UV (He 1640Å emission) and in the optical (the blue bump 4650 Å and the red bump 5808 Å). Despite these regions are excluded from the fits as the WR physics is not included in the model, the presence of such features confirms the age range recovered for the younger stellar populations.
Figure \ref{fig:fit_knots} shows the FUV and optical spectra and the best-fit models for each knot with the bottom panels illustrating the residuals as a function of wavelength and the flux uncertainties. A zoom into the P-Cygni lines N V and C IV of the FUV spectrum is shown in Figure \ref{fig:fit_zoom}.

Adopting a more extended star-formation history, such as a constant star formation rate, slightly affects the ages and the stellar feedback quantities but not at a significant level. The implications of our results on stellar population properties and feedback do not change because of a different assumption on the star formation mode. We also tested models with different numbers of stellar populations. The model with three populations is marginally better for knot A and C relative to the model with two populations (based on the $\chi^2$). However the fit for knot B is significantly improved by the addition of a third population. In the latter case three populations are essential for reproducing the optical continuum and as a consequence correctly reproduce the P-Cygni FUV lines. We decided to adopt three populations for all knots in order to keep the analysis for the individual knots consistent with each other.

\subsubsection{Cluster properties as inferred by different methods: photometry vs spectroscopy}
\label{sec:clusters_vs_pops}

In this work we have adopted two different methodologies to address the same physical problem, i.e. deriving the recent star formation history of the starburst currently ongoing within Haro11. We first study the star cluster population within the starburst by modelling the multiband photometry. We independently derive the same quantities by modelling the multi-wavelength spectroscopy extracted from three apertures centred on the three knots A, B and C. Since we modeled the spectroscopy with only three stellar populations, we have implicitly assumed that the knots have undergone three single-burst star-formation events, which was not assumed for the photometric study. Another difference is that the single cluster analysis assumed a fixed value of metallicity, whereas the spectroscopy model has the metallicity as a fitting parameter and therefore it does not have to be assumed a priori.

\begin{figure}

	\includegraphics[width=0.80\columnwidth]{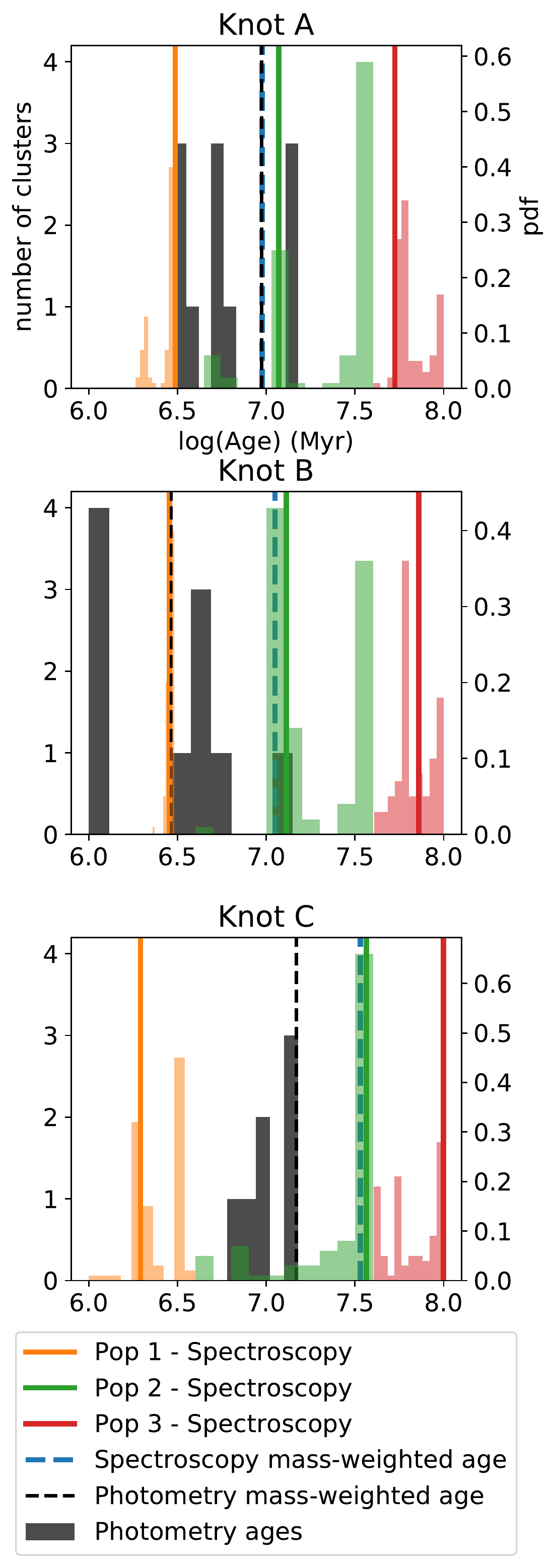}
	     
    \caption{Age histogram for the three starburst knots of Haro 11. In grey the age distribution of the single clusters detected with photometry within each knot aperture. In orange, green and red the probability density functions of the the three populations modeled using FUV and optical spectroscopy. The black dashed line shows the mass-weighted average of the cluster ages. The blue dashed line shows the mass-weighted age of the first two populations modeled with spectroscopy, which are the dominant populations in the feedback injected in the starburst region over the last 50 Myr (see text in Sec.\ref{sec:feedback}). The colored solid line show the best-fit ages of the three populations, with the same color scheme used in Figure \ref{fig:fit_knots} and \ref{fig:fit_zoom}.}
    
    \label{fig:hist_ages}
    
\end{figure}

Here we compare the ages of the clusters obtained with the photometric analysis to the ages of the spectroscopic fit. We focus on the age as it is one of the driving parameters, along with the mass, when evaluating the feedback of the starburst knots. Figure \ref{fig:hist_ages} shows, for each starburst knot, the age histogram (in grey) of the cluster population studied using photometry and the probability density functions derived from the spectroscopy fits (in colors). We include in this figure only the clusters lying within 0.5 kpc from the knot's centre (corresponding to the spectroscopic aperture), so that we could have a comparison with the results from spectroscopy. The solid lines show the best-fit age values of the spectroscopic fits. The dashed lines show the mass-weighted averages of the ages inferred with photometry and spectroscopy (in black and blue respectively). The spectroscopic weighted age has been derived by including only Pop 1 and Pop 2, since no cluster older than 20 Myr has been recovered using photometry. In knot A Pop 1 and Pop 2 are consistent (within 0.1 dex) with the youngest and oldest ages (respectively) inferred with photometry although the degeneracy of the model makes Pop 2 uncertain and possibly 0.5 dex older (see light green histogram in the top panel of Figure \ref{fig:hist_ages}). The mass-weighted age of the clusters (from photometry) in knot A is consistent with Pop 2 and it coincides with the mass-weighted age of Pop 1 and Pop 2. In knot B, the clusters are divided in three different age groups (1 Myr, 3 Myr, 10 Myr) and the two older groups agree with the spectroscopic ages of Pop 1 and Pop 2.  We also notice that the mass-weighted age for the clusters in knot B coincides with Pop 1 but it has a 0.5 dex offset from the mass-weighted age derived from spectroscopy. This offset between cluster and spectroscopic mass-weighted ages could be explained by the fact that very young clusters in knot B are obscured, thus, they are only partially detected in the FUV light (see below).  In knot C the photometry-based cluster ages lie between Pop 1 and Pop 2 and are consistent with the latter within the uncertainties. In all knots the older population Pop 3 is 1 dex older than the age of the oldest clusters detected with photometry.

\begin{figure}

	\includegraphics[width=0.92\columnwidth]{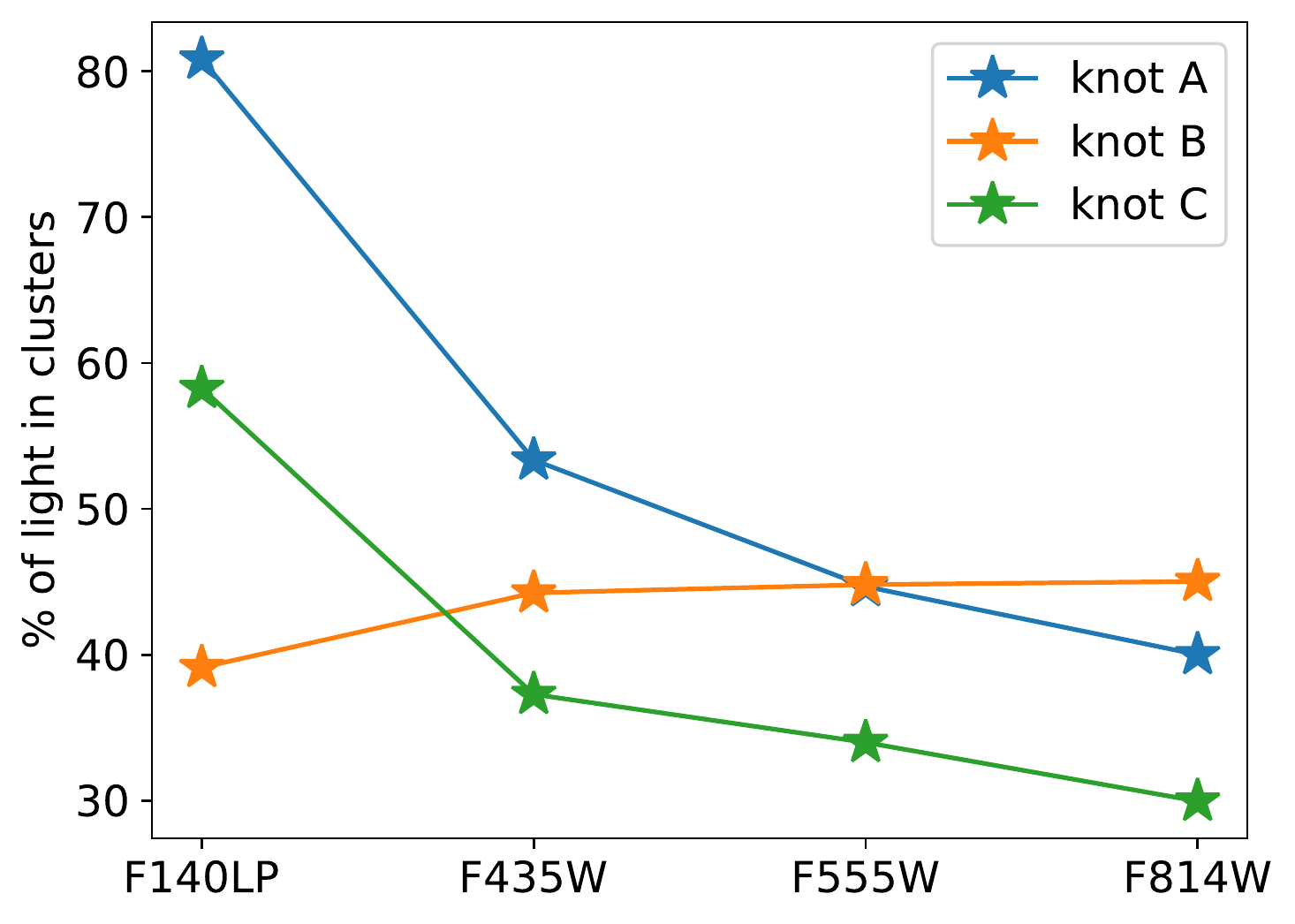}
	     
    \caption{Percentage of light measured in the cluster relative to the total light within the COS aperture, in 4 different filters, for knots A, B and C of Haro 11.}
    
    \label{fig:diffuse}
    
\end{figure}

One of the biggest differences in the two methods is the amount and type of light that is analysed: the photometric study considers only the light of the clusters, the spectroscopic study considers all the light within the COS aperture. We have measured the amount of light emitted by star clusters relative to the total light within the 1-kpc-diameter aperture centred in each knot, which includes the flux from both clustered and diffused stars. As Figure \ref{fig:diffuse} shows, star clusters appear to dominate the FUV light in knot A and C, suggesting that they are relatively young and constitute a significant fraction of the light at this wavelength.  Figure \ref{fig:diffuse} also points out a clear difference in knot B with respect to knots A and C. The percentage of FUV light in the clusters of knot B is much lower. When comparing the physical properties derived from the star cluster analysis and the spectral analysis of the region, we find that clusters in knot B are younger and more extinguished than revealed by the spectroscopic fit. We will discuss later how this difference impacts the output feedback in knot B in Section \ref{sec:outflow_results}. In the red optical filters (and also in the bluer ones for knot B) less than 50\% of the light is produced by clusters meaning that most of the light is in the diffuse component. This suggests the presence of an underlying older population that has dispersed and is confirmed by the age comparison of the best-fit ages in Figure \ref{fig:hist_ages}, where in all the knots the third population of the spectroscopy model is older than the photometry-inferred ages.

\section{Feedback budget and properties of the ionised gas}
\label{sec:feedback+outflow}

\subsection{Stellar feedback output}
\label{sec:feedback}
In our study we are mostly interested in the properties of the young star clusters and their effect on the surrounding environment. Since the stellar feedback is dominated by the star clusters in the knots, we expect to see the effect of the stellar feedback in the kinematics and physical properties of the ionised gas surrounding the clusters. The timescale of a mechanical feedback process such as the transport of energy and momentum by baryons across the starburst knots of Haro 11 is roughly a few Myr. This can be estimated (as we do later when studying the outflowing gas in Sec. \ref{sec:outflow}) by assuming a wind velocity. Therefore we are interested in the energy released by the two youngest stellar populations only (<40 Myr). We note that timescale for radiative feedback is shorter since photons can transport energy and momentum thousand of times faster than the outflow described in Sec. \ref{sec:outflow}.

We used the best-fit models described in Sec.\ref{sec:spectral_fits} to quantify the \textit{stellar} feedback in terms of photo-ionisation rate $Q_H$, power of stellar winds $P_W$, power of SNe $P_{SN}$ as well as their integrated energy since the onset of star-formation, $E_W$ and $E_{SN}$ respectively. These quantities, which we report in Table \ref{tab:feedback}, were provided by the SB99-based code and are dependent on the results of the spectroscopy fit. The uncertainties are calculated with a classic Monte Carlo approach. Table \ref{tab:feedback} also reports the \textit{cluster} feedback that we derived from the parameters of the clusters detected with photometry. In knot A, we observe that the stellar feedback produced by the star cluster population is, within the uncertainties, in agreement with the feedback values determined with spectroscopy, suggesting that in this region cluster feedback dominate the energetics. In knot B, the cluster photo-ionisation rate and stellar wind power are higher than the values derived from the spectroscopic analysis. These differences can be attributed to the large extinction values derived for these young clusters that are not contributed to the FUV light and therefore remain undetected in the spectroscopic analysis (see discussion in Sec. \ref{sec:clusters_vs_pops}). In knot C, the energetics are dominated by the stellar population probed within the COS aperture, with star clusters only contributing up to 10\% of the stellar feedback estimated in the region. We remark that the numbers in Table \ref{tab:feedback} have large errors and should be considered only as orders of magnitude estimates in our general assessment of stellar feedback acting on the ionised galactic ISM.

The feedback from young stars that we measured is quite similar in the three starburst knots of Haro 11, with a photo-ionisation rate of $10^{53}$ photons per second and a stellar-winds power of $10^{40}$ erg/s. On the other hand, the energy released by SNe depends on how many explosions have occurred and, therefore, on the intermediate age stellar population (Pop 2). The SN feedback is highest for knot C ($10^{57}$ erg), which has a massive Pop 2, and at least one order of magnitude lower for knot A and B, which have younger Pop 2. Given the SB99 assumption that one explosions releases $10^{51}$ erg, this means that in knot C, over the past 40 Myr exploded about one million SNe, implying a rate of one SN explosion every 40 years in a 1 kpc-size region.
The difference in the amount of energy released by SNe between knot C and B also reflects the different star formation history of the knots, i.e. knot C being older than knot A and B where the starburst is still very young. 

 Importantly, the large energy budget of the stellar feedback and high supernova rate of the recent starburst of Haro 11 can explain the diffuse soft X-ray emission observed with the Chandra observatory. Such emission extends up to about 6 kpc from the centre of the starburst and indicates the presence of a hot-gas galactic wind with a kinetic power of $\sim 10^{42}\rm\,erg/s$ and a total thermal energy of $10^{56}\rm\,erg$ \citep{grimes2007}, that are comparable with the energetic of the stellar feedback estimated in this work.

\begin{table*}
	\centering
	\caption{Total stellar feedback and cluster feedback in the three knots of Haro 11, derived respectively with spectroscopy and photometry.}
	\label{tab:feedback}
	\renewcommand{\arraystretch}{1.5} 
	\begin{tabular}{lccc} 
		\hline
		 & knot A & knot B & knot C\\
		 & stellar | cluster & stellar | cluster & stellar | cluster\\
		\hline
		Photo-ionisation rate, $Q_H$ (photons/s) & $2.3^{+3.1}_{-0.1}\cdot10^{53}$ | $1.9\cdot10^{53}$ & $1.6^{+0.5}_{-0.0}\cdot10^{53}$ | $2.7\cdot10^{54}$ & $5.4^{+0.3}_{-3.0}\cdot10^{53}$ | $1.8\cdot10^{51}$\\
		
		Power of stellar winds, $P_W$ (erg/s) & $4.1^{+1.6}_{-0.2}\cdot10^{40}$ | $7.5\cdot10^{40}$ & $3.1^{+4.9}_{-0.0}\cdot10^{40}$ | $1.1\cdot10^{41}$ & $6.1^{+12}_{-0.4}\cdot10^{40}$ | $1.9\cdot10^{39}$ \\
		
		Power of SNe, $P_{SN}$ (erg/s) & $2.4^{+0.3}_{-1.6}\cdot10^{41}$ | $1.2\cdot10^{41}$ & $2.3^{+5.3}_{-1.4}\cdot10^{41}$ | $8.1\cdot10^{40}$ & $8.3^{+0.2}_{-3.7}\cdot10^{41}$ | $8.8\cdot10^{40}$\\
		
		Total power, $P_{Tot}$ (erg/s) & $2.8^{+0.4}_{-1.5}\cdot10^{41}$ | $2.0\cdot10^{41}$ & $2.6^{+5.5}_{-0.4}\cdot10^{41}$ | $1.9\cdot10^{41}$ & $8.9^{+0.4}_{-3.3}\cdot10^{41}$ | $9.0\cdot10^{40}$\\

		Energy of stellar winds, $E_{W}$ (erg) & $3.7^{+4.0}_{-0.4}\cdot10^{55}$ | $4.6\cdot10^{55}$ & $3.7^{+27.4}_{-0.0}\cdot10^{55}$ | $1.3\cdot10^{56}$ & $1.7^{+1.6}_{-0.1}\cdot10^{56}$ | $1.1\cdot10^{56}$ \\	
		
		Energy of SNe, $E_{SN}$ (erg) & $7.4^{+15.9}_{-0.7}\cdot10^{55}$ | $6.2\cdot10^{55}$ & $8.4^{+13.4}_{-2.6}\cdot10^{55}$ | $2.8\cdot10^{55}$ & $1.3^{+0.1}_{-1.1}\cdot10^{57}$ | $2.3\cdot10^{56}$\\
		
		Total energy, $E_{Tot}$ (erg) & $1.1^{+6.1}_{-0.0}\cdot10^{56}$ | $1.1\cdot10^{56}$ & $1.2^{+27.5}_{-0.0}\cdot10^{56}$ | $1.6\cdot10^{56}$ & $1.5^{+1.0}_{-0.1}\cdot10^{57}$ | $3.4\cdot10^{56}$ \\			
		
		
	
		\hline
	\end{tabular}
\end{table*}

\subsection{Outflowing ionised gas}
\label{sec:outflow}

\subsubsection{Method}

MUSE spectroscopy allows us to investigate the profile of optical emission lines such as H$\alpha$, H$\beta$, [OIII], [SII], detected at very high significance and $S/N > 100$ integrated over the line. These lines appear broad and asymmetric and are best described by the sum of two Gaussian components: a narrow component centered on the peak of the line and a broader component tracing ionised gas moving along the line of sight towards the observer.

We interpret this broad blue-shifted component as a trace of a compact outflows around knots A and B (we do not see it in knot C) and discuss other possible interpretations in Section \ref{sec:interp}. In order to study the physical properties of the ionised gas outflowing from the knots we followed the steps below.
\begin{itemize}
    \item We extracted MUSE spectra from a circular region equivalent to the COS aperture (1.25\arcsec\, = 525 pc), applying the the correction for the COS sensitivity function (so that the outflow measurements are comparable with the cluster energetics) as well as correcting for the stellar absorption. 
    \item We fitted the [OIII]4959 line to determine the kinematics (central velocity and dispersion of both the narrow and the broad Gaussian components). We opted for [OIII]4959 as trace of the kinematics because H$\beta$ can be affected by stellar absorption and H$\alpha$ wings partially overlap with the [NII] ones. [OIII]5007 is stronger than [OIII]4959 but it is contaminated by another line, most prominent in knot A, which we identify as HeI (5015.68Å). We corrected the dispersion measurement taking into account the MUSE Line Spread Function and defined the maximum velocity of the ionised outflow to be $v_{max}=v_{cen}+2\sigma$, where $v_{cen}$ is the central velocity and $\sigma$ is the corrected dispersion of the broad component. This is equivalent to the velocity where 95\% of the total broad component emission is encompassed, as measured in other studies of gas kinematics \citep{rose2018,canodiaz2012} 
    \item We fixed the kinematics of all the rest of the emission lines of interest and fitted them allowing to vary only the amplitudes of the narrow and broad components, because all the emission lines are expected to be produced by the same physical gas components. 
    \item We fitted the H$\beta$ line to measure the fluxes of the two components and to determine the luminosity of the broad component $L_{H\beta,out}$, tracing outflowing ionised gas
    \item We fitted the H$\alpha$ line to measure the fluxes of the two components, calculated the H$\alpha$-H$\beta$ decrement and corrected all fluxes for reddening attenuation considering the both the narrow and the broad components separately, using the \texttt{pyneb} \citep[][]{luridiana2015} package in python and the CCM89 law.  
    \item We fitted the [SII] doublet simultaneously. We determined the density of the outflowing ionised gas $n_{e,out}$ with \texttt{pyneb} using the broad components of the [SII] doublet.
    \item We estimated the dynamical time of the outflow on the physical scale of the knots (COS aperture), which is R=525 pc, as $t_{dyn}=R/v_{max}$.
    \item We used the luminosity of the broad component of the H$\beta$ line and the density to determine the outflowing ionised gas mass.
    \item We finally derived the mass outflow rate, the kinetic power, and the momentum rate, which are the rates at which the outflow carries mass, energy and momentum out of the regions of interest (individual starburst knots). 
\end{itemize} 

The equation used to derive the outflow mass is:
\begin{equation}
    \label{eq:mass}
    M_{out} = f_{He}\,L_{H\beta,out}\,m_p / (k_B\,h\,\nu\,n_{e,out}) 
\end{equation}

where $m_p$ is the mass of the proton, $k_B$ is the case B recombination coefficient (which means that we assumed optically thick gas), $h$ is the Planck constant and $\nu$ is the H$\beta$ frequency. $f_{He}$ is a correction factor of 1.16 that accounts for two facts: (i) the electron density that we measure with the [SII] diagnostics includes electrons from He (16\% assuming all the He is doubly ionised), (ii) the ionised gas mass must include He nuclei composed by 2 protons and two neutrons. The equations used to derive the mass, energy and momentum rates of the outflow are:

\begin{equation}
    \label{eq:outflow_rates}
    \begin{aligned}
     \dot{M}_{out} = M_{out} / t_{dyn}\\
     \dot{E}_{out} = \frac{1}{2} \dot{M}_{out} v_{max}^2\\
     \dot{p}_{out} = \dot{M}_{out} v_{max}
    \end{aligned}
\end{equation}

\subsubsection{Results}
\label{sec:outflow_results}

Table \ref{tab:lines} lists the kinematical and physical properties (and derived quantities) of the ionised gas for each knot, inferred using the narrow and broad components of the emission lines, as well as their total flux. The outflowing ionised gas, traced by the broad components, has a larger dispersion velocity (up to a factor of two) and is more tenuous than the gas traced by the narrow components. The photo-ionisation rate derived from the $H\alpha$ luminosity is of the same order of the one inferred from the Starbust99-based model of the stellar populations (see Table \ref{tab:feedback}) for all the knots except B. The slightly lower value of the photo-ionisation rate derived from the H$\alpha$ flux in knot C corroborates the suggestion that this region has been evacuated from the nebular gas, see Section \ref{sec:outflow_results} and \cite{oestlin2021}. For the latter the rate of ionising photons is a factor of 5 higher than the value predicted by the spectroscopic model. The star formation rate, derived from the extinction-corrected Ha flux, is a few solar masses per year and highest for knot B. The narrow component gas in knot B is the most extinguished, whereas the gas in knot C is the least extinguished, according to the H$\alpha$-H$\beta$ decrement. The high reddening in knot B is seen also when studying the clusters with photometry and it might hide a significant fraction of UV flux from young stars that is missed by the spectroscopic analysis. This is also suggested by the fact that the E(B-V) values of the younger populations of knot B in the spectroscopic model are lower than the one inferred from the H$\alpha$-H$\beta$ decrement, which gives a lower photo-ionisation rate. This hypothesis might explain the discrepancy in knot B between the photo-ionisation rate measured in the optical spectrum and the one derived from the SB99 spectroscopy fit. In order to test this hypothesis we run SB99 simulations with a Kroupa IMF and the Padova stellar evolutionary tracks as the photometric models described in Section \ref{sec:phot_clusters}. We then used these SB99 models to infer the photo-ionisation rate of each cluster detected with photometry within the three knots A, B and C. We found $Q_H = 2\cdot10^{53}\,\rm s^{-1}$ for knot A, $Q_H = 3\cdot10^{54}\,\rm s^{-1}$ for knot B and $Q_H = 2\cdot10^{51}\rm\, s^{-1}$ for knot C. The photometry and spectroscopy values are in agreement in knot A and disagree in knot B and C. The low rate found in knot C is due to the absence of young clusters in the photometric catalogue. The disagreement in knot B instead indicates that the high reddening attenuation is clearly affecting the recovered total photo-ionisation rate in the young stellar population. Using the star clusters photometry we obtained a factor of three more photons per second than the ones that are actually ionising the gas and being seen in H$\alpha$ emission. We suggest that a large fraction of ionising photons is reprocessed by dust or escaping the region that we are studying. We conclude that the most reliable estimate of the photo-ionisation rate for knot B is the one derived using the clusters and the corresponding photometric models. The value inferred from the spectroscopic model likely underestimates the intrinsic UV flux from young stars due to a low E(B-V), whereas the value derived from H$\alpha$ flux does not include the photons absorbed and reprocessed by dust as well as potential leaking photons.
We notice that despite the E(B-V) values are calculated using the same data as in \cite{oestlin2021}, the values obtained using the total flux of the lines are slightly different due to a few differences in the calculation. We used the Gaussian model to measure the flux (instead of integrating the observed flux), we neglected the MW reddening (E(B-V) = 0.01) and used the CCM89 extinction law (rather than the one by \cite{Fitzpatrick1999}).

\begin{table*}
	\centering
	\renewcommand{\arraystretch}{1.5} 
	\begin{tabular}{lccc} 
		\hline
		      & knot A & knot B & knot C\\
		\hline
		
		$\sigma_{narrow}$ (km/s) & 94.0$\pm0.3$ & 99.0$\pm0.4$ & 90.0$\pm0.2$\\
		
		$E(B-V)_{narrow}$ & 0.58$\pm0.07$ & 0.75$\pm0.07$ & 0.46$\pm0.07$\\
		
		$n_{e,narrow}$ (cm$^{-3}$) & 548$\pm4$ & 435$\pm4$ & <10\\	
		
		\hline
		
		$\Delta v$ (km/s) & -59.0$\pm0.8$ & -18.0$\pm0.3$ & - \\
		
		$\sigma_{broad}$ (km/s) & 180.0$\pm0.6$ & 207$\pm1$ & - \\		
	
		$E(B-V)_{broad}$ & 0 & 0 & - \\

		$n_{e,broad}$ (cm$^{-3}$) & <10 & 174$\pm6$ & -\\
		
		$F_{H\beta,broad}\rm\, (erg/s/cm^2)$ & $(2.4\pm0.1)\cdot10^{-14}$ & $(5.4\pm0.3)\cdot10^{-14}$ & - \\			
		$M_{out} \rm (M_{\odot})$ & $>16 \cdot 10^6$ & $(2.2\pm0.1) \cdot 10^6$ & - \\

		\hline

		$E(B-V)_{tot}$ & $0.22\pm0.07$ & $0.43\pm0.07$ & 0.46$\pm0.07$\\

		$n_{e,tot}$ (cm$^{-3}$) & $124\pm1$ & $354\pm1$ & <10\\
		
		$F_{H\alpha,tot}\rm\, (erg/s/cm^2)$ & $(3.3\pm0.2)\cdot10^{-13}$ & $(1.20\pm0.06)\cdot10^{-12}$ & $(3.4\pm0.2)\cdot10^{-13}$\\
		
		$L_{H\alpha}$ (erg/s) & $(2.9\pm0.1)\cdot10^{41}$ & $(1.05\pm0.06)\cdot10^{42}$ & $(2.9\pm0.1)\cdot10^{41}$\\
		
		$Q_{H}(H\alpha)$ (photons/s) & $(2.3\pm0.1)\cdot10^{53}$ & $(8.2\pm0.4)\cdot10^{53}$ & $(2.3\pm0.1)\cdot10^{53}$\\
	
		$SFR(H\alpha)$ $\rm (M_{\odot}/yr)$ & $1.6$ & $5.7$ & $1.6$\\
		
		\hline

	\end{tabular}
	\caption{Physical properties derived from the optical emission lines. The table is divided in three parts. In the first part are listed the properties of the narrow components of the emission lines: dispersion velocity, reddening attenuation and electron density. In the second part are listed the properties of the broad component of the emission lines: velocity shift of the central wavelength of the broad component relative to the narrow one, dispersion velocity, reddening attenuation, electron density, H$\beta$ flux, mass of ionised gas traced by this component. In the third part are listed the properties derived from the total flux of the emission lines: reddening attenuation, electron density, H$\alpha$ flux, luminosity of the total $H\alpha$ line; photo-ionisation rate and star-formation rate both inferred from the $H\alpha$ luminosity \citep{kennicutt1998} assuming case B recombination and an electron temperature of 10000 K. The dispersion velocities are corrected for the MUSE Line Spread Function (LSF); electron densities are derived from the SII doublet diagnostics. For knot C, since the broad component of most of the emission lines is only marginally detected and undetected for [SII] we do not report the values in the second part of the table.}
\label{tab:lines}
\end{table*}

\begin{figure*}

	    \includegraphics[width=0.6\columnwidth]{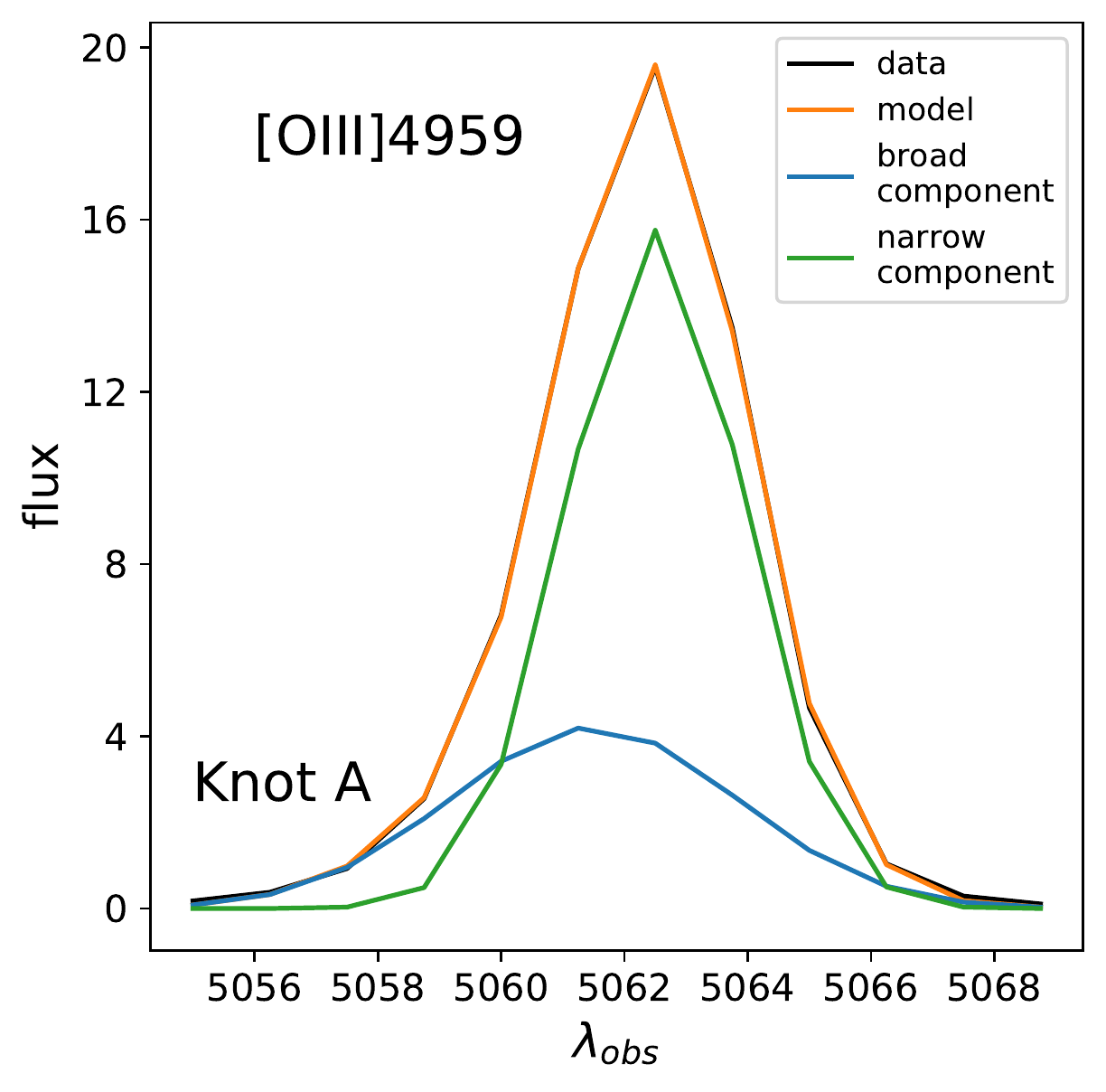}
	    \includegraphics[width=0.6\columnwidth]{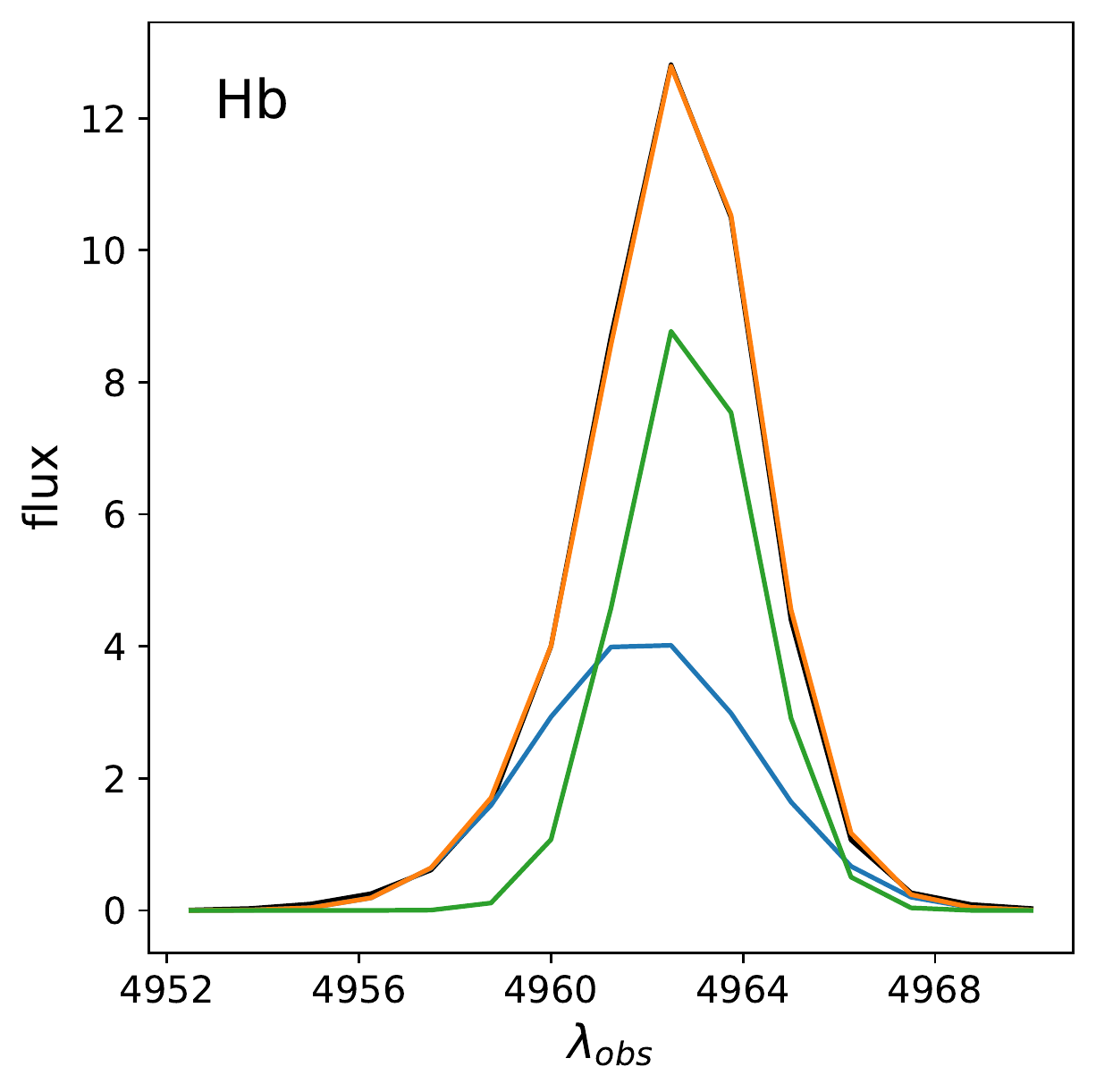}
	    \includegraphics[width=0.6\columnwidth]{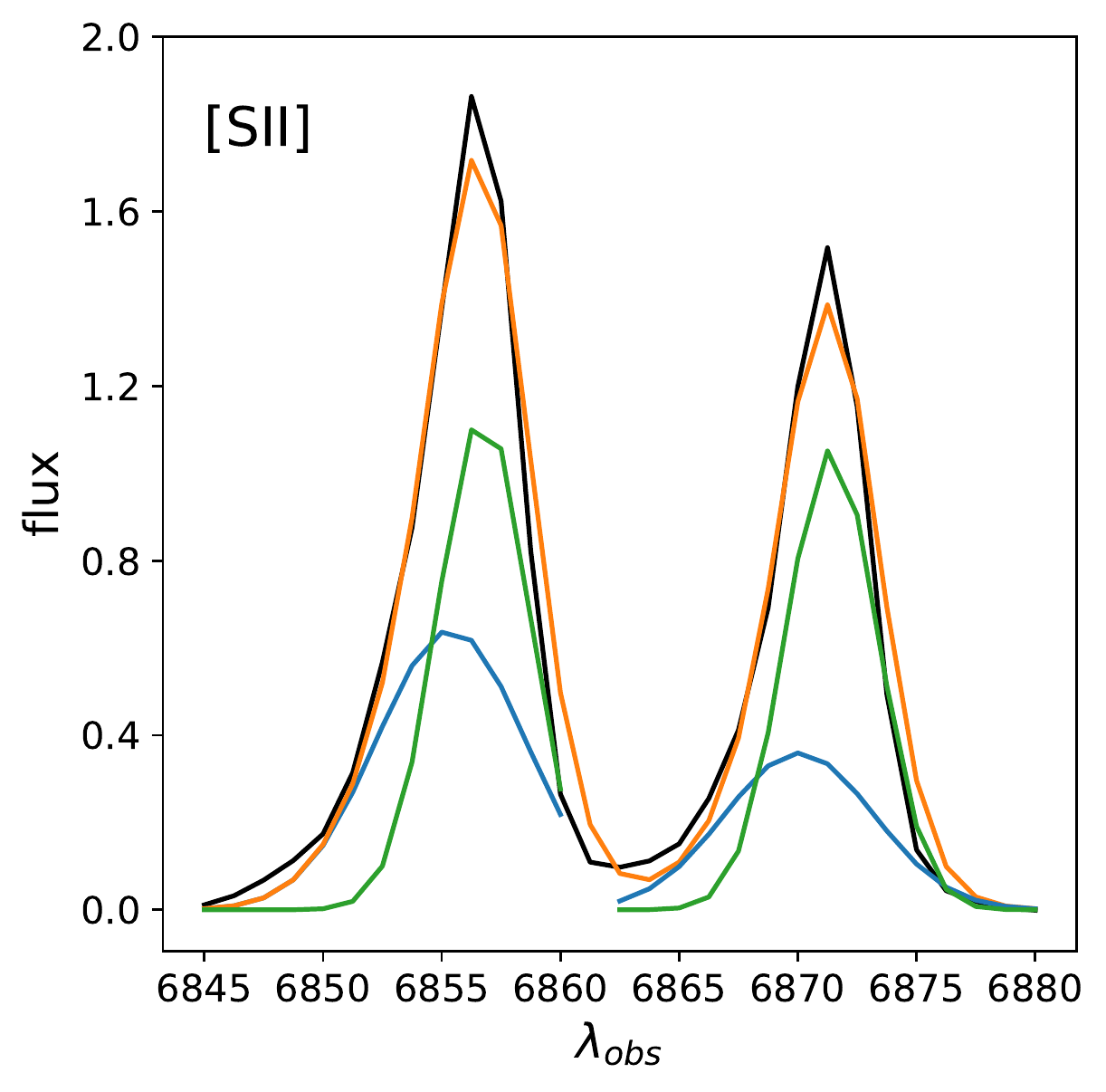}\\

	    \includegraphics[width=0.6\columnwidth]{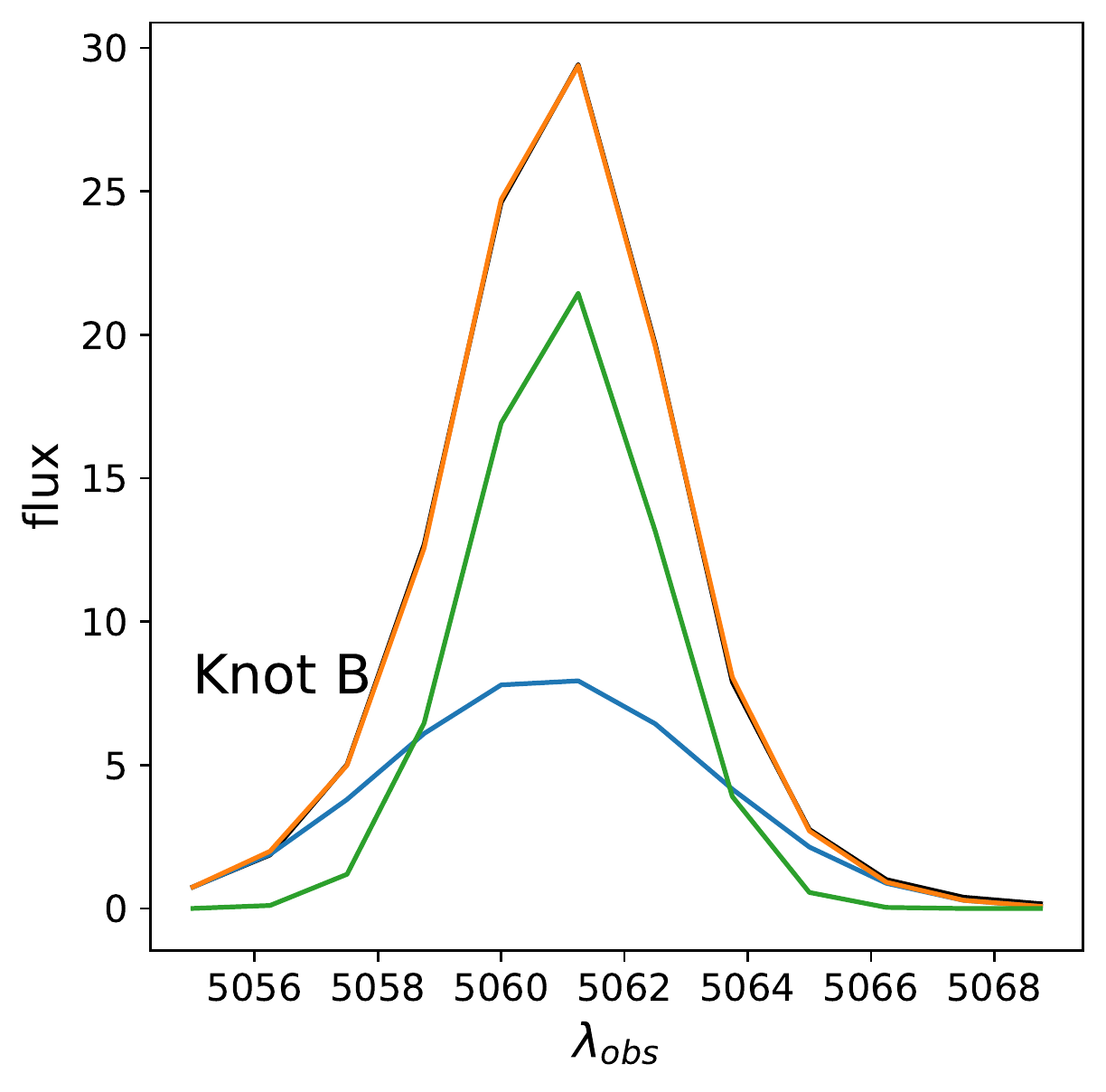}
	    \includegraphics[width=0.6\columnwidth]{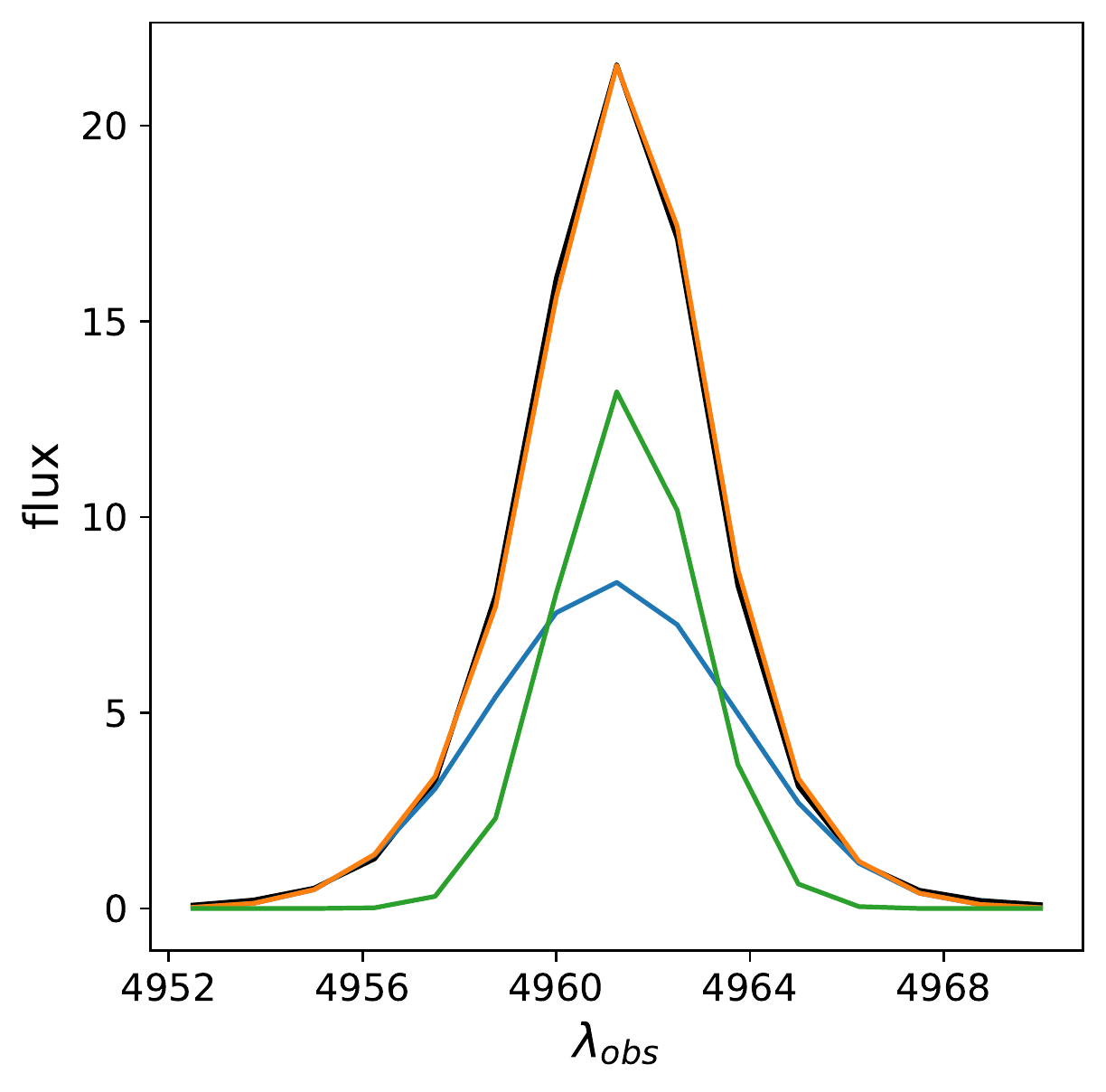}
	    \includegraphics[width=0.6\columnwidth]{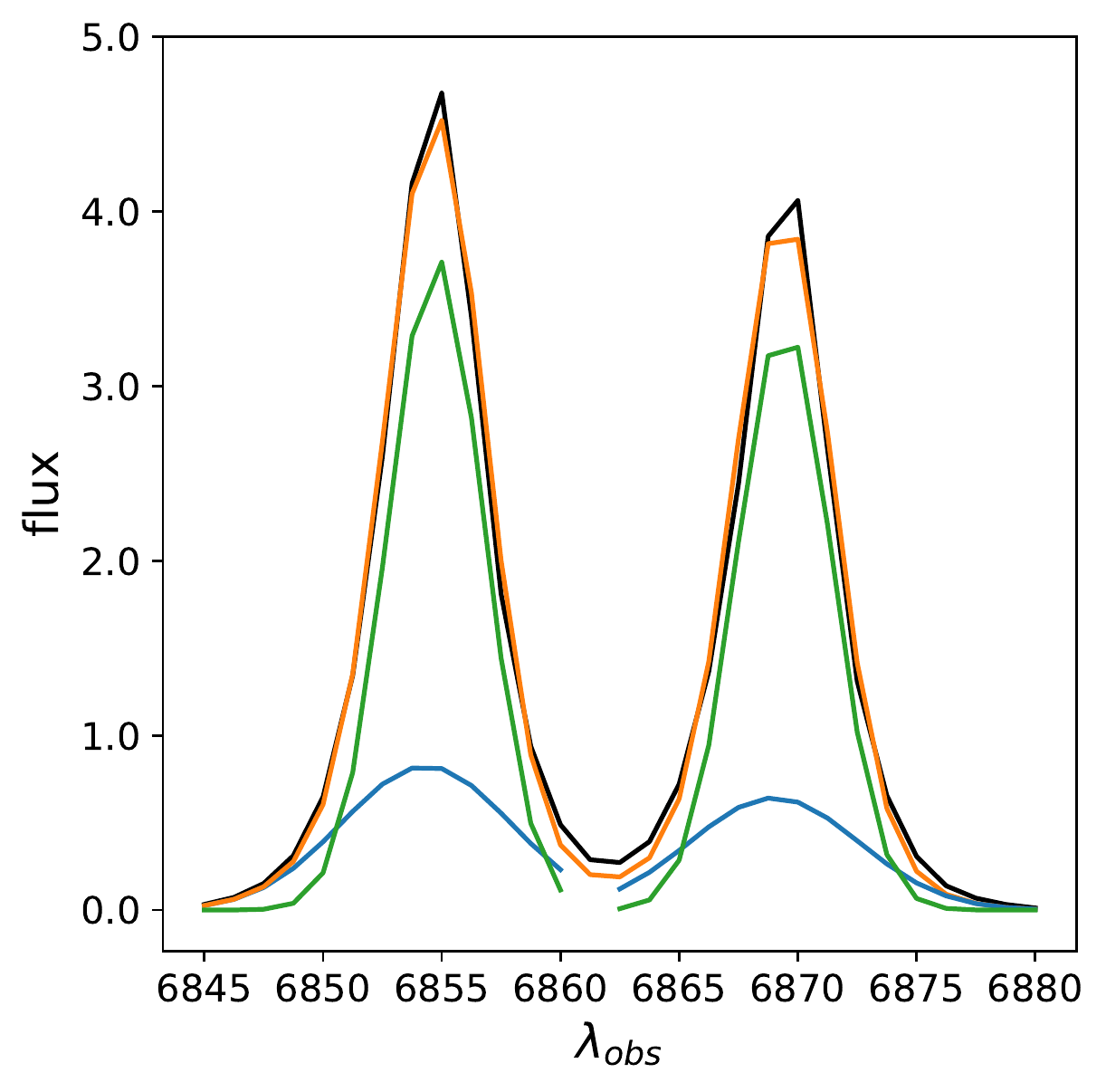}\\

	    \includegraphics[width=0.6\columnwidth]{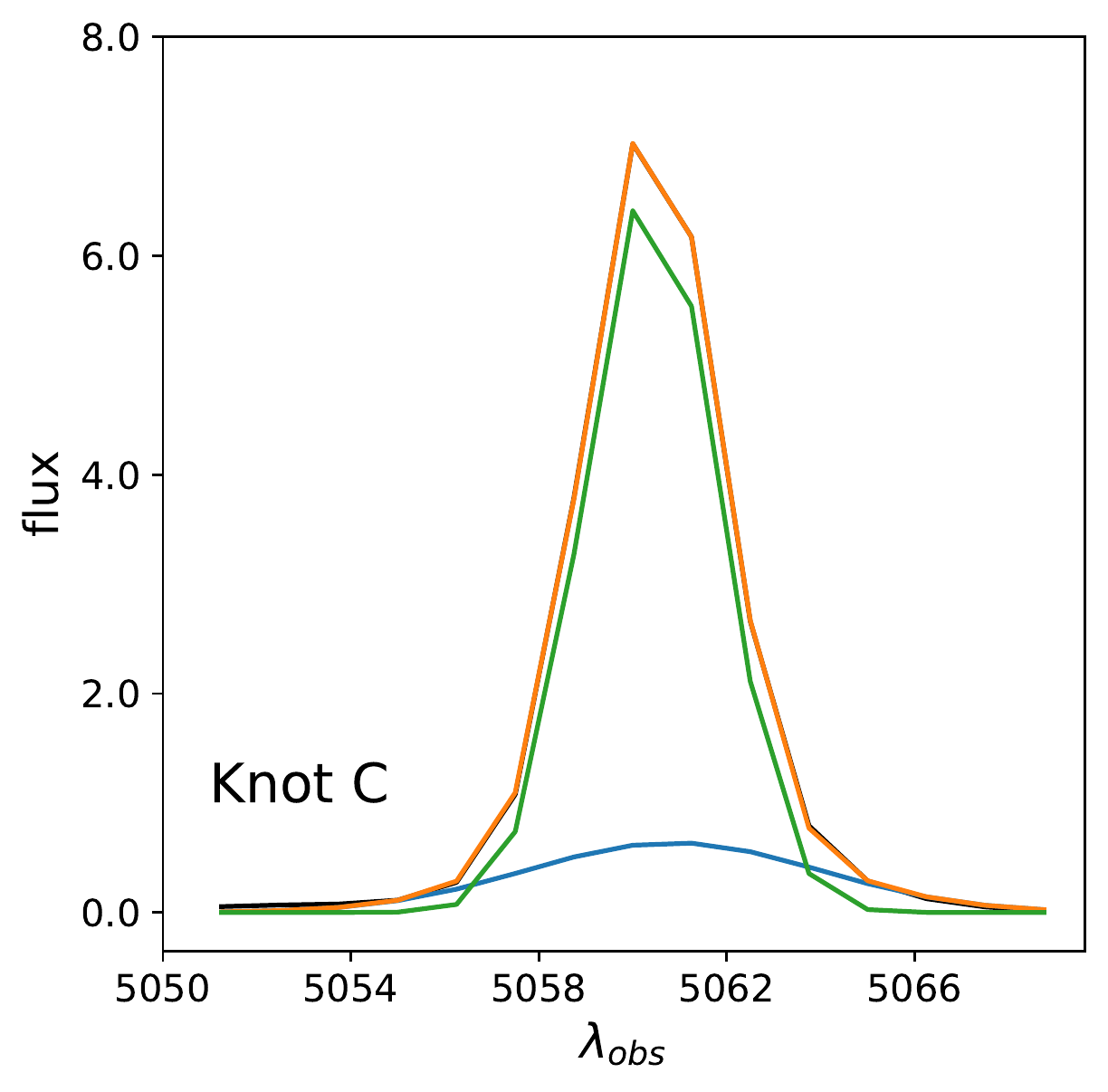}
	    \includegraphics[width=0.6\columnwidth]{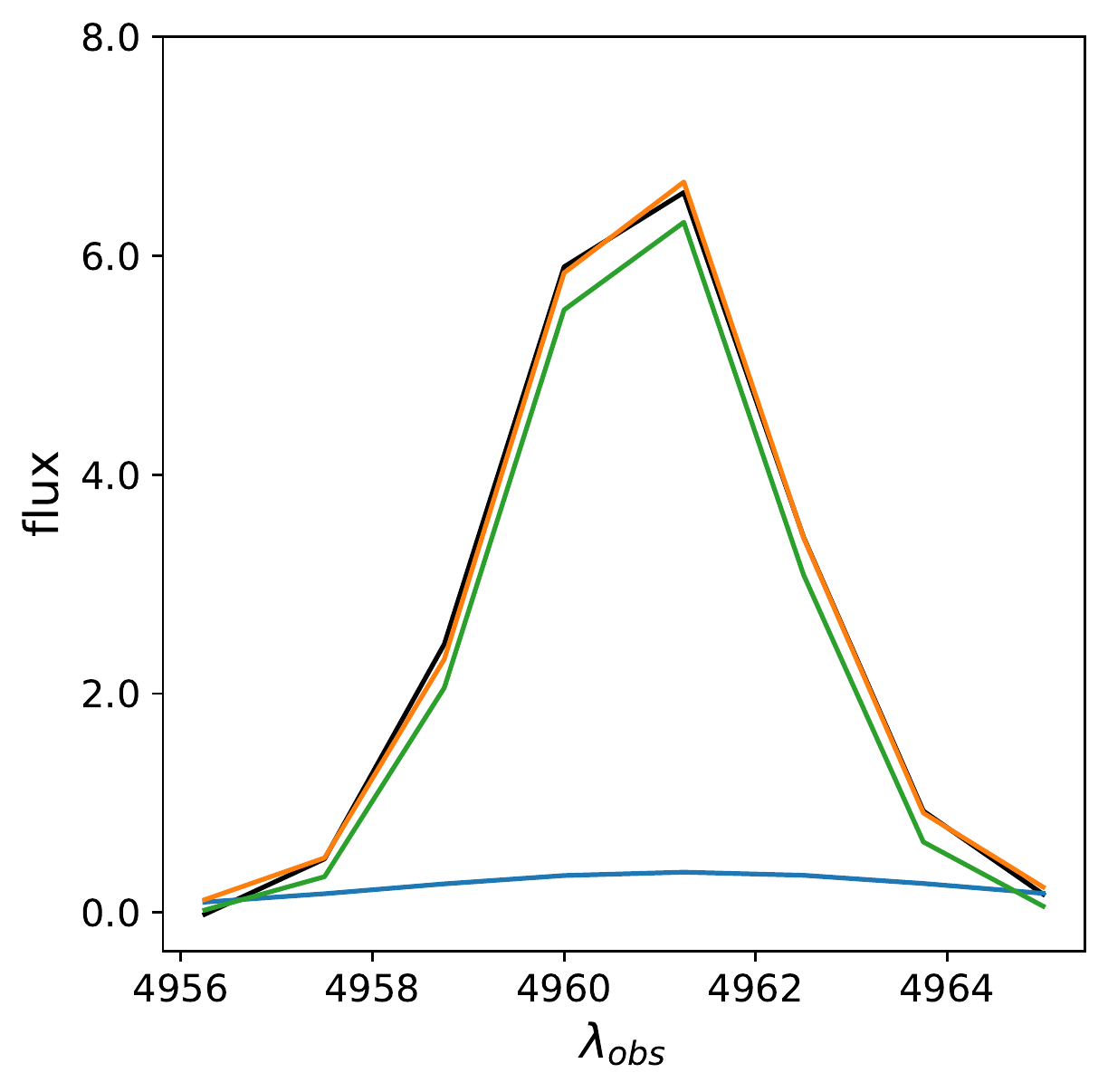}
	    \includegraphics[width=0.6\columnwidth]{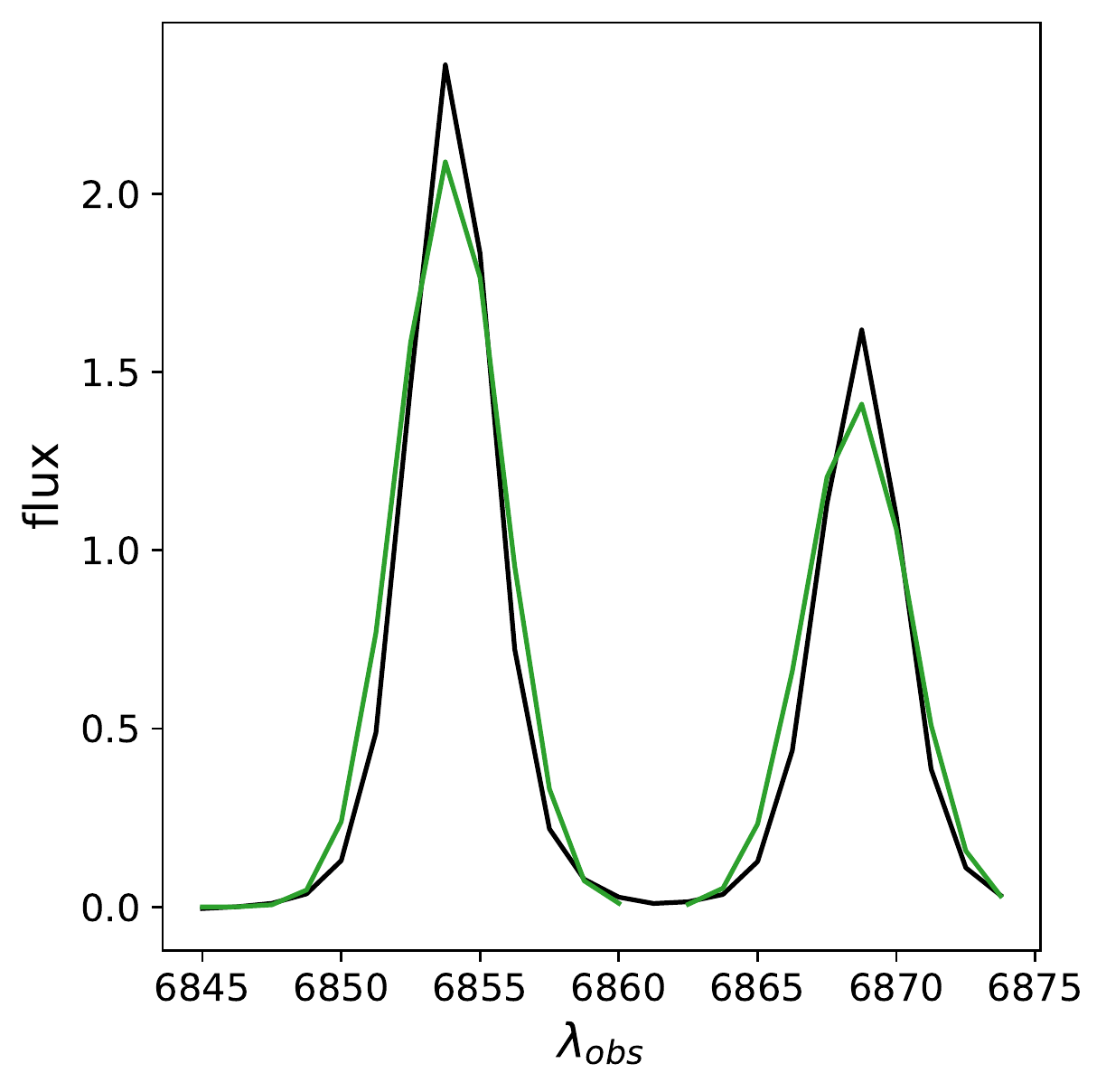}\\
	     
    \caption{Optical emission lines - continuum subtracted - ([OIII]4959, $H\beta$ and [SII] doublet) of the three knots of Haro 11 modeled with a double Gaussian fit. First, second and third row are respectively for knot A, B and C. The x-axis reports the observed wavelength in Å and the y-axis the flux intensity in units of $10^{-15}\,\rm erg/s/cm^2/$Å. The data are plotted in black, the total model is plotted in orange, the narrow component is plotted in green, the broad component is shown in blue. The kinematics (shift between the two components and dispersion of the two Gaussians) is fixed as the best-fitted for the [OIII]4959 line), so that in the fits of the $H\beta$ and [SII] doublet only the amplitudes are free parameters. The [SII] doublet in knot C is reproduced by one Gaussian component only.}
    \label{fig:fit_gauss}
    
\end{figure*}

Figure \ref{fig:fit_gauss} shows the Gaussian fits of the [OIII]4959, $H\beta$ and [SII] doublet. Both knots A and B show a broad Gaussian component that is blue-shifted with respect to the narrow component. The velocity dispersion of the broad component is more than twice as high as the one of the narrow component and has a different intrinsic dust extinction. We interpret such differences as the signature of two gas components with different physical properties, one being the outflow escaping the gravitational well of the cluster populations. The gravitational potential is estimated counting the total mass of the three stellar populations of our spectroscopic model, which dominates the mass of the ionised gas and likely of the other phases of the gas as well. Moreover, the gravitational potential around each knot is affected by the neighbor knots making it easier for the gas to escape. The same exercise repeated with the virial masses of knots A and B derived by stellar velocity dispersion \citep{oestlin2015} gives escape velocities that are lower than the maximum velocity of the outflows (reported in Table \ref{tab:ion_outflow}).  In knot C the broad component detected in the [OIII]4959 and $H\beta$ is fainter compared to the other knots and red-shifted rather than blue-shifted, whereas in the [SII] doublet the broad component is not required to fit the lines profile. The uncertainties on the kinematics derived with the OIII4959 Gaussian fit have been determined with a classic Monte Carlo approach that assumes a statistical error on the optical spectra of 0.1\%, as derived by the MUSE variance cube. The errors on the velocity shifts and widths of the Gaussian components are between 0.5-1 km/s for all knots.

Table \ref{tab:ion_outflow} summarises the values of the physical properties of the outflow derived as described in the Equations \ref{eq:mass} and \ref{eq:outflow_rates}. In knot C we do not detect a broad component in the [SII] fit, likely because the outflow is faint and not important in the region coinciding with knot C. Because we cannot measure the electron density using the [SII] doublet for knot C, we are not able to determine the mass of the outflow and all the other dependent physical quantities in this knot. The outflow in knot A is the one transporting the largest mass, which means highest mass rate, momentum rate and energy rate. These values of knot A are lower limits because they are inversely proportional to the density (see Equation \ref{eq:mass}), which is an upper limit. We set an upper limit of 10 $\rm cm^{-3}$ for the outflow density of knot A because the [SII] doublet ratio is higher than the diagnostics range (Osterbrok Ferland 2006). The error associated to the outflow properties is large due to many assumptions that are necessary for calculating them (size and geometry of the outflow). Therefore, the numbers tabulated in Table \ref{tab:ion_outflow} are meant to be understood as orders of magnitude estimates.

However, it is interesting to compare the mass outflow rates with the star-formation rates calculated for each knot: in knot A the ionised gas is transported away at rate a factor of 10 or higher than the rate at which stars are forming, based upon the $H\alpha$ flux. This indicates that in this region the starburst might stop because of negative feedback. In knot B the mass outflow rate is currently a factor of 4 lower than the star-formation rate. The energy transported by the outflows in knot A and B ($10^{55-56}$ erg) is up to a factor of 10 lower than the one released by the stellar feedback in the respective knots, when the energetics from both stellar winds and SNe are summed up. Considering that only a few percent of the energy from stellar feedback is converted into kinetic energy of the ISM gas \citep{Fierlinger2016}, this is consistent with a scenario in which both stellar winds and SNe explosions contribute to drive the outflow. On the other hand the rate at which the kinetic energy of the ionised gas is transported out of the knots ($10^{42-41}$ erg/s) is one order of magnitude larger in knot A (similar for knot B) than the power produced by stellar feedback calculated in Sec. \ref{sec:spectral_fits}. This suggests that stellar feedback in knot A might have powered the current outflow in the recent past and it is now in a deceleration phase. Instead knot B has a weaker and younger outflow likely because the stellar population is younger and thus has not yet delivered the full potential feedback output. This is consistent with a more dusty (highly extinguished) and dense (see Table \ref{tab:lines}) region that is developing a young outflow. We speculate that knot C is an evacuated region emptied by a past outflow \citep{menacho2021}. In fact the feedback output from the stellar populations in the previous 50 Myr is the strongest for knot C.

\subsubsection{Alternative interpretations of the broad-component gas}
\label{sec:interp}

We cannot exclude that the broad component of the optical emission lines is produced by the turbulence of mixing layers of ionised gas. Because the broad component is well fitted by a Gaussian, it is the sum of numerous individual contribution, many of which are outflowing and many others of which may not be outflowing and have instead a random velocity field. However, the shift toward the bluer wavelengths of the broad component favours the outflow interpretation. This scenario is also supported by (i) the previously published high-velocity and kpc-scale kinematics of ionised gas \citep{Menacho+2019}, (ii) the presence of an outflow in the neutral phase \citep{oestlin2021} (see comparison of ionised and neutral gas kinematics in Section \ref{sec:multi-phase}), (iii) diffuse X-ray emission tracing a hot X-ray wind around the starburst knots \citep{grimes2007}.

We considered the hypothesis that the broad components of the emission lines of the ionised gas in the Haro 11 starburst are due to the presence of a hidden AGN. \citet{gross2021} reviewed the line emission diagnostics of the Haro 11 literature finding that the two regions (knots B and C) where they detect X-ray emission are more consistent with AGN/composite classification. We performed a line emission diagnostics separately for the narrow and the broad component using a BPT diagram. We found that for both components and for all knots the data points lie within the errors on or below the separation line between starburst and AGN ionisation. Hence, as already discussed in previous works \citep[][]{menacho2021,rivera2015}, the outflows detected in these regions are mainly driven by stellar feedback.
\begin{table}
	\centering
	\caption{Physical properties of the ionised gas outflowing from the knots of Haro 11}
	\label{tab:ion_outflow}
	\renewcommand{\arraystretch}{1.5} 
	\begin{tabular}{lccc} 
		\hline
		      & knot A & knot B & knot C\\
		\hline
		$v_{max}$ (km/s) & 420 & 433 & -\\
		
		$t_{dyn}$ (Myr) & 1.2 & 1.2 & -\\
		
		$\dot{M}_{out}\, \rm (M_{\odot}/yr)$ & >13 & 1.8 & -\\
	
		$\dot{p}_{out}\, \rm (g\,cm\,s^{-2})$ & >4 $\cdot 10^{34}$ & $5 \cdot 10^{33}$ & -\\

		$\dot{E}_{out}\, \rm (erg/s)$ & >1$\cdot 10^{42}$ & $2\cdot 10^{41}$ & -\\

		$E_{out}\, \rm (erg)$ & >6$\cdot 10^{55}$ & $8\cdot 10^{54}$ & -\\		
		\hline
	\end{tabular}
\end{table}

\section{Discussion}
\label{sec:discuss}

\subsection{Multiphase outflows: the neutral and the ionised phases}
\label{sec:multi-phase}

We have studied the kinematics of the ionised gas within the starburst knots of Haro 11 using several optical emission lines. More information about the kinematics of ionised gas as well as other phases of the interstellar gas can be derived from the FUV spectroscopy. Multiple FUV absorption lines are produced by the intervening gas along the line of sight of the FUV continuum source, and therefore can be used to study the motion of that gas as was done by \citet{oestlin2021}. The disadvantage of using FUV absorption lines for studying the properties of the outflowing gas is that is not possible to measure the spatial extension as the absorption occurs only where there is a background, generally not extended, continuum source. However, a big advantage of this method relies on the wide variety of FUV absorption lines, that can trace different phases of the gas: neutral hydrogen and gas with different degrees of ionisation \citep[see][]{Emil2017,oestlin2021}. We compare below the kinematics of the ionised and neutral phase gas as traced by optical emission lines (narrow plus broad component) and FUV absorption lines, respectively.

\begin{figure}

	\includegraphics[width=0.92\columnwidth]{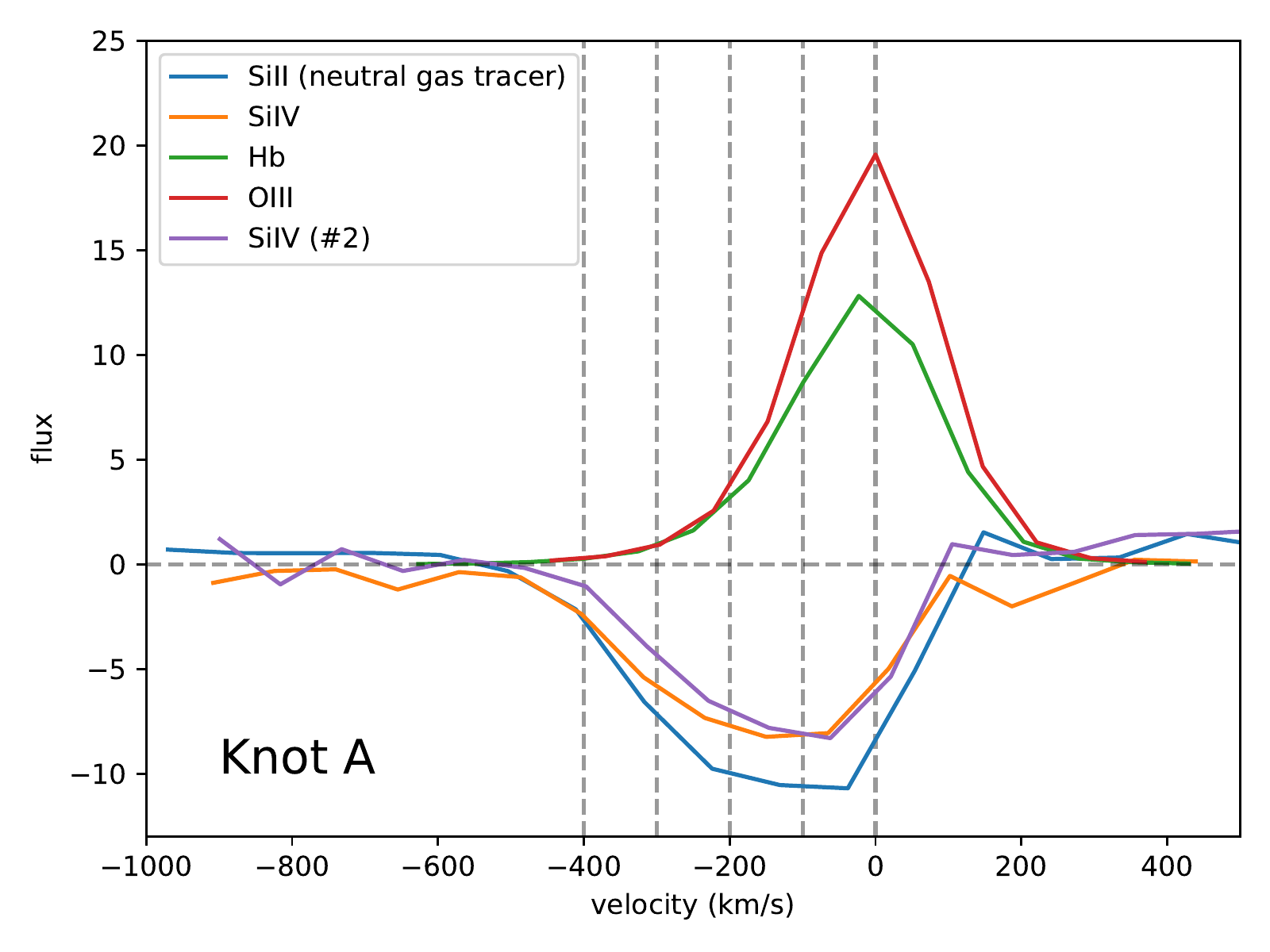}
	    
	\includegraphics[width=0.92\columnwidth]{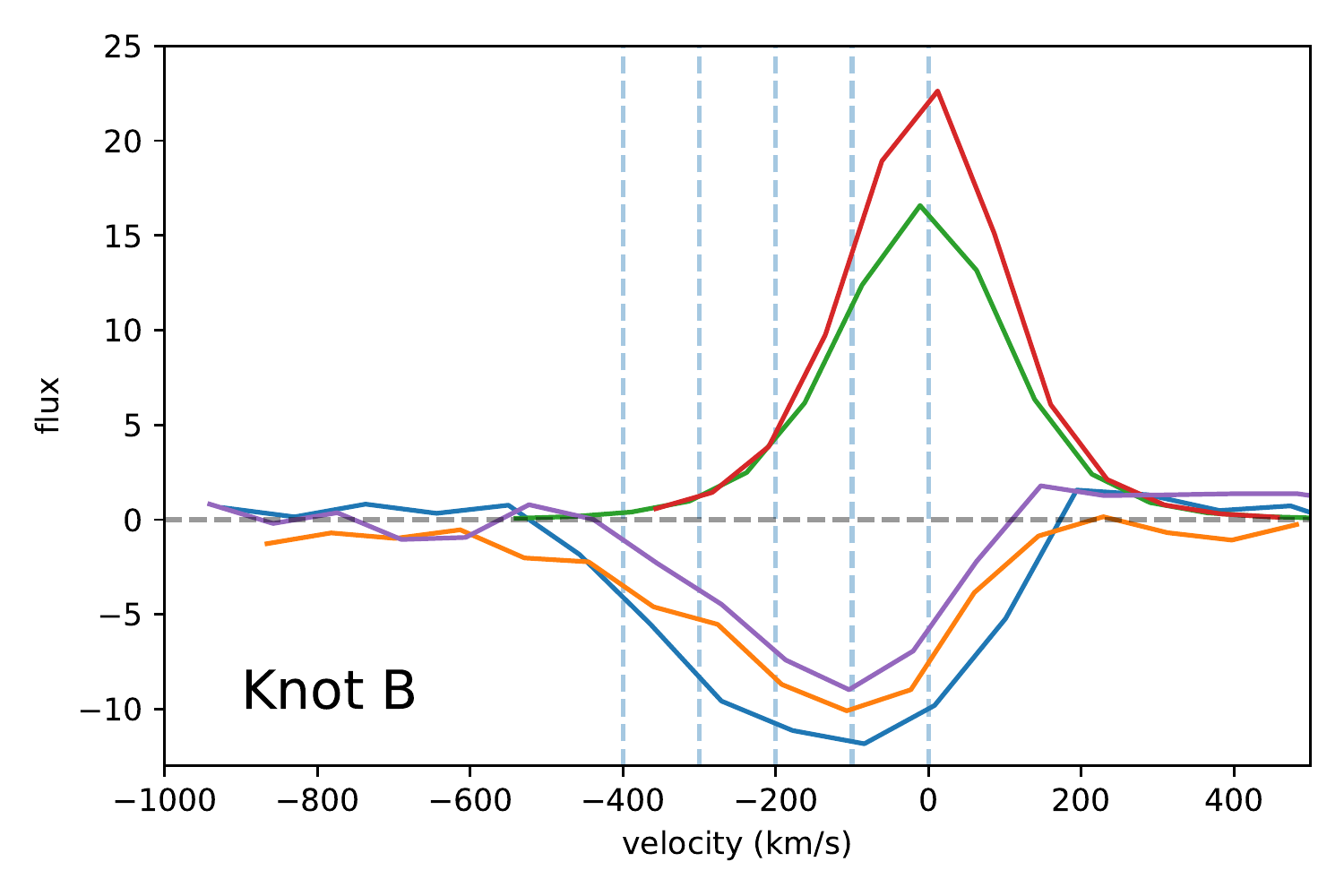}
	
    \caption{Optical emission lines -continuum subtracted- compared to UV absorption profiles on a common velocity axis with a zero-velocity determined by the galaxy redshift of z=0.0206. Top panel is knot A, bottom panel is knot B. The flux intensity of the lines is re-scaled and plotted with arbitrary units for an easier visual comparison. The vertical dashed lines define a grid of velocities spaced by 100 km/s.}
    
    \label{fig:multi-phase}
    
\end{figure}

In Figure \ref{fig:multi-phase} we compare the emission and absorption profiles of different optical emission lines and of different FUV absorption lines, on a common velocity axis. Most interesting among the UV absorption lines,  is the SiII (plotted in blue) which mostly traces the neutral hydrogen gas. Notice that the spectral profile of the emission and absorption lines is intrinsically different. However, it is meaningful to compare a property such as the maximum velocity of the outflow, inferred from such profiles. The two different phases of the outflows in knot A and B, the neutral gas and the ionised gas, share similar kinematics both reaching maximum velocities up to about 400 km/s. This means that the outflows of knots A and B (and likely the putative past outflow in knot C) are all multi-phase, bringing a significant fraction of both neutral and ionised gas out of the starburst region. 

\subsection{Caveats and limitations of this work}
Our analysis is split in two parts: we first study the physical properties of the clusters and stellar populations in the starburst knots of Haro 11, secondly we measure the physical properties of the ionised gas within the same regions. For the first part of the analysis, a major limitation is the age-extinction degeneracy when studying stellar population properties in the optical and UV. FUV spectroscopy has the advantage of having age sensitive P-Cygni lines, however dust extinguishes a large fraction of the FUV radiation, making the analysis more challenging. Both photometry and spectroscopy models do not include the physics of binary systems. 
Since multiple stellar systems are common among massive stars, we defer the examination of more sophisticated models such as the Binary Population and Spectral Synthesis code \citep[BPASS, ][]{eldridge2017} to a future paper. However, a major limitation when applying these models to COS FUV spectra is the poor spectral resolution of the theoretical spectra, which is insufficient to properly model the P-Cygni lines.

The second part of our analysis is perhaps even more challenging as it ultimately aims at quantifying the impact of the stellar feedback on the gas surrounding the clusters. Regarding the optical emission line analysis, the properties of the gas are measured without any restrictive assumption. Afterwards we interpret the broad Gaussian component of each line to be tracing the outflowing gas escaping the gravitational well of the knots. In order to estimate the energetics of such outflows we assume:
\begin{itemize}
    \item the size of the outflows is equal to the aperture that we use to extract the spectroscopy (i.e. 1 kpc in diameter); larger outflow radii would imply that the outflows of different knots merge to form a single large-scale homogeneous outflow but the absence of outflowing gas in knot C argues against this hypothesis.
    \item the maximum velocity of the outflow is equivalent to the velocity that is 2$\sigma$ away (in the blue direction) from the centre of the broad component; as discussed in \cite{rose2018}, velocities that encompass a smaller fraction might raise inaccuracies due to flux errors; we do not correct for projection effects because we assume that the line broadening is partly due to intersecting different projections of the velocity vector field.
    \item the outflow timescale is given by the radius of the outflow divided by its maximum velocity; the underlying assumption here is that the gas has been pushed outward across the knots' region at a constant speed. 
\end{itemize}

\subsection{Implications of our work: feedback physics and the high-redshift Universe}


Our work can be an important reference for similar future studies that will aim at understanding the role of stellar feedback, especially from young star clusters, combining the information of stellar population and kinematics of the gas around the major sources of feedback. The spatially-resolved analysis of the stellar feedback in Haro 11, combined with the gas physical properties allows us to discuss the effect of the feedback on the ISM of the galaxy. For example, knot C appears as an evacuated region, probably due to the large SN explosion rate that the region has undergone. Because of these properties, \cite{oestlin2021} argue that knot C is most likely the source of the escaping Lyman continuum radiation detected in Haro 11.

\citet{Naidu2021} propose a simple picture for describing the escape of Ly$\alpha$ and LyC radiation during the evolution of star forming regions in galaxies. In phase I stars form in a dense cloud in which Ly$\alpha$ and LyC photons are trapped; in phase II the first stars produce feedback dispersing gas and creating holes through which Ly$\alpha$ and LyC photons can escape; in phase III the feedback weakens, the gas becomes more dusty and thicker and as a result Ly$\alpha$ photons scatter out but LyC photons are absorbed; in phase IV dust and gas assembly continues, the LyC production drops and fewer Ly$\alpha$ photons escape. In Haro 11, knot C hosts the oldest (> 10 Myr)  population of star clusters, shows strong LyA emission, EW ~ 20 Å \citep{hayes2007}, and low-density gas, which is consistent with a star-forming region between phase II and III. Knot B hosts the youngest cluster population (< 3 Myr) that is highly extinguished by the surrounding dense gas, which falls in the phase I. Knot A hosts a cluster population with a large range of ages (3 - 15 Myr) that makes it difficult to fit this region in a schematic picture as the one described above. We will have further insights on the escaping of LyC photons thanks to new ongoing HST observations (PI M.S.Oey). 

Spectroscopy tools to derive stellar physics are fundamental as they can be applied to a sample of galaxies to study in a more statistically meaningful way the properties of young populations of stars. This work is useful also for comparing two independent observational techniques, i.e. spectroscopy and photometry, applied over the same range of wavelengths to derive similar physical properties. Both techniques are generally used by the community and it is important to know their limitations, their accuracy, and how to interpret their results.

In the literature there are a few other studies similar to this one. \citet{chisholm2019} inferred the properties of massive star populations using FUV stellar continua of 61 star-forming galaxies at $z\sim2$. They used a similar approached based on Starburst99 libraries to model the spectroscopy with the difference that their data cover a larger physical spatial scale and therefore represent stellar properties averaged over substantial portions of the target galaxies. They provide simple prescriptions to determine the ionising photon production by massive stars, which is relevant in the context of reionisation of the Universe. The advantage in our approach to study a starburst process is given by adding the optical spectrum to the UV as well as by having access to the high-spatial resolution imaging to resolve single clusters. \citet{Vanzella2021} found in the Sunburst lensed galaxy at z=2.37 a LyC emitting knot (called knot 5.1 in their work), believed to be the youngest (3 Myr) and most massive star cluster ($10^7 \msun$) of the Sunburst galaxy. Another similarity of this galaxy with Haro 11 is the outflowing gas velocity (> 300 km/s).

Future studies will likely make use of multi-wavelength data sets and combine new observations with archived data. Our work paves the way in the direction of maximising the scientific output by combining different observations with multiple telescopes, instruments and wavelengths. Moreover, future studies will exploit the capabilities of the next generation facilities that will allow to study the first galaxies that formed in our Universe and have a more complete view on the formation and evolution of galaxies. As Haro 11 is high-redshift analogue galaxy, it is particularly interesting to look at our analysis in the context of clustered star formation in the young Universe, a window that will be opened by upcoming and future facilities. Recently a few studies have investigated the rest-frame UV/optical spectroscopy of galaxies at $z\sim2-3$. \cite{strom2018} used photo-ionisation models to constrain the gas chemistry and nebular ionisation and excitation conditions for $\sim150$ galaxies from the Keck Baryonic Structure Survey (KBSS). Near-infrared IFU observations taken with SINFONI at the VLT allowed \cite{davies2019} to measure the properties of kpc-scale star-formation driven outflows in a sample of 35 galaxies at $z\sim2-2.6$. They found that the outflow velocities scales with the star-formation rate of the galaxies suggesting, similarly to our work, that these kpc-scale winds are driven by a combination of mechanical energy released by stellar winds, SN explosions as well as radiation pressure on dust grains. However, in order to really understand the driving mechanism of such outflows and their effects on the surrounding, the accuracy on the outflow rates and energy balance need to improve significantly.

Previous works by \cite{chisholm2015} investigated scaling relations between galactic outflows and the properties of the host galaxies such as stellar mass, star-formation rate and circular velocity. They measured the kinematics of the warm gas of the outflows using absorption lines in UV COS spectroscopy, similarly to how it was done for Haro 11. Earlier, \cite{martin2009} analysed near-UV and optical spectroscopy to infer the starburst outflow properties in a sample of ultra-luminous infra-red galaxies (ULIRGs).

What can be improved in these studies is to resolve galaxies at smaller scales, identify the source of such outflows and shed light on their interaction with the interstellar medium. Our work attempts to connect the small scale cluster physics with the larger-scale ($\sim$ kpc) kinematics of ionised gas. Even though this is currently possible only with nearby targets and a few lensed galaxies, future high-resolution IFUs such as ELT or JWST will help the scientific community in this direction. For galaxies at early epochs (z>2) the star cluster scales (e.g. <20-30 pc), will be resolved by adding moderate lensing magnification even with the next generation observatories. Another important improvement in future studies will be to include more typical and less extreme galaxies (e.g. not as star-forming as the ULIRGs).

\section{Conclusions}
\label{sec:conclude}


We have studied the nuclear starburst regions of Haro 11, which is a highly star-forming blue compact galaxy and the closest confirmed Lyman continuum leaker in the local Universe (z=0.021
). We have expanded and updated the study of the star cluster population based on multi-band photometry, which clearly shows that the galaxy is mainly forming stars in three knots called A, B and C, each with a size of no more than 1 kpc in diameter. Our analysis shows that this clustered star-formation propagates from the eastern side of the galaxy, where the older clusters have been detected, to the western more dusty side where young clusters are forming.

We combined the FUV COS spectroscopy with the optical MUSE spectra, both extracted from the same aperture of diameter 2.5\arcsec\, = 1 kpc, and modeled the combined spectrum using a code based on Starburst 99. Each knot is best modeled by three stellar populations in the age ranges 1-4 Myr, 4-40 Myr, 40-100 Myr. The spectroscopy fits provide not only the physical properties of the stellar populations (age, metallicites, masses and intrinsic reddening attenuations) but also the feedback output in terms of momentum, energy and power from stellar winds and SNe, as well as photo-ionisation rates.

The cluster light dominates the FUV light of the knots, but not at the redder wavelenghts confirming that cluster are very young while there is an underlying older diffuse stellar component that contributes to the recent star-formation history of the central starburst of the galaxy. Very massive clusters (M$>10^5 M_{\odot}$) have formed in the last 20 Myr within the three knots. The cluster mass distribution is flatter than typically observed in local spiral galaxies ($\alpha \sim$ -1.4) suggesting that the formation of very massive clusters has been the favored mode. The most massive and youngest clusters are currently detected in knot B and the dusty arm, while knot C appears to be the most evolved region. Star formation has been proceeding in knot A throughout the starburst phase, as suggested by the wide age range detected in the clusters of this knot.

The photometry of the clusters and the spectroscopy of the total region probe different age ranges, which can be understood given the assumptions and intrinsic differences of the two methodologies. In fact we proved the diffuse light (that is only included in the spectra) to be more than half of the total light in the red filter. This explains the presence of a third old (> 40 Myr) stellar population in the spectroscopic model.

We have investigated the kinematics of the ionised gas in the three knots using the MUSE optical spectra. The emission lines  H$\alpha$, H$\beta$, [OIII], [SII] are best reproduced by the sum of two Gaussian components, a narrow and a broad ($\sigma\sim$ 200 km/s) component. For knot A and B, the broad component is blue-shifted and has an intensity comparable to the narrow component, which we interpret as the presence of ionised gas outflowing away from the two starburst knots. In knot C instead the broad component is red-shifted and too faint for allowing a measurement of the outflow properties. Therefore we focused on the knots A and B and estimated the physical properties of the outflowing gas such as maximum velocity, mass rate and kinetic energy rate. The outflow is also traced by the UV absorption lines SiII, SiIV, the former being a tracer of the neutral hydrogen. This allows to study the kinematics of different phases of the gas. We showed that the neutral gas and the ionised gas move with similar maximum velocities ($\sim$ 400 km/s). The strongest outflow is the one in knot A, which is removing gas mass at a rate 10 times faster than the star-formation rate. This suggests that the star-formation activity of this region will be quenched by stellar feedback. The ionised outflow energetics of both knots A and B are comparable with the feedback output calculated by modelling the stellar populations, which indicates that the main driver of the outflow are stellar winds and SN explosions. Even though knot C does not currently show any trace of outflowing ionised gas, the SN feedback from this region has been very strong in the past 40 Myr and has likely evacuated already the knot. The high-rate of SN explosions measured in this region in our work can explain the energetics of the hot gas wind around the knots traced by the observed diffuse X-ray emission.

\section*{Acknowledgements}

We thank Jens Melinder for sharing part of the reduced data with the authors. A.A. acknowledges the support of the Swedish Research Council, Vetenskapsr\aa{}det and the Swedish National Space Agency (SNSA).
M.S.O. acknowledges support from NASA grant HST-GO-15352.002


\section*{Data availability}
The data underlying this work are available in the Mikulski Archive for Space Telescopes (programs 15352 and 13017) and in the ESO Science Archive Facility (program 096.B-0923(A)). 



\bibliographystyle{mnras}
\bibliography{haro11_draft} 

\begin{thebibliography}{}
\makeatletter
\relax
\def\mn@urlcharsother{\let\do\@makeother \do\$\do\&\do\#\do\^\do\_\do\%\do\~}
\def\mn@doi{\begingroup\mn@urlcharsother \@ifnextchar [ {\mn@doi@}
  {\mn@doi@[]}}
\def\mn@doi@[#1]#2{\def\@tempa{#1}\ifx\@tempa\@empty \href
  {http://dx.doi.org/#2} {doi:#2}\else \href {http://dx.doi.org/#2} {#1}\fi
  \endgroup}
\def\mn@eprint#1#2{\mn@eprint@#1:#2::\@nil}
\def\mn@eprint@arXiv#1{\href {http://arxiv.org/abs/#1} {{\tt arXiv:#1}}}
\def\mn@eprint@dblp#1{\href {http://dblp.uni-trier.de/rec/bibtex/#1.xml}
  {dblp:#1}}
\def\mn@eprint@#1:#2:#3:#4\@nil{\def\@tempa {#1}\def\@tempb {#2}\def\@tempc
  {#3}\ifx \@tempc \@empty \let \@tempc \@tempb \let \@tempb \@tempa \fi \ifx
  \@tempb \@empty \def\@tempb {arXiv}\fi \@ifundefined
  {mn@eprint@\@tempb}{\@tempb:\@tempc}{\expandafter \expandafter \csname
  mn@eprint@\@tempb\endcsname \expandafter{\@tempc}}}

\bibitem[\protect\citeauthoryear{{Adamo}, {{\"O}stlin}, {Zackrisson}, {Hayes},
  {Cumming}  \& {Micheva}}{{Adamo} et~al.}{2010}]{adamo2010}
{Adamo} A.,  {{\"O}stlin} G.,  {Zackrisson} E.,  {Hayes} M.,  {Cumming} R.~J.,
   {Micheva} G.,  2010, \mn@doi [\mnras] {10.1111/j.1365-2966.2010.16983.x},
  \href {https://ui.adsabs.harvard.edu/abs/2010MNRAS.407..870A} {407, 870}

\bibitem[\protect\citeauthoryear{{Adamo} et~al.,}{{Adamo}
  et~al.}{2017}]{adamo2017}
{Adamo} A.,  et~al., 2017, \mn@doi [\apj] {10.3847/1538-4357/aa7132}, \href
  {https://ui.adsabs.harvard.edu/abs/2017ApJ...841..131A} {841, 131}

\bibitem[\protect\citeauthoryear{{Adamo} et~al.,}{{Adamo}
  et~al.}{2020a}]{adamo2020a}
{Adamo} A.,  et~al., 2020a, \mn@doi [\ssr] {10.1007/s11214-020-00690-x}, \href
  {https://ui.adsabs.harvard.edu/abs/2020SSRv..216...69A} {216, 69}

\bibitem[\protect\citeauthoryear{{Adamo} et~al.,}{{Adamo}
  et~al.}{2020b}]{adamo2020b}
{Adamo} A.,  et~al., 2020b, \mn@doi [\mnras] {10.1093/mnras/staa2380}, \href
  {https://ui.adsabs.harvard.edu/abs/2020MNRAS.499.3267A} {499, 3267}

\bibitem[\protect\citeauthoryear{{Annibali}, {Tosi}, {Aloisi}  \& {van der
  Marel}}{{Annibali} et~al.}{2011}]{annibali2011}
{Annibali} F.,  {Tosi} M.,  {Aloisi} A.,   {van der Marel} R.~P.,  2011,
  \mn@doi [\aj] {10.1088/0004-6256/142/4/129}, \href
  {https://ui.adsabs.harvard.edu/abs/2011AJ....142..129A} {142, 129}

\bibitem[\protect\citeauthoryear{{Avila}, {Bohlin}, {Hathi}, {Lockwood}, {Lim}
  \& {De La Pena}}{{Avila} et~al.}{2019}]{avila2019}
{Avila} R.~J.,  {Bohlin} R.,  {Hathi} N.,  {Lockwood} S.,  {Lim} P.~L.,   {De
  La Pena} M.,  2019, {SBC Absolute Flux Calibration}, Instrument Science
  Report ACS 2019-5

\bibitem[\protect\citeauthoryear{{Bacon} et~al.,}{{Bacon}
  et~al.}{2010}]{bacon2010}
{Bacon} R.,  et~al., 2010, in {McLean} I.~S.,  {Ramsay} S.~K.,   {Takami} H.,
  eds,  Society of Photo-Optical Instrumentation Engineers (SPIE) Conference
  Series Vol. 7735, Ground-based and Airborne Instrumentation for Astronomy
  III. p. 773508, \mn@doi{10.1117/12.856027}

\bibitem[\protect\citeauthoryear{{Bacon} et~al.,}{{Bacon}
  et~al.}{2017}]{Bacon+2017}
{Bacon} R.,  et~al., 2017, \mn@doi [\aap] {10.1051/0004-6361/201730833}, \href
  {https://ui.adsabs.harvard.edu/abs/2017A&A...608A...1B} {608, A1}

\bibitem[\protect\citeauthoryear{{Bastian} et~al.,}{{Bastian}
  et~al.}{2012}]{bastian2012}
{Bastian} N.,  et~al., 2012, \mn@doi [\mnras]
  {10.1111/j.1365-2966.2011.19909.x}, \href
  {https://ui.adsabs.harvard.edu/abs/2012MNRAS.419.2606B} {419, 2606}

\bibitem[\protect\citeauthoryear{{Bergvall} \& {{\"O}stlin}}{{Bergvall} \&
  {{\"O}stlin}}{2002}]{bergvall2002}
{Bergvall} N.,  {{\"O}stlin} G.,  2002, \mn@doi [\aap]
  {10.1051/0004-6361:20020759}, \href
  {https://ui.adsabs.harvard.edu/abs/2002A&A...390..891B} {390, 891}

\bibitem[\protect\citeauthoryear{{Bergvall}, {Zackrisson}, {Andersson},
  {Arnberg}, {Masegosa}  \& {{\"O}stlin}}{{Bergvall}
  et~al.}{2006}]{bergvall2006}
{Bergvall} N.,  {Zackrisson} E.,  {Andersson} B.~G.,  {Arnberg} D.,  {Masegosa}
  J.,   {{\"O}stlin} G.,  2006, \mn@doi [\aap] {10.1051/0004-6361:20053788},
  \href {https://ui.adsabs.harvard.edu/abs/2006A&A...448..513B} {448, 513}

\bibitem[\protect\citeauthoryear{{Bertin} \& {Arnouts}}{{Bertin} \&
  {Arnouts}}{1996}]{sextractor}
{Bertin} E.,  {Arnouts} S.,  1996, \mn@doi [\aaps] {10.1051/aas:1996164}, \href
  {https://ui.adsabs.harvard.edu/abs/1996A&AS..117..393B} {117, 393}

\bibitem[\protect\citeauthoryear{{Calzetti}, {Armus}, {Bohlin}, {Kinney},
  {Koornneef}  \& {Storchi-Bergmann}}{{Calzetti} et~al.}{2000}]{calzetti2000}
{Calzetti} D.,  {Armus} L.,  {Bohlin} R.~C.,  {Kinney} A.~L.,  {Koornneef} J.,
   {Storchi-Bergmann} T.,  2000, \mn@doi [\apj] {10.1086/308692}, \href
  {https://ui.adsabs.harvard.edu/abs/2000ApJ...533..682C} {533, 682}

\bibitem[\protect\citeauthoryear{{Cano-D{\'\i}az}, {Maiolino}, {Marconi},
  {Netzer}, {Shemmer}  \& {Cresci}}{{Cano-D{\'\i}az}
  et~al.}{2012}]{canodiaz2012}
{Cano-D{\'\i}az} M.,  {Maiolino} R.,  {Marconi} A.,  {Netzer} H.,  {Shemmer}
  O.,   {Cresci} G.,  2012, \mn@doi [\aap] {10.1051/0004-6361/201118358}, \href
  {https://ui.adsabs.harvard.edu/abs/2012A&A...537L...8C} {537, L8}

\bibitem[\protect\citeauthoryear{{Cardelli}, {Clayton}  \& {Mathis}}{{Cardelli}
  et~al.}{1989}]{Cardelli1989}
{Cardelli} J.~A.,  {Clayton} G.~C.,   {Mathis} J.~S.,  1989, \mn@doi [\apj]
  {10.1086/167900}, \href
  {https://ui.adsabs.harvard.edu/abs/1989ApJ...345..245C} {345, 245}

\bibitem[\protect\citeauthoryear{{Chisholm}, {Tremonti}, {Leitherer}, {Chen},
  {Wofford}  \& {Lundgren}}{{Chisholm} et~al.}{2015}]{chisholm2015}
{Chisholm} J.,  {Tremonti} C.~A.,  {Leitherer} C.,  {Chen} Y.,  {Wofford} A.,
  {Lundgren} B.,  2015, \mn@doi [\apj] {10.1088/0004-637X/811/2/149}, \href
  {https://ui.adsabs.harvard.edu/abs/2015ApJ...811..149C} {811, 149}

\bibitem[\protect\citeauthoryear{{Chisholm}, {Rigby}, {Bayliss}, {Berg},
  {Dahle}, {Gladders}  \& {Sharon}}{{Chisholm} et~al.}{2019}]{chisholm2019}
{Chisholm} J.,  {Rigby} J.~R.,  {Bayliss} M.,  {Berg} D.~A.,  {Dahle} H.,
  {Gladders} M.,   {Sharon} K.,  2019, \mn@doi [\apj]
  {10.3847/1538-4357/ab3104}, \href
  {https://ui.adsabs.harvard.edu/abs/2019ApJ...882..182C} {882, 182}

\bibitem[\protect\citeauthoryear{{Davies} et~al.,}{{Davies}
  et~al.}{2019}]{davies2019}
{Davies} R.~L.,  et~al., 2019, \mn@doi [\apj] {10.3847/1538-4357/ab06f1}, \href
  {https://ui.adsabs.harvard.edu/abs/2019ApJ...873..122D} {873, 122}

\bibitem[\protect\citeauthoryear{{Dessauges-Zavadsky} \&
  {Adamo}}{{Dessauges-Zavadsky} \& {Adamo}}{2018}]{Dessauges2018}
{Dessauges-Zavadsky} M.,  {Adamo} A.,  2018, \mn@doi [\mnras]
  {10.1093/mnrasl/sly112}, \href
  {https://ui.adsabs.harvard.edu/abs/2018MNRAS.479L.118D} {479, L118}

\bibitem[\protect\citeauthoryear{{Eldridge}, {Stanway}, {Xiao}, {McClelland},
  {Taylor}, {Ng}, {Greis}  \& {Bray}}{{Eldridge} et~al.}{2017}]{eldridge2017}
{Eldridge} J.~J.,  {Stanway} E.~R.,  {Xiao} L.,  {McClelland} L.~A.~S.,
  {Taylor} G.,  {Ng} M.,  {Greis} S.~M.~L.,   {Bray} J.~C.,  2017, \mn@doi
  [\pasa] {10.1017/pasa.2017.51}, \href
  {https://ui.adsabs.harvard.edu/abs/2017PASA...34...58E} {34, e058}

\bibitem[\protect\citeauthoryear{{Fierlinger}, {Burkert}, {Ntormousi},
  {Fierlinger}, {Schartmann}, {Ballone}, {Krause}  \& {Diehl}}{{Fierlinger}
  et~al.}{2016}]{Fierlinger2016}
{Fierlinger} K.~M.,  {Burkert} A.,  {Ntormousi} E.,  {Fierlinger} P.,
  {Schartmann} M.,  {Ballone} A.,  {Krause} M. G.~H.,   {Diehl} R.,  2016,
  \mn@doi [\mnras] {10.1093/mnras/stv2699}, \href
  {https://ui.adsabs.harvard.edu/abs/2016MNRAS.456..710F} {456, 710}

\bibitem[\protect\citeauthoryear{{Fitzpatrick}}{{Fitzpatrick}}{1999}]{Fitzpatrick1999}
{Fitzpatrick} E.~L.,  1999, \mn@doi [\pasp] {10.1086/316293}, \href
  {https://ui.adsabs.harvard.edu/abs/1999PASP..111...63F} {111, 63}

\bibitem[\protect\citeauthoryear{{Grimes} et~al.,}{{Grimes}
  et~al.}{2007}]{grimes2007}
{Grimes} J.~P.,  et~al., 2007, \mn@doi [\apj] {10.1086/521353}, \href
  {https://ui.adsabs.harvard.edu/abs/2007ApJ...668..891G} {668, 891}

\bibitem[\protect\citeauthoryear{{Gross}, {Prestwich}  \& {Kaaret}}{{Gross}
  et~al.}{2021}]{gross2021}
{Gross} A.~C.,  {Prestwich} A.,   {Kaaret} P.,  2021, \mn@doi [\mnras]
  {10.1093/mnras/stab1331}, \href
  {https://ui.adsabs.harvard.edu/abs/2021MNRAS.505..610G} {505, 610}

\bibitem[\protect\citeauthoryear{{Hayes}, {{\"O}stlin}, {Atek}, {Kunth},
  {Mas-Hesse}, {Leitherer}, {Jim{\'e}nez-Bail{\'o}n}  \& {Adamo}}{{Hayes}
  et~al.}{2007}]{hayes2007}
{Hayes} M.,  {{\"O}stlin} G.,  {Atek} H.,  {Kunth} D.,  {Mas-Hesse} J.~M.,
  {Leitherer} C.,  {Jim{\'e}nez-Bail{\'o}n} E.,   {Adamo} A.,  2007, \mn@doi
  [\mnras] {10.1111/j.1365-2966.2007.12482.x}, \href
  {https://ui.adsabs.harvard.edu/abs/2007MNRAS.382.1465H} {382, 1465}

\bibitem[\protect\citeauthoryear{{Henry}, {Scarlata}, {Martin}  \&
  {Erb}}{{Henry} et~al.}{2015}]{Henry2015}
{Henry} A.,  {Scarlata} C.,  {Martin} C.~L.,   {Erb} D.,  2015, \mn@doi [\apj]
  {10.1088/0004-637X/809/1/19}, \href
  {https://ui.adsabs.harvard.edu/abs/2015ApJ...809...19H} {809, 19}

\bibitem[\protect\citeauthoryear{{Kennicutt}}{{Kennicutt}}{1998}]{kennicutt1998}
{Kennicutt} Robert~C. J.,  1998, \mn@doi [\apj] {10.1086/305588}, \href
  {https://ui.adsabs.harvard.edu/abs/1998ApJ...498..541K} {498, 541}

\bibitem[\protect\citeauthoryear{{Kim}, {Kim}, {Ostriker}  \& {Skinner}}{{Kim}
  et~al.}{2017}]{kim2017}
{Kim} J.-G.,  {Kim} W.-T.,  {Ostriker} E.~C.,   {Skinner} M.~A.,  2017, \mn@doi
  [\apj] {10.3847/1538-4357/aa9b80}, \href
  {https://ui.adsabs.harvard.edu/abs/2017ApJ...851...93K} {851, 93}

\bibitem[\protect\citeauthoryear{{Krumholz} et~al.,}{{Krumholz}
  et~al.}{2014}]{krumholz2014}
{Krumholz} M.~R.,  et~al., 2014, in {Beuther} H.,  {Klessen} R.~S.,
  {Dullemond} C.~P.,   {Henning} T.,  eds, Protostars and Planets VI. p.~243
  (\mn@eprint {arXiv} {1401.2473}),
  \mn@doi{10.2458/azu\_uapress\_9780816531240-ch011}

\bibitem[\protect\citeauthoryear{{Kunth}, {Leitherer}, {Mas-Hesse},
  {{\"O}stlin}  \& {Petrosian}}{{Kunth} et~al.}{2003}]{kunth2003}
{Kunth} D.,  {Leitherer} C.,  {Mas-Hesse} J.~M.,  {{\"O}stlin} G.,
  {Petrosian} A.,  2003, \mn@doi [\apj] {10.1086/378396}, \href
  {https://ui.adsabs.harvard.edu/abs/2003ApJ...597..263K} {597, 263}

\bibitem[\protect\citeauthoryear{{Leitet}, {Bergvall}, {Piskunov}  \&
  {Andersson}}{{Leitet} et~al.}{2011}]{leitet2011}
{Leitet} E.,  {Bergvall} N.,  {Piskunov} N.,   {Andersson} B.~G.,  2011,
  \mn@doi [\aap] {10.1051/0004-6361/201015654}, \href
  {https://ui.adsabs.harvard.edu/abs/2011A&A...532A.107L} {532, A107}

\bibitem[\protect\citeauthoryear{{Leitherer} et~al.,}{{Leitherer}
  et~al.}{1999}]{leitherer1999}
{Leitherer} C.,  et~al., 1999, \mn@doi [\apjs] {10.1086/313233}, \href
  {https://ui.adsabs.harvard.edu/abs/1999ApJS..123....3L} {123, 3}

\bibitem[\protect\citeauthoryear{{Leitherer}, {Ekstr{\"o}m}, {Meynet},
  {Schaerer}, {Agienko}  \& {Levesque}}{{Leitherer}
  et~al.}{2014}]{leitherer2014}
{Leitherer} C.,  {Ekstr{\"o}m} S.,  {Meynet} G.,  {Schaerer} D.,  {Agienko}
  K.~B.,   {Levesque} E.~M.,  2014, \mn@doi [\apjs]
  {10.1088/0067-0049/212/1/14}, \href
  {https://ui.adsabs.harvard.edu/abs/2014ApJS..212...14L} {212, 14}

\bibitem[\protect\citeauthoryear{{Luridiana}, {Morisset}  \&
  {Shaw}}{{Luridiana} et~al.}{2015}]{luridiana2015}
{Luridiana} V.,  {Morisset} C.,   {Shaw} R.~A.,  2015, \mn@doi [\aap]
  {10.1051/0004-6361/201323152}, \href
  {https://ui.adsabs.harvard.edu/abs/2015A&A...573A..42L} {573, A42}

\bibitem[\protect\citeauthoryear{{Martin} \& {Bouch{\'e}}}{{Martin} \&
  {Bouch{\'e}}}{2009}]{martin2009}
{Martin} C.~L.,  {Bouch{\'e}} N.,  2009, \mn@doi [\apj]
  {10.1088/0004-637X/703/2/1394}, \href
  {https://ui.adsabs.harvard.edu/abs/2009ApJ...703.1394M} {703, 1394}

\bibitem[\protect\citeauthoryear{{Menacho} et~al.,}{{Menacho}
  et~al.}{2019}]{Menacho+2019}
{Menacho} V.,  et~al., 2019, \mn@doi [\mnras] {10.1093/mnras/stz1414}, \href
  {https://ui.adsabs.harvard.edu/abs/2019MNRAS.487.3183M} {487, 3183}

\bibitem[\protect\citeauthoryear{{Menacho} et~al.,}{{Menacho}
  et~al.}{2021}]{menacho2021}
{Menacho} V.,  et~al., 2021, \mn@doi [\mnras] {10.1093/mnras/stab1491}, \href
  {https://ui.adsabs.harvard.edu/abs/2021MNRAS.506.1777M} {506, 1777}

\bibitem[\protect\citeauthoryear{{Naidu} et~al.,}{{Naidu}
  et~al.}{2021}]{Naidu2021}
{Naidu} R.~P.,  et~al., 2021, \mn@doi [\mnras] {10.1093/mnras/stab3601}, \href
  {https://ui.adsabs.harvard.edu/abs/2021MNRAS.tmp.3287N} {}

\bibitem[\protect\citeauthoryear{{Oey}, {King}  \& {Parker}}{{Oey}
  et~al.}{2004}]{oey2004}
{Oey} M.~S.,  {King} N.~L.,   {Parker} J.~W.,  2004, \mn@doi [\aj]
  {10.1086/381926}, \href
  {https://ui.adsabs.harvard.edu/abs/2004AJ....127.1632O} {127, 1632}

\bibitem[\protect\citeauthoryear{{{\"O}stlin}, {Marquart}, {Cumming}, {Fathi},
  {Bergvall}, {Adamo}, {Amram}  \& {Hayes}}{{{\"O}stlin}
  et~al.}{2015}]{oestlin2015}
{{\"O}stlin} G.,  {Marquart} T.,  {Cumming} R.~J.,  {Fathi} K.,  {Bergvall} N.,
   {Adamo} A.,  {Amram} P.,   {Hayes} M.,  2015, \mn@doi [\aap]
  {10.1051/0004-6361/201323233}, \href
  {https://ui.adsabs.harvard.edu/abs/2015A&A...583A..55O} {583, A55}

\bibitem[\protect\citeauthoryear{{{\"O}stlin} et~al.,}{{{\"O}stlin}
  et~al.}{2021}]{oestlin2021}
{{\"O}stlin} G.,  et~al., 2021, \mn@doi [\apj] {10.3847/1538-4357/abf1e8},
  \href {https://ui.adsabs.harvard.edu/abs/2021ApJ...912..155O} {912, 155}

\bibitem[\protect\citeauthoryear{{Overzier}}{{Overzier}}{2006}]{overzier2006}
{Overzier} R.~A.,  2006, \mn@doi [Astronomische Nachrichten]
  {10.1002/asna.200510507}, \href
  {https://ui.adsabs.harvard.edu/abs/2006AN....327..204O} {327, 204}

\bibitem[\protect\citeauthoryear{{Renaud}, {Bournaud}, {Kraljic}  \&
  {Duc}}{{Renaud} et~al.}{2014}]{Renaud2014}
{Renaud} F.,  {Bournaud} F.,  {Kraljic} K.,   {Duc} P.~A.,  2014, \mn@doi
  [\mnras] {10.1093/mnrasl/slu050}, \href
  {https://ui.adsabs.harvard.edu/abs/2014MNRAS.442L..33R} {442, L33}

\bibitem[\protect\citeauthoryear{{Rivera-Thorsen} et~al.,}{{Rivera-Thorsen}
  et~al.}{2015}]{rivera2015}
{Rivera-Thorsen} T.~E.,  et~al., 2015, \mn@doi [\apj]
  {10.1088/0004-637X/805/1/14}, \href
  {https://ui.adsabs.harvard.edu/abs/2015ApJ...805...14R} {805, 14}

\bibitem[\protect\citeauthoryear{{Rivera-Thorsen}, {{\"O}stlin}, {Hayes}  \&
  {Puschnig}}{{Rivera-Thorsen} et~al.}{2017}]{Emil2017}
{Rivera-Thorsen} T.~E.,  {{\"O}stlin} G.,  {Hayes} M.,   {Puschnig} J.,  2017,
  \mn@doi [\apj] {10.3847/1538-4357/aa5d0a}, \href
  {https://ui.adsabs.harvard.edu/abs/2017ApJ...837...29R} {837, 29}

\bibitem[\protect\citeauthoryear{{Rose}, {Tadhunter}, {Ramos Almeida},
  {Rodr{\'\i}guez Zaur{\'\i}n}, {Santoro}  \& {Spence}}{{Rose}
  et~al.}{2018}]{rose2018}
{Rose} M.,  {Tadhunter} C.,  {Ramos Almeida} C.,  {Rodr{\'\i}guez Zaur{\'\i}n}
  J.,  {Santoro} F.,   {Spence} R.,  2018, \mn@doi [\mnras]
  {10.1093/mnras/stx2590}, \href
  {https://ui.adsabs.harvard.edu/abs/2018MNRAS.474..128R} {474, 128}

\bibitem[\protect\citeauthoryear{{Schlafly} \& {Finkbeiner}}{{Schlafly} \&
  {Finkbeiner}}{2011}]{schlafly2011}
{Schlafly} E.~F.,  {Finkbeiner} D.~P.,  2011, \mn@doi [\apj]
  {10.1088/0004-637X/737/2/103}, \href
  {https://ui.adsabs.harvard.edu/abs/2011ApJ...737..103S} {737, 103}

\bibitem[\protect\citeauthoryear{{Senchyna} et~al.,}{{Senchyna}
  et~al.}{2021}]{Senchyna2021}
{Senchyna} P.,  et~al., 2021, arXiv e-prints, \href
  {https://ui.adsabs.harvard.edu/abs/2021arXiv211111508S} {p. arXiv:2111.11508}

\bibitem[\protect\citeauthoryear{{Strom}, {Steidel}, {Rudie}, {Trainor}  \&
  {Pettini}}{{Strom} et~al.}{2018}]{strom2018}
{Strom} A.~L.,  {Steidel} C.~C.,  {Rudie} G.~C.,  {Trainor} R.~F.,   {Pettini}
  M.,  2018, \mn@doi [\apj] {10.3847/1538-4357/aae1a5}, \href
  {https://ui.adsabs.harvard.edu/abs/2018ApJ...868..117S} {868, 117}

\bibitem[\protect\citeauthoryear{{Vader}, {Frogel}, {Terndrup}  \&
  {Heisler}}{{Vader} et~al.}{1993}]{vader1993}
{Vader} J.~P.,  {Frogel} J.~A.,  {Terndrup} D.~M.,   {Heisler} C.~A.,  1993,
  \mn@doi [\aj] {10.1086/116762}, \href
  {https://ui.adsabs.harvard.edu/abs/1993AJ....106.1743V} {106, 1743}

\bibitem[\protect\citeauthoryear{{Vanzella} et~al.,}{{Vanzella}
  et~al.}{2021}]{Vanzella2021}
{Vanzella} E.,  et~al., 2021, arXiv e-prints, \href
  {https://ui.adsabs.harvard.edu/abs/2021arXiv210610280V} {p. arXiv:2106.10280}

\bibitem[\protect\citeauthoryear{{Whitmore} et~al.,}{{Whitmore}
  et~al.}{2010}]{whitmore2010}
{Whitmore} B.~C.,  et~al., 2010, \mn@doi [\aj] {10.1088/0004-6256/140/1/75},
  \href {https://ui.adsabs.harvard.edu/abs/2010AJ....140...75W} {140, 75}

\bibitem[\protect\citeauthoryear{{Zackrisson}, {Rydberg}, {Schaerer},
  {{\"O}stlin}  \& {Tuli}}{{Zackrisson} et~al.}{2011}]{zackrisson2011}
{Zackrisson} E.,  {Rydberg} C.-E.,  {Schaerer} D.,  {{\"O}stlin} G.,   {Tuli}
  M.,  2011, \mn@doi [\apj] {10.1088/0004-637X/740/1/13}, \href
  {https://ui.adsabs.harvard.edu/abs/2011ApJ...740...13Z} {740, 13}

\makeatother
\end{thebibliography}




\appendix

\section{Analysis of the cluster mass function in Haro 11}

\begin{figure}
	\includegraphics[width=0.96\columnwidth]{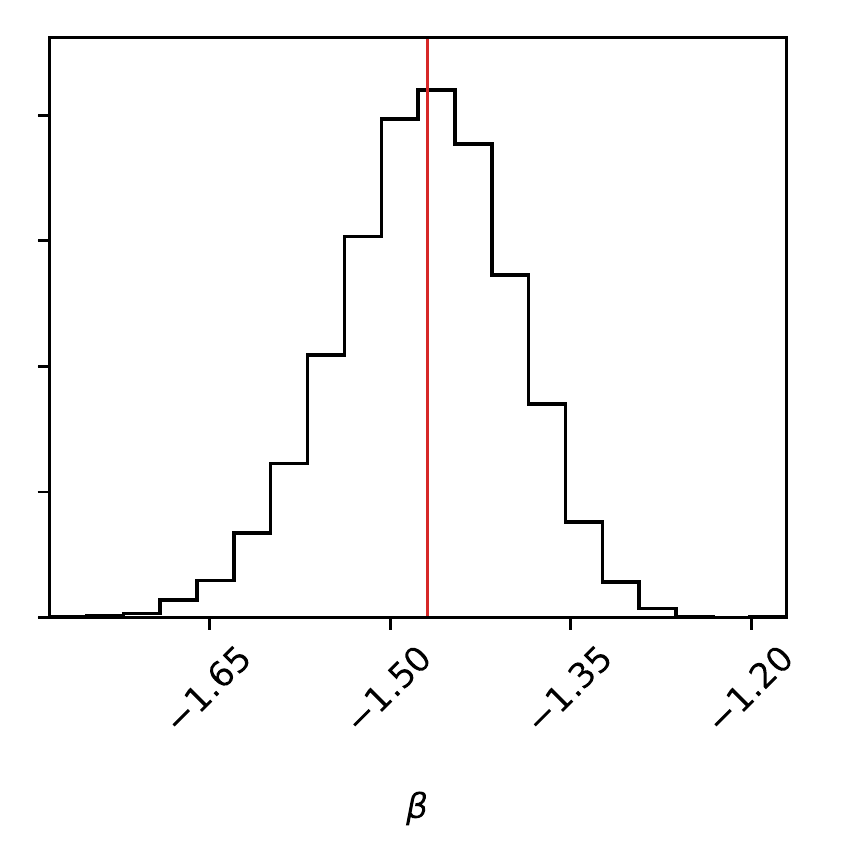}
	\caption{Posterior probability of the power-law slope of the Haro 11 cluster mass function. We used clusters masses above $10^5$ \msun\, and age range 1–20 Myr. The red vertical line marks the median value of the probability distribution.}
    \label{fig:PDF_PL}
\end{figure}

\begin{figure}
	\includegraphics[width=0.96\columnwidth]{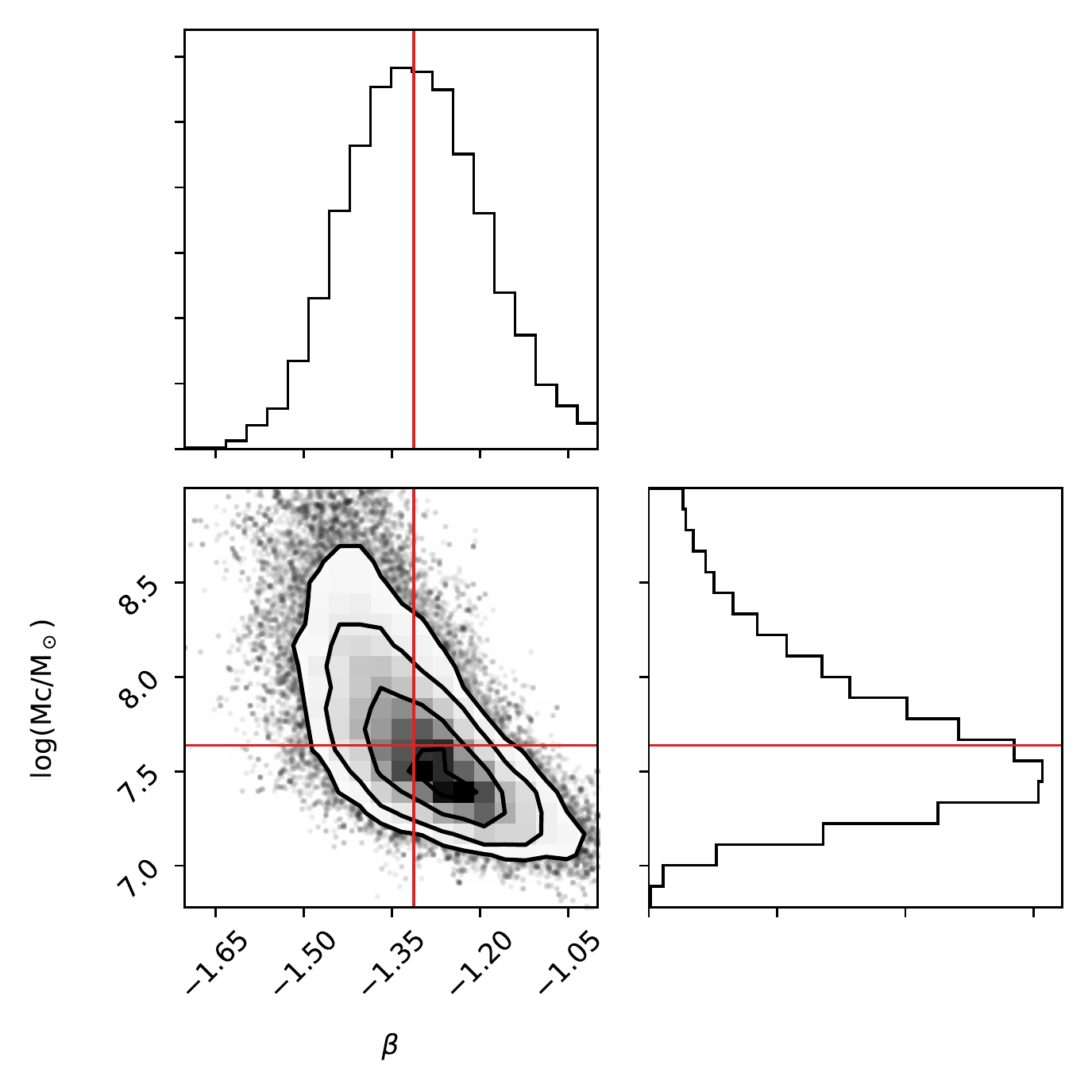}
	\caption{Corner plots of the Bayesian fit with a truncated power-law to describe the cluster mass function of Haro 11. The vertical red line marks the median value of the power-law slope, the horizontal red line marks the median of the truncation mass.}
    \label{fig:PDF_Sch}
\end{figure}

\begin{figure}
	\includegraphics[width=0.96\columnwidth]{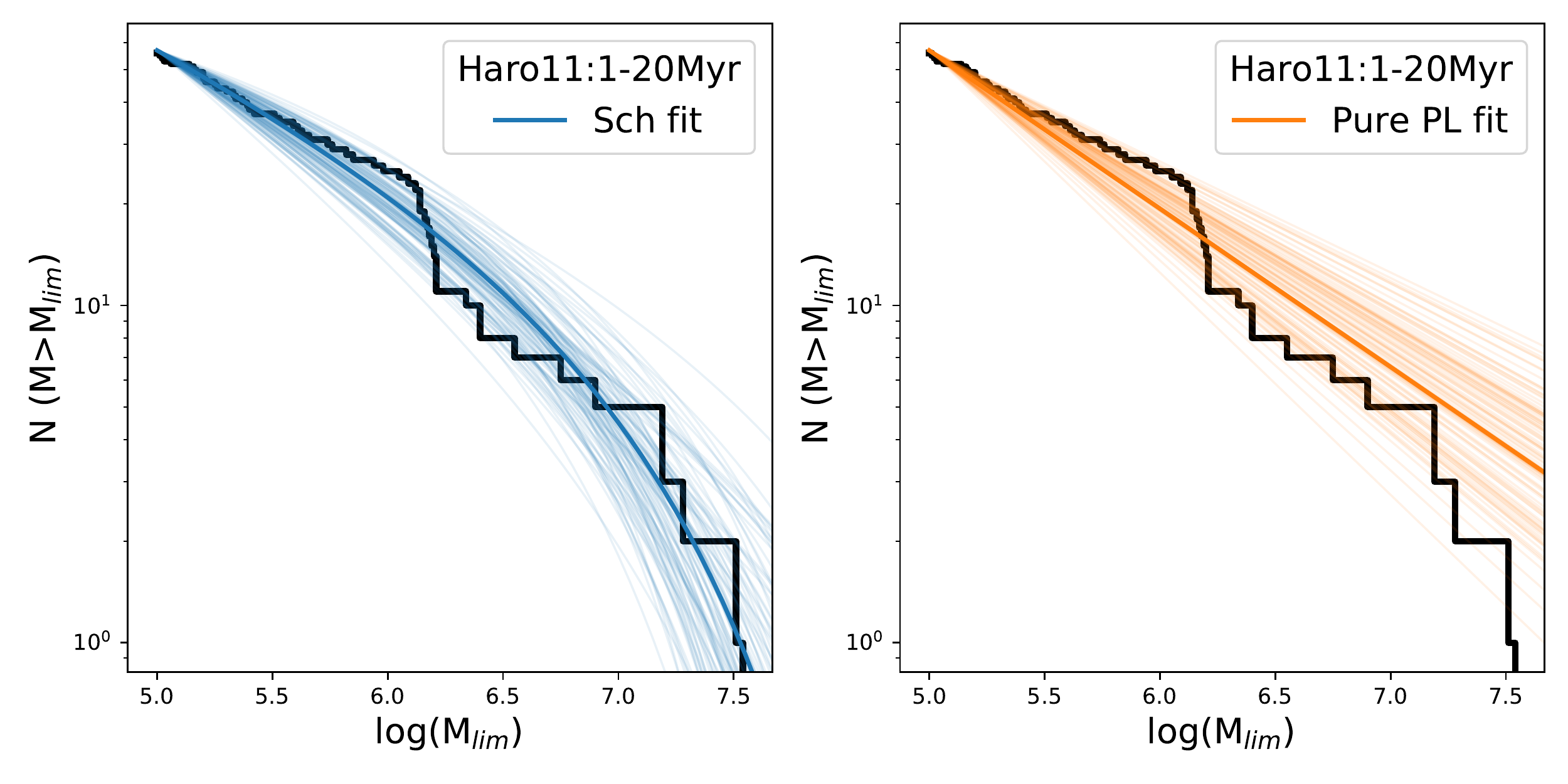}
	\caption{Cumulative mass distribution observed for clusters with masses above the minimum fitted mass ($10^5$ \msun) plotted together with a Schechter function (solid blue line) on the left panel, with a power-law (solid orange line) on the right panel.}
    \label{fig:cumulative_MF}
\end{figure}

We report here the details of the analysis of the cluster mass function of Haro 11. In Figure~\ref{fig:PDF_PL} we show the marginalized posterior probability distribution function of the slope of a single power-law. The median value is indicated by the vertical solid line. In Figure~\ref{fig:PDF_Sch}, we show the results of a similar Bayesian analysis where we try to determine two parameters that describe a Schechter function, i.e., a power-law slope and a truncation mass, which correspond to an exponential decline at the high mass end. The corner plot shows the parameter space visited by the Markov Chain Monte Carlo analysis to fit the cluster mass function. The contours represent the 1$\sigma$ , 2$\sigma$ , and 3$\sigma$ values of the density distribution around the two fitted parameters, M$_c$ and $\beta$. Overall, we observe that there is convergence in the fit, but we notice that we do not find any cluster in the galaxy with mass larger than the derived M$_c$. This points toward a poor constrain of M$_c$. In Figure~\ref{fig:cumulative_MF} we plot the observed cumulative mass distribution of clusters with masses above the minimum fitted mass ($10^5$ \msun). In the same plot we overlay the family of solutions recovered within 1$\sigma$ confidence level (thin solid lines) and their median value (thick solid line) for both the Schechter funtion and power-law function, respectively. We see that the latter functional form overestimate the number of massive clusters at the high-mass end but it can have solutions reproducing the observed shape of the mass function.

 More details about the analysis can be found in the main text.

\section{Modelling of the Lyman-alpha absorption in Haro 11}

\begin{figure}
	\includegraphics[width=0.96\columnwidth]{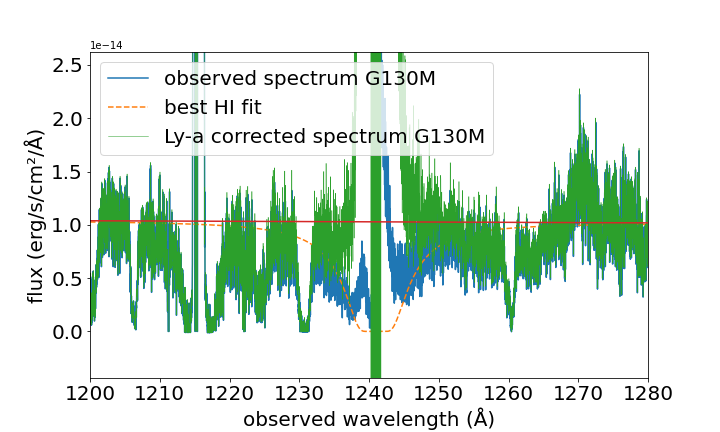}
	
	\includegraphics[width=0.96\columnwidth]{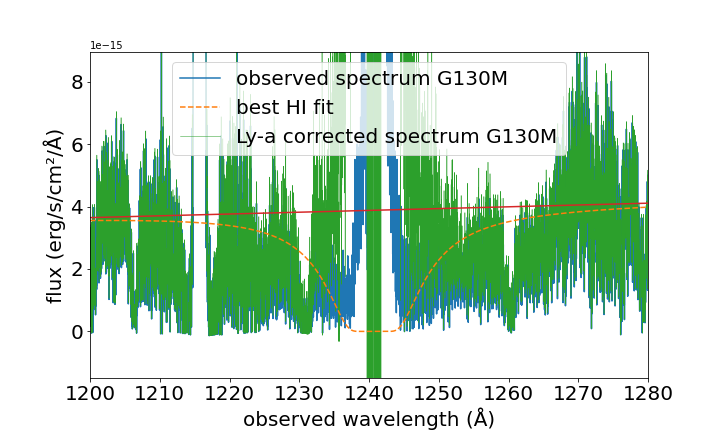}

  	\includegraphics[width=0.96\columnwidth]{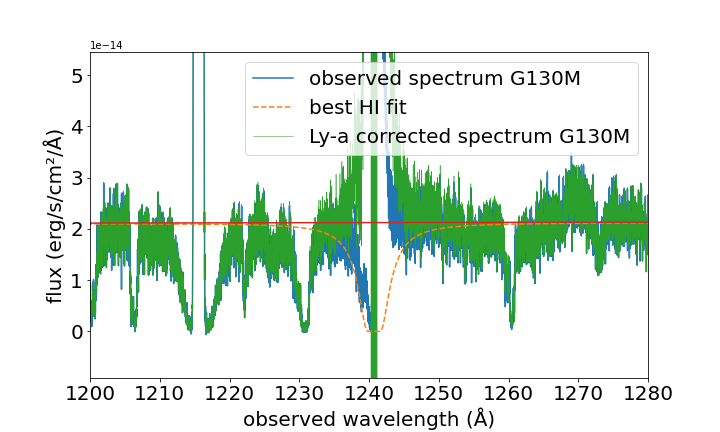}
  
	\caption{Lyman-$\alpha$ absorption fit for the knots A, B and C of Haro 11. The blue line is the observed spectrum, the dashed orange line is the Voigt model, the red line is the continuum level and the green line is the rectified spectrum corrected for the Lyman-$\alpha$ absorption.}
    \label{fig:lyman-A}
\end{figure}

We report here a more detailed illustration of how we modelled the HI absorption to rectify the spectrum before fitting a stellar population model to the FUV and optical spectroscopy. We used a Voigt function to model the absorption wings around the Ly-$\alpha$. We did this with an interactive tool that allows one to select separately spectral regions for evaluating the continuum and regions for evaluating the Voigt profile. We repeated this procedure for each knot and corrected the observed spectrum of each knot according to best Voigt model found. Figure \ref{fig:lyman-A} shows the best-fit model and the spectrum of each knot, before and after the correction.

\bsp	
\label{lastpage}
\end{document}